\documentclass[10pt,a4paper]{article}
\usepackage{jheppub}

\makeatletter
\renewcommand{\@fpheader}{}
\makeatother

\usepackage[utf8]{inputenc}
\usepackage{mathtools}
\usepackage{amsfonts}
\usepackage{mathrsfs}
\usepackage{physics}
\usepackage{slashed}
\usepackage{tensor}
\usepackage{dsfont}
\usepackage{bbm}
\usepackage{amsmath,bm}
\usepackage{tabularx}

\usepackage{graphicx}
\usepackage{color, float}
\usepackage{array}
\usepackage[abs]{overpic}

\usepackage{tikz}
\usetikzlibrary{decorations.pathmorphing}
\usepackage{tikz}
\usetikzlibrary{arrows.meta}
\usetikzlibrary{calc}
\tikzset{propagator/.style={thick}}
\usepackage{xcolor}
\usetikzlibrary{arrows.meta,calc,decorations.markings}

\definecolor{axiscol}{RGB}{45,48,52}
\definecolor{propcol}{RGB}{150,85,60}
\definecolor{odcol}{RGB}{38,94,150}
\definecolor{guidecol}{RGB}{88,94,101}

\usepackage{placeins}
\usepackage{makecell}
\usepackage{subcaption}

\usepackage{xspace}
\usepackage{siunitx}
\usepackage{xfrac}
\usepackage{hyperref}
\usepackage[nameinlink]{cleveref}
\usepackage{appendix}
\usepackage{adjustbox}

\usepackage{xifthen}
\usepackage{xcolor}
\hypersetup{
	colorlinks,
	linkcolor={red!75!black},
	citecolor={blue!75!black},
	urlcolor={blue!75!black}
}

\hypersetup{
	bookmarksopen=true,
	bookmarksopenlevel=2,
	bookmarksnumbered=true
}

\usepackage{amsmath,amssymb,amsfonts}
\usepackage{bm}
\usepackage{booktabs}
\usepackage{hyperref}
\usepackage{array}
\usepackage{tikz}
\usetikzlibrary{decorations.markings,calc}
\newcommand{\ii}{\mathrm i}

\newcommand{\Dbar}{\widebar D}

\newcommand{\intx}{\int \dd^{d+1}x}

\newcommand{\eps}{\epsilon}

\usepackage{comment}

\newcommand{\widebar}[1]{\overline{\mkern-1.4mu #1\mkern-1.4mu}}
\usepackage{xcolor}

\definecolor{EGPink}{HTML}{FF1493}
\definecolor{EGBg}{HTML}{FFE6F2}

\definecolor{ScienceBlue}{HTML}{0C5DA5}
\definecolor{ScienceGreen}{HTML}{00B945}
\definecolor{ScienceOrange}{HTML}{FF9500}
\definecolor{ScienceRedRed}{HTML}{FF2C00}
\definecolor{SciencePurple}{HTML}{845B97}
\definecolor{ScienceDarkGray}{HTML}{474747}
\definecolor{ScienceGray}{HTML}{9E9E9E}

\definecolor{LBPink}{HTML}{00897B} 
\definecolor{LBBg}{HTML}{E0F2F1}

\title{Critical dynamics of a scalar field near four spatial dimensions}

\author[a]{Laura Batini}
\author[b]{and Eduardo Grossi}

\affiliation[a]{
Institut für Theoretische Physik, ETH Zürich,\\
Wolfgang-Pauli-Strasse 27, 8093 Zürich, Switzerland
}

\affiliation[b]{
Dipartimento di Fisica, Università di Firenze and INFN Sezione di Firenze,\\
Via G. Sansone 1, 50019 Sesto Fiorentino, Italy
}

\emailAdd{lbatini@ethz.ch}
\emailAdd{eduardo.grossi@unifi.it}

\abstract{
The critical dynamics of a non-conserved order parameter is generally expected to become overdamped at long distances, even when propagating modes occur at microscopic or intermediate scales. We investigate the critical dynamics of a scalar field theory in thermal equilibrium which, in addition to local friction and noise, also contains a time-dependent second-order kinetic term. We show how to build a supersymmetric field-theory formulation. Using a two-loop expansion about four spatial dimensions, we show that the propagating and strictly overdamped limits share the same static Gaussian and Wilson--Fisher fixed points but realize distinct dynamical scaling regimes. The overdamped limit reproduces Model A. On the surface where local friction and noise vanish, the theory instead supports an interacting propagating fixed point whose dynamic exponent receives corrections at two loops. We demonstrate that coarse-graining does not generate a local dissipative operator on this surface, which therefore remains invariant under the RG flow. Local dissipation is nevertheless relevant at the propagating fixed point: an arbitrarily small equilibrium friction--noise perturbation drives the flow away from propagating scaling. Propagating critical dynamics thus defines a consistent but fine-tuned regime that is unstable to local equilibrium dissipation.

}

\begin{document}

\maketitle
\flushbottom

\section{Introduction}
\label{sec:introduction}
The real-time dynamics of critical phenomena is a central problem across statistical physics and quantum field theory, especially when infrared structures---such as dissipation, conservation laws, or long-range temporal
correlations---compete to determine the dynamical universality class under coarse graining
\cite{HohenbergHalperin1977,FolkMoser2006,Tauber2014,ZengZhong2023}.
Such transitions are ubiquitous in physics: the same framework
governs relaxation in magnets and fluids near their critical points \cite{HohenbergHalperin1977}, in
ultracold and driven--dissipative quantum systems
\cite{SiebererBuchholdDiehl2016,SorienteEtAl2021,KhedriHornZilberberg2022},
in the chiral transition of QCD probed by heavy-ion collisions
\cite{Rajagopal:1992qz,FlorioGrossiSolovievTeaney2022}, and  in the early universe \cite{Berges2015,Moreau:2019jpn}.

Continuous phase transitions are commonly described by Ginzburg--Landau effective models for the order parameter. Dimensionality and symmetry determine the static universality class and, ultimately, the critical exponents.
Their real-time dynamics depends additionally on how long-wavelength fluctuations relax and which variables remain slow. Near criticality, the central quantity is the relaxation timescale of the order parameter
$\tau_{\rm rel}(p)\sim p^{-z}$,
where \(z\) is the dynamic critical exponent. 
The value of \(z\) is not fixed by the static
universality class alone, but depends on the relevant slow variables, their
conservation laws, and the low-frequency structures allowed in the effective theory.

The standard framework is the Hohenberg--Halperin classification of dynamical critical phenomena
\cite{HohenbergHalperin1977,FolkMoser2006,Tauber2014}. For a non-conserved scalar order parameter with no coupling to additional conserved densities,
the equilibrium infrared dynamics in the presence of local dissipation is described by Model~A. Its coarse-grained evolution is relaxational and first order in time \cite{BauschJanssenWagner1976}, and the dynamic scaling exponent
\(z=2\). 
For some models, the underlying microscopic theory is a relativistic quantum field theory with $z=1$.  
The physical question is whether propagating dynamics can survive and eventually transform into the relaxational dynamics dictated by the critical infrared regime.

This competition between propagation and  relaxation has been approached from several complementary directions. An renormalization-group (RG) analysis of the $O(N)$ Ginzburg--Landau model with the canonical momentum retained as a dynamical variable found both propagating and diffusive fixed points, with the former unstable to dissipative perturbations \cite{OhnishiKunihiro2006}. A finite-temperature relativistic scalar theory likewise exhibits an interacting propagating scaling regime whose exponent remains close to the ballistic value \cite{BoyanovskyDeVega2002, SchweitzerSchlichtingVonSmekal2020}. 

These questions become especially important when a system crosses or passes near a critical region within a finite time, as occurs in settings ranging from ultracold atomic gases and cosmological phase transitions to heavy-ion collisions. Under a quench, the driving timescale must be compared with the critical relaxation time $\tau_{\rm rel}$, which grows upon approaching criticality. Once relaxation can no longer keep pace with the evolution of the system, the dynamics falls out of equilibrium, as described by the Kibble--Zurek mechanism \cite{Kibble:1976sj,Kibble:1980mv,Zurek:1985qw}. This loss of adiabaticity limits the growth of long-wavelength correlations and causes fluctuation observables to lag behind their equilibrium values.

In heavy-ion physics, identifying which dynamical fixed point controls this relaxation is essential for translating equilibrium critical behavior into experimentally accessible signatures of the chiral transition or a possible QCD critical endpoint \cite{Son:2004iv}. Recent numerical developments have made it possible to investigate critical
real-time dynamics directly in stochastic field theories and fluctuating
hydrodynamics. Simulations of Models A, B, F, and H have explored critical slowing
down, finite-rate evolution, propagating collective modes, and the scaling of
transport coefficients in systems with conserved densities and reversible
mode couplings
\cite{
Chattopadhyay:2023jfm,
Chattopadhyay:2024jlh,
Chattopadhyay:2025uqo, Chattopadhyay:2025zac,
Chattopadhyay:2026dyd, Sieke:2026ozy}.

Related numerical studies of the $O(4)$ chiral transition in QCD have
investigated Model G dynamics both in and out of equilibrium
\cite{FlorioGrossiSolovievTeaney2022,
FlorioGrossiTeaney2024,
FlorioEtAl2025Quenching,
FlorioEtAl2025Goldstones}.
In equilibrium, these studies exhibit the characteristic interplay between
propagating pion modes and the diffusion of conserved chiral charges, with
dynamical scaling consistent with the Model G exponent $z=3/2$. Quenches
from the symmetric to the broken phase further reveal a period of nonlinear
growth and a long-lived enhancement of soft Goldstone fluctuations. Together,
these results illustrate that propagation, diffusion, and relaxation may
coexist near a critical point, and that their relative importance is
determined by the conserved variables and reversible mode couplings of the
dynamical universality class.

Complementary progress has been made using the real-time functional
renormalization group (FRG) \cite{Dupuis:2020fhh, RothVonSmekal2023}. A closed-time-path FRG formulation has been developed
for several dynamical universality classes and used to calculate critical
spectral functions and dynamical exponents in Models A, B, and C
\cite{Batini:2023nan, RothVonSmekal2023}. More recently, this framework has been extended to
systems with reversible mode couplings. Its application to Model G reproduces
the strong-scaling relation $z=d/2$ and yields the universal scaling function
of the conserved-charge diffusion coefficient
\cite{RothEtAl2025ModelG}. A parallel analysis of Models G and H has also
clarified similarities and differences between the dynamical fixed points
relevant to the two-flavor chiral transition and the finite-density QCD
critical point
\cite{RothEtAl2025ModelsGH}.

A broader body of work addresses how dissipation arises in effective descriptions. Damping and stochastic dynamics can emerge through scattering, resummation, or coarse graining in relativistic field theories \cite{SaitoFujiiItakuraMorimatsu2015,BorsanyiEtAl2000,BatiniChatrchyanBerges2024}. Related RG reductions of kinetic theory show how irreversible kinetic and hydrodynamic equations arise after projection onto slow variables \cite{HattaKunihiro,TsumuraKunihiroOhnishi}. More generally, Schwinger--Keldysh and real-time functional approaches provide a systematic language for encoding dissipation, noise, and spectral information in effective actions \cite{kamenev2005manybodytheorynonequilibriumsystems,SiebererBuchholdDiehl2016,Floerchinger_2012,Floerchinger_2016,Braun:2022mgx,Frangi:2025xss,StoetzelFloerchinger2025,Canet:2011wf, Floerchinger:2026pwi}. Studies of driven open systems further demonstrate that dissipation can reorganize excitation spectra and produce underdamped-to-overdamped transitions, although such systems need not obey the equilibrium fluctuation--dissipation relation \cite{SorienteEtAl2021,KhedriHornZilberberg2022}.

These developments leave an important distinction to be made. If a propagating theory is exactly dissipationless, does coarse graining generate a local dissipative operator? And if such an operator is already present, however small, is it relevant at the propagating fixed point? The first question asks whether the dissipationless theory is an invariant subspace of the RG flow; the second asks whether that subspace is stable. These questions are logically separate, but are often conflated in discussions of overdamping near criticality.

To address these questions, in this work we consider a minimal scalar dynamical theory in which both propagation and dissipation are allowed. The real scalar field
\(\phi(t,\bm{x})\) represents the critical order-parameter fluctuation, whose static properties are governed by the usual \(\mathbb{Z}_2\)-symmetric Ginzburg--Landau functional \(\mathcal H[\phi]\). Its dynamics is described by
\begin{equation}
    \frac{1}{c^2}\partial_t^2\phi
    +
    X\partial_t\phi
    =
    -\frac{\delta \mathcal H}{\delta\phi}
    +
    \xi ,
    \label{eq:intro_damped_langevin}
\end{equation}
where \(c^{-2}\) multiplies the inertial operator and \(X\) controls local friction. Thermal equilibrium requires
\begin{equation}
    \bigl\langle
        \xi(\bm{x},t)\xi(\bm{x}',t')
    \bigr\rangle
    =
    2XT\,
    \delta^{(d)}(\bm{x}-\bm{x}')\delta(t-t').
    \label{eq:noise_intro}
\end{equation}
Consequently, \(X\) characterizes the entire equilibrium friction-noise sector. The limits $X=0$ and $c^{-2}=0$ describe, respectively, dissipationless propagation and strictly overdamped Model~A dynamics.

Rather than deriving $X$ from a particular microscopic theory, we determine its RG role once it is included among the operators allowed by equilibrium \cite{AronBiroliCugliandolo2010,Marguet_2021,Gao:2017bqf,Canet:2011wf}. We use the supersymmetric formulation \cite{MartinSiggiaRose1973,Janssen1976,DeDominicis1976,Parisi:1979ka,Kurchan1992}, in which inertia and dissipation remain distinct while the fluctuation--dissipation relation is built into the action. First, in Section~\ref{sec:general_setup}, we present the associated equilibrium, Becchi–Rouet–Stora–Tyutin (BRST) \cite{Tyutin:1975qk, DeDominicis:1976, HaehlLoganayagamRangamani2017}, and supersymmetric structures that constrain the effective theory.
Section~\ref{sec:effective_action} develops the
effective action, superpropagators, and the two Gaussian limits, and introduces the relevant two-loop diagrams.
After reviewing, in Section~\ref{sec:static},  the
common static Gaussian and Wilson--Fisher sectors, 
we analyze both dynamical limits in Sections~\ref{sec:overdamped} and \ref{sec:dissipationless}. 
Respectively, they present the overdamped Model A and dissipationless propagating regimes.
Our main result, presented in Section~\ref{sec:X-perturbation}, is that the exactly dissipationless surface $X=0$ is perturbatively invariant, but not infrared stable: any nonzero $X$ is a relevant perturbation and drives the flow away from the propagating fixed point. 
 Section \ref{sec:fixed-point-summary} combines the results in the four-fixed-point phase diagram. Finally, we conclude in Sec.~\ref{sec:conclusions}.

\section{General setup}
\label{sec:general_setup}
In this section, we establish the general framework used throughout the paper. In Subsec.~\ref{sub:partition function}, we construct the dynamical partition function associated with the stochastic equations introduced above. In Subsecs.~\ref{KMSsymmetry} and \ref{subsec:symmetries}, we discuss the role of the equilibrium boundary conditions and present the Kubo–Martin–Schwinger (KMS) \cite{Kubo:1957mj, Martin:1959jp}, ghost-number, and BRST symmetries of the theory. In Subsec.~\ref{subsec:24}, we organize these symmetries into a compact superspace formulation and derive the corresponding Ward identities. Finally, in Subsec.~\ref{eq:Subsection canonical dim}, we determine the canonical dimensions and identify the Gaussian scaling regimes that provide the starting point for the renormalization-group analysis.

\subsection{Partition function}
\label{sub:partition function}
The static properties of the non-conserved scalar order parameter
$\phi(t,\bm{x})$ are governed by the Landau--Ginzburg functional
\begin{equation}
    \mathcal H[\phi]
    =
    \int \dd^d x\,
    \left[
        \frac{1}{2}(\boldsymbol{\nabla}\phi)^2
        +U(\phi)
    \right].
    \label{eq:H}
\end{equation}
Thus
\(
-\delta \mathcal H/\delta\phi=\boldsymbol{\nabla}^2\phi-U'(\phi)
\).
It is convenient to write the stochastic equation as
\begin{equation}
    \mathcal E_X[\phi]=\xi,
    \qquad
    \mathcal E_X[\phi]
    \equiv
    \frac{1}{c^2}\partial_t^2\phi
    +X\partial_t\phi
    -\boldsymbol{\nabla}^2\phi
    +U'(\phi).
    \label{eq:EX}
\end{equation}
Separating the dissipationless part, we define 
\begin{equation}
    \mathcal E_0[\phi]
    \equiv
    \frac{1}{c^2}\partial_t^2\phi
    -\boldsymbol{\nabla}^2\phi
    +U'(\phi).
    \label{eq:E0}
\end{equation}
The equilibrium noise covariance is given in
Eq.~\eqref{eq:noise_intro}. We keep the temperature $T$ explicit in this
section to display the fluctuation--dissipation relation and the equilibrium
initial ensemble; it will subsequently be set to unity.

    One must also specify the initial
	field and initial velocity,
	\begin{equation}
		\phi_I
		\equiv
		\phi(t_I,\bm x),
		\qquad
		\dot\phi_I
		\equiv
		\partial_t\phi(t_I,\bm x).
		\label{eq:initial_data_MSRJD}
	\end{equation}
	We denote their probability distribution by
	\(P_I[\phi_I,\dot\phi_I]\). For fixed initial data, the notation
	\(\int_{\phi_I,\dot\phi_I}{\cal D}\phi\) denotes an integral over field
	histories satisfying Eq.~\eqref{eq:initial_data_MSRJD}. The initial data are
	integrated independently only after this conditional history integral has
	been defined. For \(X>0\), the covariance in
	Eq.~\eqref{eq:noise_intro} corresponds to the Gaussian probability
	functional
	\begin{equation}
		P_\xi[\xi]
		=
		{\cal N}_\xi
		\exp\left[
		-\frac{1}{4XT}
		\int_{t_I}^{t_F}\dd t\,\dd^d x\,
		\xi^2(t,\bm x)
		\right],
		\label{eq:noise_probability_MSRJD}
	\end{equation}
	where \({\cal N}_\xi\) is fixed by
	\(\int{\cal D}\xi\,P_\xi[\xi]=1\). In the strict dissipationless limit,
	\(X=0\), the covariance vanishes and the noise measure reduces to
	\begin{equation}
		P_\xi[\xi]=\delta[\xi].
		\label{eq:noise_measure_X_zero}
	\end{equation}
	
	At fixed \((\phi_I,\dot\phi_I)\), the Langevin equation defines a change of
	variables from the noise history \(\xi\) to the field history \(\phi\).
	The corresponding functional Jacobian is
	\begin{equation}
		J[\phi]
		=
		\det M_X[\phi],
		\qquad
		M_X[\phi](t, \bm x;t', \bm y)
		\equiv
		\frac{\delta{\cal E}_X[\phi](t, \bm x)}
		{\delta\phi(t', \bm y)}.
		\label{eq:Jacobian_MSRJD}
	\end{equation}
	In particular, for the present scalar theory,
	\begin{equation}
		\begin{aligned}
			M_X[\phi](t, \bm x;t', \bm y)
			&=
			\left[
			\frac{1}{c^2}\partial_t^2
			+
			X\partial_t
			-
			\boldsymbol{\nabla}^2
			+
			U''\bigl(\phi(t,\bm x)\bigr)
			\right]
			\delta(t-t')\,
			\delta^{(d)}(\bm x-\bm y),
		\end{aligned}
		\label{eq:Jacobian_operator_MSRJD}
	\end{equation}
	where the derivatives in Eq.~\eqref{eq:Jacobian_operator_MSRJD} act on the
	\((t,\bm x)\) arguments. The determinant is defined with the causal
	prescription associated with the fixed initial data. With this prescription,
	the change of variables may be written as
	\begin{equation}
		1
		=
		\int_{\phi_I,\dot\phi_I}{\cal D}\phi\,
		\delta[{\cal E}_X[\phi]-\xi]\,
		J[\phi].
		\label{eq:functional_identity_MSRJD}
	\end{equation}
	Equivalently, the conditional probability of a field history is
	\begin{equation}
		P_{\rm hist}[\phi\mid\phi_I,\dot\phi_I]
		=
		P_\xi[{\cal E}_X[\phi]]\,J[\phi].
		\label{eq:history_probability_MSRJD}
	\end{equation}
	For \(X>0\), this gives the Onsager--Machlup representation
	\begin{equation}
		P_{\rm hist}[\phi\mid\phi_I,\dot\phi_I]
		=
		{\cal N}_\xi
		\exp\left[
		-\frac{1}{4XT}
		\int_{t_I}^{t_F}\dd t\,\dd^d x\,
		{\cal E}_X[\phi]^2
		\right]
		J[\phi].
		\label{eq:Onsager_Machlup_MSRJD}
	\end{equation}
	This form is useful conceptually, but the response-field representation is
	more convenient for perturbation theory and for the supersymmetric
	formulation, which will be introduced later on. We exponentiate the functional constraint by introducing the response field
	\(\phi_a\):
	\begin{equation}
		\delta[{\cal E}_X[\phi]-\xi]
		\propto
		\int_{\cal C}{\cal D}\phi_a\,
		\exp\left[
		-\int_{t_I}^{t_F}\dd t\,\dd^d x\,
		\phi_a
		\bigl(
		{\cal E}_X[\phi]-\xi
		\bigr)
		\right].
		\label{eq:response_field_representation}
	\end{equation}
	Here \({\cal C}\) is the standard MSRJD response-field contour, parallel to
	the imaginary axis. Overall field-independent normalization factors are left
	implicit. The causal Jacobian is represented by Grassmann fields \(c\) and \(\bar c\):
	\begin{equation}
		J[\phi]
		=
		\int{\cal D}\bar c\,{\cal D}c\,
		\exp\left[
		\int_{t_I}^{t_F}\dd t\,\dd^d x\,
		 c\,M_X[\phi] \bar c
		\right],
		\label{eq:ghost_representation_MSRJD}
	\end{equation}
	with the ghost and antighost fields obeying the
	initial- and final-time conditions
	\begin{equation}
		\bar c(t_I)=\partial_t{\bar c}(t_I)=0,
		\qquad
		 c(t_F)=\partial_t{ c}(t_F)=0.
		\label{eq:ghost_boundary_conditions_MSRJD}
	\end{equation}
	These conditions implement the causal prescription in
	Eq.~\eqref{eq:Jacobian_MSRJD} and remove the temporal surface terms that
	arise when the ghost operator is integrated by parts.
	
	For \(X>0\), the remaining noise integral is Gaussian:
	\begin{equation}
		\begin{aligned}
			&\int{\cal D}\xi\,
			\exp\left[
			-\frac{1}{4XT}
			\int\dd t\,\dd^d x\,\xi^2
			+
			\int\dd t\,\dd^d x\,\phi_a\xi
			\right]
			\propto \exp\left[
			XT
			\int\dd t\,\dd^d x\,\phi_a^2
			\right],
		\end{aligned}
		\label{eq:noise_Gaussian_integral_MSRJD}
	\end{equation}
	because
	\begin{equation}
		-\frac{\xi^2}{4XT}+\phi_a\xi
		=
		-\frac{(\xi-2XT\phi_a)^2}{4XT}
		+
		XT\phi_a^2.
		\label{eq:noise_completing_square_MSRJD}
	\end{equation}
	The proportionality factor in
	Eq.~\eqref{eq:noise_Gaussian_integral_MSRJD} is independent of all dynamical
	fields and is absorbed into the normalization of the generating functional.
	For \(X=0\), the same result follows directly from
	Eq.~\eqref{eq:noise_measure_X_zero}: the noise integral imposes \(\xi=0\),
	and the term proportional to \(\phi_a^2\) is absent.
	
	Combining these ingredients gives the source-free MSRJD generating
	functional
	\begin{equation}
		\begin{aligned}
			Z
			&=
			\int
			{\cal D}\phi_I\,
			{\cal D}\dot\phi_I\,
			\int_{\phi_I,\dot\phi_I}
			{\cal D}\phi\,
			{\cal D}\phi_a\,
			{\cal D}\bar c\,
			{\cal D}c\,
			e^{-S-S_I},
		\end{aligned}
		\label{eq:MSRJD_generating_functional}
	\end{equation}
	where
\begin{equation}
    S_I[\phi_I,\dot\phi_I]
		\equiv
		-\log P_I[\phi_I,\dot\phi_I],
        \label{eq:SI}
\end{equation}
	and the bulk action is
	\begin{equation}
		S
		=
		S_{\rm det}
		+
		S_{\rm diss}
		+
		S_{\rm ghost},
		\label{eq:MSRJD_action_decomposition}
	\end{equation}
	with
	\begin{equation}
		\begin{aligned}
			S_{\rm det}
			&\equiv
			\int_{t_I}^{t_F}\dd t\,\dd^d x\,
			\phi_a\,{\cal E}_0[\phi],
			\\
			S_{\rm diss}
			&\equiv
			X
			\int_{t_I}^{t_F}\dd t\,\dd^d x\,
			\phi_a
			\left(
			\partial_t\phi-T\phi_a
			\right),
			\\
			S_{\rm ghost}
			&\equiv
			-\int_{t_I}^{t_F}\dd t\,\dd^d x\,
			 c\,M_X[\phi] \bar c.
		\end{aligned}
		\label{eq:MSRJD_action_terms}
	\end{equation}

\subsection{Boundary term and equilibrium KMS symmetry} \label{KMSsymmetry}

The initial distribution plays an essential role in the equilibrium
time-reversal symmetry of the finite-time MSRJD functional.  
 On a general
interval \(t\in[t_I,t_F]\), time reversal is implemented as reflection about
the midpoint,
\begin{equation}
	\bar t \equiv t_I+t_F-t.
	\label{eq:time_reflection}
\end{equation}
The KMS transformation acts on the physical and response fields as
\begin{equation}
\begin{aligned}
\Theta\phi(t, \bm x)&=\phi(\bar t, \bm x),
\label{eq:KMS_trafo}
\\
\Theta \phi_a(t, \bm x)
&=
\phi_a(\bar t, \bm x)
+
\beta\partial_t\phi(\bar t, \bm x)=
\phi_a(\bar t, \bm x)
-
\beta\partial_{\bar t}\phi(\bar t, \bm x),
\end{aligned}
\end{equation}
where \(\beta=T^{-1}\). 
Applying this transformation to the deterministic contribution ${S}_{\rm det}$ gives
\begin{align}
\Theta{ S}_{\rm det}
=
S_{\rm det}
-
\beta
\int_{t_I}^{t_F}\dd t\,\dd^dx\,
\dot\phi(\bar t, \bm x)\mathcal E_0[\phi](\bar t, \bm x).
\end{align}
The last term  is a total derivative:
\begin{align}
\int_{t_I}^{t_F}\dd t\,\dd^dx\,
\dot\phi(\bar t )\,\mathcal{E}_0[\phi(\bar t )]
=
\int_{t_I}^{t_F}\dd \bar t\,\frac{\dd { E_{\rm tot}}[\phi,\dot \phi]}{\dd \bar t}
=
{ E_{\rm tot}}[\phi_F,\dot\phi_F]
-
{ E_{\rm tot}}[\phi_{I},\dot\phi_{I}],
\end{align}
  where we have defined the total energy
\begin{equation}
{E_{\rm tot}}[\phi,\dot\phi]
=
\int \dd^dx
\left[
\frac{\dot\phi^2}{2c^2}
+
\frac12(\boldsymbol{\nabla}\phi)^2
+
U(\phi)
\right]. 
\label{eq:energy_functional}
\end{equation}
Therefore
\begin{align}
\Theta{ S}_{\rm det}
-
{ S}_{\rm det}
=
&
-
\beta
({ E_{\rm tot}}[\phi_F,\dot\phi_F]
-
{ E_{\rm tot}}[\phi_{I},\dot\phi_{I}]).
\label{eq:Sdet_boundary_variation}
\end{align}
Under time reversal, the initial field is mapped to the final field, while
the velocity changes sign:
\begin{equation}
	(\Theta\phi)(t_I)=\phi(t_F),
	\qquad
	\partial_t(\Theta\phi)(t_I)=-\dot\phi(t_F).
\end{equation}
It follows that
\begin{equation}
   \Theta S_I -S_I= -\log
\frac{
P_I[\phi_F,-\dot\phi_F]
}{
P_I[\phi_{I},\dot\phi_{I}]
}
\label{eq:initial_action_transformation}
\end{equation} 
For a generic initial distribution,
Eq.~\eqref{eq:initial_action_transformation} does not cancel the boundary
term in Eq.~\eqref{eq:Sdet_boundary_variation}; the finite-time generating
functional is then not invariant under dynamical KMS time reversal. In equilibrium, the initial fields are instead sampled from the Gibbs
distribution
\begin{equation}
P_I[\phi,\dot\phi]=P_{\rm eq}=Z_{\rm eq}^{-1}\exp[-\beta{ E_{\rm tot}[\phi,\dot\phi]}].
\end{equation}
Since the mechanical energy is even under velocity reversal,
$(
	E_{\rm tot}[\phi,-\dot\phi]
	=
	E_{\rm tot}[\phi,\dot\phi]
)$, 
Eq.~\eqref{eq:initial_action_transformation} becomes
\begin{equation}
	\Theta S_I-S_I
	=
	\beta ({ E_{\rm tot}}[\phi_F,\dot\phi_F]
-
{ E_{\rm tot}}[\phi_{I},\dot\phi_{I}]).
	\label{eq:SI_boundary_variation}
\end{equation}
The variations in Eqs.~\eqref{eq:Sdet_boundary_variation} and
\eqref{eq:SI_boundary_variation} cancel, and hence
\begin{equation}
	\Theta\left(S_{\rm det}+S_I\right)
	=
	S_{\rm det}+S_I.
	\label{eq:KMS_deterministic_boundary_invariance}
\end{equation}
Thus, on a finite time interval, \(S_{\rm det}\) is not invariant by itself. Its variation is precisely cancelled by the Gibbs boundary
contribution.

 We next consider the dissipative part of the action $S_{\rm diss}$. It is also invariant:  
\begin{align}
\Theta{ S}_{\rm diss}
=&
X\int \dd \bar t\,\dd^dx\,
\left(\phi_a-\beta\dot\phi\right)
\left[
-T\left(\phi_a-\beta\dot\phi\right)-\dot\phi
\right]
\nonumber\\
=
&
X\int \dd \bar t\,\dd^dx\,
\left(\phi_a-\beta\dot\phi\right)
(-T \phi_a)
\nonumber\\
=
&
X\int \dd \bar t\,\dd^dx\,
\phi_a
\left(
\dot\phi-T \phi_a
\right)
=
{S}_{\rm diss}.
\label{eq:Sdiss_KMS_invariance}
\end{align}
Here all fields in the intermediate expressions are evaluated at
\((\bar t,\bm x)\). 
The invariance of \(S_{\rm diss}\) therefore relies on the
fluctuation--dissipation relation: the same coefficient \(X\) must determine
both friction and noise.

The fermionic sector must be transformed at the same time as
\begin{equation}
\Theta c(t, \bm x)=\bar c(\bar t, \bm x),
\qquad
\Theta\bar c(t, \bm x)=-c(\bar t, \bm x).
\end{equation}
 Since $\partial_t\to-\partial_{\bar t}$,
the operator \(M_X\) is mapped to
\begin{equation}
M_X^{\rm R}[\phi]
=
\frac{1}{c^2}\partial_t^2
-
X\partial_t
-
\boldsymbol{\nabla}^2
+
U''(\phi).
\end{equation}
It follows that
\begin{equation}
\Theta S_{\rm ghost}
=
S_{\rm ghost}.
\label{eq:Sghost_KMS_invariance}
\end{equation}
Finally, the KMS transformation is linear and has a field-independent determinant \cite{AronBiroliCugliandolo2010},
which can be absorbed into the normalization of the path integral. The
functional measure is therefore also invariant. Combining
Eqs.~\eqref{eq:KMS_deterministic_boundary_invariance},
\eqref{eq:Sdiss_KMS_invariance}, and
\eqref{eq:Sghost_KMS_invariance}, we conclude that the complete equilibrium
action
\begin{equation}
	S_{\rm eq}
	\equiv
	S_{\rm det}
	+
	S_{\rm diss}
	+
	S_{\rm ghost}
	+
	S_I
	\label{eq:complete_equilibrium_action}
\end{equation}
obeys
$
	\Theta S_{\rm eq}=S_{\rm eq}.
$
Dynamical KMS symmetry is therefore a property of the complete finite-time
equilibrium functional. It requires a conservative, time-independent
deterministic force, the fluctuation--dissipation relation between friction
and noise, and a Gibbs initial ensemble at the same temperature. If any of
these conditions is violated, the finite-time KMS symmetry is broken.

The argument also includes the dissipationless limit. When the dissipative
coefficient is set to zero, both damping and noise disappear and
\(S_{\rm diss}=0\). Nevertheless, the deterministic bulk action still
produces the boundary term in
Eq.~\eqref{eq:Sdet_boundary_variation}, so the Gibbs boundary action remains
necessary for the finite-time equilibrium functional to be KMS invariant.

In the rest of the paper we will consider the limit $t_I\to -\infty,t_F\to +\infty $.

\subsection{Symmetries of the dynamical action}
\label{subsec:symmetries}
In this subsection, we collect the symmetries that constrain the dynamical action. Ghost-number and BRST symmetries follow from the functional representation of the stochastic dynamics and hold independently of equilibrium. When the fluctuation--dissipation relation and the Gibbs initial ensemble are imposed, dynamical KMS invariance provides an additional equilibrium symmetry. Conjugating the BRST transformation by dynamical KMS then produces a second nilpotent symmetry; together, the two nilpotent transformations generate the supersymmetry algebra developed in Subsec.~\ref{subsec:24}.

\subsubsection{Ghost-number symmetry}

The ghost fields \(c\) and \(\bar c\) enter \(S_{\rm ghost}\) only through
the bilinear combination \(c M_X[\phi] \bar c\). The action is therefore
invariant under the global transformation
\begin{equation}
    c\longrightarrow e^\alpha c,
    \qquad
    \bar c\longrightarrow e^{-\alpha}\bar c,
    \qquad
    \phi\longrightarrow\phi,
    \qquad
    \phi_a\longrightarrow\phi_a,
    \label{eq:ghost_number}
\end{equation}
where \(\alpha\) is a constant Grassmann-even parameter. The opposite
rescalings also leave the Grassmann measure
\(\mathcal D\bar c\,\mathcal Dc\) invariant. This symmetry defines the
assignments
\(\operatorname{gh}(c)=+1\),
\(\operatorname{gh}(\bar c)=-1\), and zero ghost number for the bosonic
fields. It is consequently inherited by the effective action: perturbative
and RG corrections can generate only operators with vanishing total ghost
number, provided the regularization preserves the symmetry. Because the
argument relies solely on the determinant structure of the Jacobian,
ghost-number symmetry holds independently of whether the dynamics is
dissipative or dissipationless.

\subsubsection{BRST symmetry}
The bulk action $S$ possesses a nilpotent BRST symmetry
\cite{10.1093/oso/9780198834625.001.0001,Marguet_2021}, that expresses the consistency between the functional constraint
that imposes the Langevin equation and the ghost determinant representing
its Jacobian. 
It is a structural property of the MSRJD construction and
does not require equilibrium.  Its action on the component fields is defined by 
\begin{equation} 
\delta_{\epsilon}\phi =\epsilon\bar c, \qquad \delta_{\epsilon} \bar c =0, \qquad \delta_{\epsilon} c=\epsilon\phi_a, \qquad
\delta_{\epsilon} \phi_a=0. \label{eq:BRST-component-transformations} 
\end{equation}
This transformation is nilpotent,  $\delta_\eps^2 =0$. 
Using this definition we can write 
\begin{equation}
\delta_{\epsilon}{\cal E}_X[\phi]
=
M_X[\phi] \epsilon\bar c.
\label{eq:sEX}
\end{equation}
The variation of the bosonic constraint is cancelled by the variation of the
ghost action, so this implies
$
    \delta_{\epsilon}S
    =0,
$  which proves the invariance.

\subsubsection{Equilibrium dynamical KMS symmetry}

As established in Subsec.~\ref{KMSsymmetry}, when the fluctuation--dissipation relation
holds and the initial fields are sampled from the Gibbs ensemble, the
complete finite-time action is invariant under the dynamical KMS
transformation $\Theta$. This symmetry is the
field-theoretic expression of equilibrium detailed balance, and is the
stochastic counterpart of the dynamical KMS symmetry of Schwinger--Keldysh
effective actions \cite{GloriosoCrossleyLiu2017}.

For the construction below, we choose the midpoint of the time interval
as the origin, so that $\bar t=-t$. Moreover, with the ghost conventions of
Subsec.~\ref{KMSsymmetry}, one has $\Theta^{-1}=\Theta$ on the bosonic fields and
$\Theta^{-1}=-\Theta$ on the ghost fields.
\subsubsection{KMS-conjugate BRST symmetry}
In equilibrium, the BRST transformation can be conjugated by dynamical
KMS time reversal to define a second charge \cite{10.1093/oso/9780198834625.001.0001,Gao:2017bqf,Gao:2018bxz},
\begin{equation}
    \bar \delta_{\bar \epsilon} 
    \equiv
    \Theta \delta_{\bar \epsilon} \Theta^{-1}.
    \label{eq:second_charge_definition}
\end{equation}
Since dynamical KMS symmetry also gives
\(\Theta S_{\rm eq}=S_{\rm eq}\), the conjugate charge is a symmetry of the
complete equilibrium functional:
\begin{equation}
\bar \delta_{ \bar \epsilon} S_{\rm eq}
=
\Theta \delta_{\bar \epsilon}\Theta^{-1}S_{\rm eq}
=
0.
\label{eq:Qbar_invariance}
\end{equation}
Its action on the component fields is
\begin{equation}
\begin{aligned}
    \bar \delta_{\bar \epsilon }\phi=c\bar \epsilon ,
    \qquad
  \bar   \delta_{\bar \epsilon}c=0,
    \qquad
   \bar \delta_{\bar \epsilon} \bar c=(\phi_a-\beta\partial_t\phi)\bar \epsilon ,
    \qquad
    \bar \delta_{\bar \epsilon} \phi_a=\beta\partial_t c\bar \epsilon .
\end{aligned}
\label{eq:second_charge_transformations}
\end{equation}

\subsection{Superspace formulation}
\label{subsec:24}
In Subsec.~\ref{subsec:symmetries} we identified two nilpotent symmetries. The first is the BRST symmetry, which is a general property of the stochastic functional integral. In equilibrium, the KMS transformation provides a second, conjugate symmetry. Together, these two transformations form the equilibrium supersymmetry \cite{Parisi:1979ka, Intriligator1997, Canet:2011wf,AronBiroliCugliandolo2010,10.1093/oso/9780198834625.001.0001, Crossley:2015evo, Hertz_2016, Gao:2017bqf,Rychkov:2023rgq}. The same pair of nilpotent charges underlies the superspace formulation of Schwinger--Keldysh effective field theories, where it organizes the construction of dissipative hydrodynamic actions \cite{HaehlLoganayagamRangamani2017,HaehlLoganayagamRangamani2017b,JensenPinzaniFokeevaYarom2018,JensenMarjiehPinzaniFokeevaYarom2018}. This structure becomes particularly simple in superspace.
 We set $\beta=1$, and introduce two anticommuting Grassmann coordinates
$\theta$ and $\bar\theta$\footnote{$
    \{\theta,\bar{\theta}\}
    =
    \theta^2
    =
    \bar{\theta}^{\,2}
    =
    0 .
$
The integrals over these variables are defined by
$
    \int d\theta
    =
    \int d\bar{\theta}
    =
    0,
    \int d\theta\,\theta
    =
    \int d\bar{\theta}\,\bar{\theta}
    =
    1 .
$}
and collect the physical field, response
field, and ghosts into the real superfield
\begin{equation}
    \Phi(t,\bm{x},\theta,\bar\theta)
    =
    \phi(t,\bm{x})
    +\theta\bar c(t,\bm{x})
    +c(t,\bm{x})\bar\theta
    +\theta\bar\theta\,\phi_a(t,\bm{x}).
    \label{eq:real-superfield}
\end{equation}
The BRST symmetry and its KMS conjugate are represented on superspace by the operators 
\begin{equation}
    Q=\partial_\theta,
    \qquad
    \widebar Q
    =
    \partial_{\bar\theta}
    +\theta\partial_t ,
    \label{eq:superspace-generators}
\end{equation}
respectively. Acting with $Q$ or $\widebar Q$ on the superfield reproduces the corresponding transformations of its component fields. One can verify it by computing the infinitesimal supersymmetry transformation 
$
    \delta\Phi
    =
    \left(
        \epsilon Q
        +\bar\epsilon\widebar Q
    \right)\Phi
    $,
where \(\epsilon\) and \(\bar \epsilon\) are Grassmann parameters. These operators satisfy
\begin{equation}
    Q^2=\widebar Q^{\,2}=0,
    \qquad
    \{Q,\widebar Q\}
    =
   \partial_t .
    \label{eq:superspace-algebra}
\end{equation}
Thus, the two supersymmetry transformations close on an ordinary time translation.

We introduce the supersymmetric covariant derivatives by
\begin{equation}
    \widebar D=\partial_{\bar\theta},
    \qquad
    D=\partial_\theta-\bar\theta\partial_t, \quad 
    \text{with}\quad  D^2=\widebar D^2=0,
    \qquad
    \{D,\widebar D\}=-\partial_t.
    \label{eq:superspace-covariant-derivatives}
\end{equation}
With these definitions, the dynamical action can be written compactly as
\begin{equation}
    S[\Phi]
    =
    \intx \mathrm{d} \bar \theta \mathrm{d} \theta
    \left[
        -\frac{1}{c^2}(D\Dbar\Phi)(D\Dbar\Phi)
        +X\Dbar\Phi D\Phi
        +A(\Phi)
    \right],
    \label{eq:superspace-action}
\end{equation}
where we specify the potential term as 
\begin{equation}
    A(\Phi)=\frac12(\boldsymbol{\nabla}\Phi)^2+\frac12m^2\Phi^2+\frac{1}{4!}g \Phi^4.
\end{equation}
The two superspace kinetic operators of \eqref{eq:superspace-action} have different physical meanings. The
term
$
\sim c^{-2}
$
is the inertial, propagating operator. In components it generates the
second-order time derivative in the Langevin equation. By contrast, the term
$
\sim X
$
is the dissipative operator. It contains both the response term
$\sim \phi_a\partial_t\phi$ and the noise term $\sim \phi_a^2$, together with their
ghost partners. Therefore the coefficient $X$ controls the whole equilibrium
dissipative multiplet.

\subsubsection{Supersymmetry}
\label{subsec:ward-identities}

The supersymmetry introduced above implies Ward identities that constrain the Grassmann structure and time dependence of correlation functions. We begin with a scalar function \(f(Z)\) of a single superpoint \(Z=(t,\bm{x},\theta,\bar\theta)\). Supersymmetry requires
\begin{equation}
    Qf(Z)=   \widebar Qf(Z)=0.
    \label{eq:one-point-susy-ward}
\end{equation} 
The first identity implies that \(f\) is independent of \(\theta\), and hence it can be written as
\begin{equation}
    f(Z)
    =
    A(t,\bm{x})
    +
    \bar\theta B(t,\bm{x}).
\end{equation}
Acting with
\(\widebar Q\), we obtain
\begin{equation}
\begin{aligned}
    0
    =
    \widebar Q f
    =
    B(t,\bm{x})
    +
    \theta\,\partial_t A(t,\bm{x})
    +
    \theta\bar\theta\,\partial_tB(t,\bm{x}).
\end{aligned}
\end{equation}
The independence of the different Grassmann structures therefore gives \(B=0\) and \(\partial_tA=0\). Thus a supersymmetric scalar one-point function is independent of \(t,\theta,\bar\theta\), i.e.,
$
    f(Z)=f(\bm{x}).
$

We next consider a scalar function \(F(Z_1,Z_2)\) of two superpoints. Its Ward identities are 
\begin{equation}
    (Q_1+Q_2)F=0,
    \qquad
    (\widebar Q_1+\widebar Q_2)F=0.
    \label{eq:two-point-susy-ward}
\end{equation}
Since
\begin{equation}
    \left\{
        Q_1+Q_2,
        \widebar Q_1+\widebar Q_2
    \right\}
    =
    \partial_{t_1}+\partial_{t_2},
\end{equation}
these identities imply invariance under a common time translation. Consequently, the time dependence can occur only through
\begin{equation}
    t_{12}\equiv t_1-t_2.
\end{equation}
It is useful to introduce
\begin{equation}
    \theta_{12}=\theta_1-\theta_2,
    \qquad
    \bar\theta_{12}
    =
    \bar\theta_1-\bar\theta_2,
    \qquad
    \bar\theta^{+}_{12}
    =
    \bar\theta_1+\bar\theta_2.
\end{equation}
When acting on a time-translation-invariant function, the total supersymmetry generators become
\begin{equation}
    Q_1+Q_2
=\partial_{\theta_1}+\partial_{\theta_2},
\qquad
    \widebar Q_1+\widebar Q_2
    =
    \partial_{\bar\theta_1}
    +\partial_{\bar\theta_2}
    +\theta_{12}\partial_{t_{12}}.
    \label{eq:relative-barQ}
\end{equation}
The first identity restricts the dependence on \(\theta_1,\theta_2\) to the difference \(\theta_{12}\). The second mixes \(t_{12}\) with \(\bar\theta_{12}^{+}\). The invariant supersymmetric time separation is therefore
\begin{equation}
    \mathsf{t}_{12}
    =
    t_{12}
    +
    \frac12\theta_{12}\bar\theta^+_{12}.
    \label{eq:supersymmetric-time-interval}
\end{equation}
Indeed,
\begin{equation}
    (Q_1+Q_2)\mathsf t_{12}=0,
\end{equation}
while, using left Grassmann derivatives,
\begin{equation}
\begin{aligned}
    (\widebar Q_1+\widebar Q_2)\mathsf t_{12}
    &=
    \theta_{12}
    +
    \left(\partial_{\bar\theta_1}
      +\partial_{\bar\theta_2}\right)
    \left(
        \frac12\theta_{12}\bar\theta^+_{12}
    \right)
    =
    \theta_{12}-\theta_{12}
    =0.
\end{aligned}
\end{equation}
The Grassmann variables
\(\theta_{12}\) and \(\bar\theta_{12}\) are separately invariant:
\begin{equation}
\begin{aligned}
    (Q_1+Q_2)\theta_{12}
    &=
    (\widebar Q_1+\widebar Q_2)\theta_{12}
    =0,
    \\
    (Q_1+Q_2)\bar\theta_{12}
    &=
    (\widebar Q_1+\widebar Q_2)\bar\theta_{12}
    =0.
\end{aligned}
\end{equation}
The most general solution of the two Ward identities can consequently be written as
\begin{equation}
    F(Z_1,Z_2)
    =
    F\left(
        \mathsf t_{12},
        \bm{x}_{1},  \bm{x}_{2};
        \theta_{12},
        \bar\theta_{12}
    \right).
    \label{eq:general-two-point-invariant}
\end{equation}
Since the Grassmann variables are nilpotent, the general invariant function in Eq.~\eqref{eq:general-two-point-invariant} admits the expansion
\begin{equation}
\begin{aligned}
    F
    =
    F_1(\mathsf t_{12})
    +\theta_{12}
        F_\theta(\mathsf t_{12})
    +\bar\theta_{12}
        F_{\bar\theta}(\mathsf t_{12})
    +
    \theta_{12}\bar\theta_{12}\,
        F_3(\mathsf t_{12}),
\end{aligned}
\label{eq:even-Grassmann-expansion}
\end{equation}
where, for simplicity, we have suppressed the spatial arguments.

Usually one is interested in the two-point functions that are Grassmann-even and have vanishing ghost number. The terms proportional to only \(\theta_{12}\) or only
\(\bar\theta_{12}\) are therefore absent, and the expansion reduces to
\begin{equation}
    F(Z_1,Z_2)
    =
    F_1(\mathsf t_{12})
    +
    \theta_{12}\bar\theta_{12}\,
    F_3(\mathsf t_{12}).
    \label{eq:general-supersymmetric-two-point}
\end{equation}
Expanding the nilpotent shift contained in \(\mathsf{t}_{12}\), the first term in Eq.~\eqref{eq:even-Grassmann-expansion} becomes
\begin{equation}
    F_1(\mathsf t_{12})
    =
    F_1(t_{12})
    +
    \frac12\theta_{12}\bar\theta^+_{12}\,
    \partial_{t_{12}}F_1(t_{12}).
\end{equation}
Therefore,
\begin{equation}
\begin{aligned}
    F(Z_1,Z_2)
    =
    F_1(t_{12})
    +
    \theta_{12}
    \left[
        \frac12\bar\theta^+_{12}\,
        \partial_{t_{12}}F_1(t_{12})
        +
        \bar\theta_{12}F_3(t_{12})
    \right].
    \label{eq:expanded-supersymmetric-two-point}
\end{aligned}
\end{equation}
It is convenient to write the general Grassmann-even decomposition as
\begin{equation} F(Z_1,Z_2) = F_1(t_{12}) + \theta_{12}\bar\theta_{12}^{+}F_2(t_{12}) + \theta_{12}\bar\theta_{12}F_3(t_{12}). \label{eq:F123-decomposition} \end{equation}
Comparing Eqs.~\eqref{eq:expanded-supersymmetric-two-point} and \eqref{eq:F123-decomposition}, we obtain the supersymmetric Ward identity
\begin{equation}
        F_2(t)
        =
        \frac12\partial_{t}F_1(t),
    \label{eq:G2-G1-susy-relation}
\end{equation}
which is the two-point manifestation of the equilibrium
supersymmetry Ward identities. The corresponding closed-form two-point
superspace function, obtained by imposing in addition zero ghost number and
causality, is standard
\cite{Kurchan1992,10.1093/oso/9780198834625.001.0001}; its
extension to arbitrary $n$-point functions is given in
Appendix~\ref{app:n-pointkms}.
The function \(F_3\) in Eq. \eqref{eq:F123-decomposition} multiplies an independently supersymmetric structure and is therefore not fixed by the Ward identities.

The same construction extends to arbitrary
connected correlation functions and, since the symmetry transformations are
linear, to the vertices of the effective action. In particular, the
Grassmann components of an \(n\)-point function are not independent: the
components involving one of the reference Grassmann coordinates are fixed
by derivatives of the remaining components with respect to the relative
times. The general solution of the corresponding Ward identities is derived in
Appendix~\ref{app:n-pointkms}. 

We can apply the previous identity to the superpropagator, 
\begin{equation}
\begin{aligned}
    \Delta(Z_1,Z_2)  \equiv \langle \Phi(Z_1) \Phi(Z_2) \rangle= & F(t_{12}) + \theta_1 \bar \theta_1   G_{\mathrm A}(t_{12})
     +\theta_2 \bar \theta_2  G_{\mathrm R}(t_{12})  + \theta_1 \bar \theta_1 \theta_2 \bar \theta_2 \langle\phi_a(t_1) \phi_a(t_2) \rangle 
    \\
    &+ \theta_1 \bar \theta_2 \langle \bar c(t_1)  c(t_2) \rangle  +\bar \theta_1     \theta_2\langle c(t_1) \bar c(t_2) \rangle,
\end{aligned}
\label{eq:superpop_expanded}
\end{equation}
where 
\begin{equation}
    F(t_{12})
    \equiv
    \langle\phi(t_1)\phi(t_2)\rangle,
    \qquad
    G_{\mathrm R}(t_{12})
    \equiv
    \langle\phi(t_1)\phi_{a}(t_2)\rangle,
    \qquad
    G_{\mathrm A}(t_{12})
    \equiv
    \langle\phi_{a}(t_1)\phi(t_2)\rangle,
    \label{eq:Ward-component-definitions}
\end{equation}
and we assume that terms odd in the Grassmann variables vanish. 
The spatial arguments have again been suppressed.

Correlation functions containing only response fields vanish as a consequence of the BRST Ward identity\footnote{%
With the supersymmetry generator $Q$ defined in Eq.~\eqref{eq:superspace-generators}, the component BRST transformations imply
$
Qc=\phi_a,
Q\phi_a=0.
$
Consequently, any product containing only response fields is $Q$-exact:
\begin{equation*}
\prod_{j=1}^{n}\phi_a(t_j)
Q\left[
c(t_1)\prod_{j=2}^{n}\phi_a(t_j)
\right],
\qquad n\geq 1.
\label{eq:pure_response_Q_exact}
\end{equation*}
Since the functional measure and the causal boundary conditions are BRST invariant, the Ward identity $\langle Q\mathcal O\rangle=0$ gives
$
\left\langle
\prod_{j=1}^{n}\phi_a(t_j)
\right\rangle=0.
$
In particular, $\langle\phi_a(t_1)\phi_a(t_2)\rangle=0$, so the corresponding fourth-order Grassmann term in Eq.~\eqref{eq:superpop_expanded}
is absent.%
 }.
    Equivalently, the
coefficient of \(\theta_1\bar\theta_1\theta_2\bar\theta_2\) in the superfield two-point
function vanishes, which is why this fourth-order Grassmann structure is absent
from the decomposition used below.  According to Eq.~\eqref{eq:F123-decomposition}, the supersymmetric superpropagator can be written as \begin{equation} \Delta(Z_1,Z_2) = \Delta_1(t_{12}) + \theta_{12}\bar\theta_{12}^{+}\Delta_2(t_{12}) + \theta_{12}\bar\theta_{12}\Delta_3(t_{12}). \label{eq:Delta123-decomposition}
\end{equation}
Expanding Eq.~\eqref{eq:Delta123-decomposition} in the original Grassmann coordinates and comparing with Eq.~\eqref{eq:superpop_expanded}, we find
\begin{equation}
    \begin{aligned}
       \Delta_2(t_{12}) +\Delta_3(t_{12})   &= G_A (t_{12}), \\
       \Delta_2(t_{12})  -\Delta_3(t_{12})  &= \langle \bar c(t_1) c(t_2) \rangle ,\\
       -\Delta_2(t_{12})  -\Delta_3(t_{12})  &=-\langle  c(t_1) \bar c(t_2) \rangle ,\\
       -\Delta_2(t_{12})  +\Delta_3(t_{12})  &= G_R(t_{12}) ,
    \end{aligned}
\end{equation}
and this implies
\begin{equation}
    \langle \bar c(t_1) c(t_2) \rangle =- G_R(t_{12}),\quad \langle  c(t_1) \bar c(t_2) \rangle = G_A(t_{12}),
\end{equation}
and taking the difference between the retarded and advanced responses gives
\begin{equation}
    G_R(t_{12}) -G_A(t_{12})  = -2 \Delta_2(t_{12})  = -\partial_{t_{12}} \Delta_1(t_{12}) .
    \label{eq:FDR}
\end{equation}
Equation~\eqref{eq:FDR} is the classical
fluctuation--dissipation relation \cite{AronBiroliCugliandolo2010,Gao:2017bqf}.  The remaining function $\Delta_3$ is fixed by the causal support of the
response functions. For $t_{12}>0$, the advanced response vanishes:
\begin{equation}
    0=G_A(t_{12})
    =
    \Delta_2(t_{12})+\Delta_3(t_{12}),
    \qquad t_{12}>0,
\end{equation}
so that $\Delta_3=-\Delta_2$. For $t_{12}<0$, the retarded response vanishes,
and therefore
\begin{equation}
    0=G_R(t_{12})
    =
    -\Delta_2(t_{12})+\Delta_3(t_{12}),
    \qquad t_{12}<0,
\end{equation}
which gives $\Delta_3=\Delta_2$. The two time domains can be combined as
\begin{equation}
    \Delta_3(t_{12})
    =
    -\operatorname{sgn}(t_{12})\Delta_2(t_{12})
    =
    -\frac{1}{2}\operatorname{sgn}(t_{12})
        \partial_{t_{12}}\Delta_1(t_{12}),
    \qquad t_{12}\neq0.
    \label{eq:delta3-from-causality}
\end{equation}
Equivalently, the response functions take the explicitly causal form
\begin{equation}
    G_R(t_{12})
    =
    -\Theta(t_{12})\partial_{t_{12}}\Delta_{1}(t_{12}),
    \qquad
    G_A(t_{12})
    =
    \Theta(-t_{12})\partial_{t_{12}}\Delta_1(t_{12}),
    \label{eq:causal-response-from-delta1}
\end{equation}
where \(\Theta\) denotes the Heaviside function. The prescription at coincident times is fixed by the causal discretization used in the MSRJD functional and does not affect the relations at noncoincident times. For later use in the two-loop calculation, it is convenient to rewrite the
propagator in frequency space using a positive-time transform.
Since \(\Delta_1(t)\) is even in time, we define its half-sided Fourier
transform by
\begin{equation}
B(\omega)
\equiv
\int_0^\infty \dd t\,
e^{  \ii \omega t}\Delta_1(t).
\label{eq:half_sided_Delta}
\end{equation}
Its full Fourier transform is therefore
\begin{equation}
\Delta_1(\omega)
=
B(\omega)+B(-\omega).
\label{eq:Delta1_full_transform}
\end{equation}
Substituting these Fourier transforms into the Grassmann decomposition of the superpropagator gives
\begin{align}
\Delta_{ 0}(\omega,\bm p;\theta_{12},\bar \theta_{12},\bar \theta^{+}_{12})
=&
\Delta_1(\omega)
+
\theta_{12}
\Biggl\{
-\frac{  \ii \omega}{2}
\Delta_1(\omega)
\bar \theta^+_{12}
+
\left[
\Delta_1(t=0)
+\frac{  \ii \omega}{2}
[B(\omega)-B(-\omega)]
\right]
\bar \theta_{12}
\Biggr\}.
\label{eq:superpropagator_half_sided_compact}
\end{align} 
\subsection{Power counting and canonical dimensions}
\label{eq:Subsection canonical dim}
We conclude the general setup by determining the canonical dimensions of
the fields, parameters, and superspace coordinates. These dimensions provide
the tree-level power counting used in the subsequent renormalization-group
analysis.

Under scaling $\bm x\to b \bm x$ and $t\to b^z t$,
we have $
    [\boldsymbol{\nabla}]=1,
 [\partial_t]=z.
$ 
The bosonic spacetime measure therefore has dimension
\begin{equation}
    [\mathrm{d}^d x\,\mathrm{d}t] = -d-z,
\end{equation}
so the bosonic action density must have dimension \(d+z\). From the terms
\(\phi_a(-\boldsymbol{\nabla}^2+m^2+\partial_t^2/c^2)\phi\), \(g\,\phi_a\phi^3\), and the
dissipative contributions \(X\phi_a^2\) and \(X\phi_a\partial_t\phi\), we obtain
\begin{align}
    [\phi_a]+[\phi] &=d +z-2,\\
    [g]+[\phi_a]+3[\phi] & =d+z,\\
    [m^2] & =2,\\
    \left[\frac{1}{c^2}\right] & =2-2z ,
\end{align}
as well as
\begin{equation}
    [X]+2[\phi_a]=d+z,\qquad
    [X]+[\phi_a]+[\phi]=d.
\end{equation}
Therefore
\begin{equation}
    [\phi]=\frac{d-2}{2},\qquad
    [\phi_a]=\frac{d +2 z -2}{2},\qquad
    [X]=-z+2,\qquad
    [g]=4-d. \label{eq:canonicaldim}
\end{equation}
In particular, the upper critical spatial dimension remains \(d_c=4\),
independently of the dynamic exponent.
 For the ghosts, the quadratic terms only fix the sum of the ghost and antighost
dimensions:
\begin{equation}
    [\bar c]+[c]=d+z-2.
\end{equation}
As a consequence, higher-order operators are canonically irrelevant whenever \(d+z>4\), and in
particular near \(d=4\) for any positive \(z\). 
Since the lowest
component of the superfield is \(\phi\),
\begin{equation}
    [\Phi]=[\phi]=\frac{d-2}{2},
\end{equation}
and the component expansion implies
\begin{equation}
    [\theta]+[\bar\theta]=-z,\qquad
    [\theta]+[\bar c]=\frac{d-2}{2},\qquad
    [\bar\theta]+[c]=\frac{d-2}{2}.
\end{equation}
Since $
    [\mathrm{d}\theta \mathrm{d} \bar \theta]=z ,$
the Grassmann derivatives and covariant derivatives satisfy
\begin{equation}
    [\partial_\theta]=-[\theta],\qquad
    [\partial_{\bar\theta}]=-[\bar\theta],\qquad
    [D]=-[\theta]=[\bar\theta]+z,\qquad
    [\widebar D]=-[\bar\theta].
\end{equation}
With these assignments, the superspace kinetic term
$
    X \widebar D \Phi D \Phi
$
has dimension \(d\), as required by the measure
$[\mathrm{d} t \mathrm{d}^{d}x \,\mathrm{d}\theta \mathrm{d}\bar \theta ]= -d $.

The two natural Gaussian scalings are now transparent. At the propagating
Gaussian fixed point, keeping the inertial coefficient dimensionless fixes
\(z_{\mathrm{prop},\mathrm{G}}=1\). Equation~\eqref{eq:canonicaldim} then gives
$	\left[1/{c^2}\right]=0,
	[X]=1,
$
so dissipation is already a relevant perturbation at tree level. At the
overdamped Gaussian fixed point, keeping the dissipative coefficient
dimensionless fixes \(z_{\mathrm{od},\mathrm{G}}=2\), for which
$
	[X]=0,
	\left[1/{c^2}\right]=-2.
$
Thus the inertial operator is canonically irrelevant in the diffusive
scaling regime. Loop corrections modify these tree-level conclusions through
anomalous dimensions, which will be determined below.

\section{Effective action }
\label{sec:effective_action}

Having established the superspace formulation and its Ward identities in the previous section, we now develop the perturbative framework used in the remainder of the paper. We begin by introducing the effective action and its one-particle-irreducible vertices \cite{10.1093/oso/9780198834625.001.0001, Floerchinger:2021uyo}; the field-theoretic renormalization of critical dynamics in this framework was developed in Refs.~\cite{BauschJanssenWagner1976,DeDominicisPeliti1978}. The construction is especially compact in superspace, where the physical, response, and ghost fields are treated simultaneously. In Subsec.~\ref{subsec31}, we organize the effective action through two-loop order, while in Subsec.~\ref{subsec:superprop} we construct the free superpropagator and examine its overdamped and dissipationless limits. Finally, in Subsec.~\ref{sec:two_loop}, we use these ingredients to derive the two-point vertex through two loops. 
These results provide the starting point for the static and dynamical
renormalization analyses developed in the following sections.

We start by coupling \(\Psi\) to a bosonic superfield source \(\mathcal{J}\) and define\footnote{To distinguish the integration variable from its expectation value, in this section we denote the microscopic superfield introduced in Sec.~\ref{sec:general_setup} by \(\Psi(Z)\), reserving \(\Phi(Z)\) for the corresponding mean superfield.}
\begin{equation}
Z[{\cal J}]
=
\int {\cal D}\Psi\,
\exp\left[
-
S[\Psi]
+
\int_Z {\cal J}(Z)\Psi(Z)
\right].
\label{eq:Z_super_source}
\end{equation}
The generating functional of connected correlation functions is
$
W[{\cal J}]
=
\log Z[{\cal J}] ,
$
where \(\int_Z\) denotes integration over spacetime and Grassmann coordinates.
The expectation value of the superfield in the presence of the source is
\begin{equation}
\Phi(Z)
=
\frac{\delta W[{\cal J}]}{\delta{\cal J}(Z)}=\langle\Psi(Z)\rangle_{\mathcal J} .
\label{eq:PhI_3l_def}
\end{equation}
The effective action is obtained by a Legendre transform
\begin{equation}
\Gamma[\Phi]
=
\int_Z {\cal J}(Z)\Phi(Z)
-
W[{\cal J}] ,
\label{eq:Gamma_legendre}
\end{equation}
with the source understood as a functional of \(\Phi\). It follows that
\begin{equation}
\frac{\delta\Gamma[\Phi]}{\delta\Phi(Z)}
=
{\cal J}(Z).
\label{eq:Gamma_source_relation}
\end{equation}
Functional derivatives of \(\Gamma\) generate the 1PI vertices,
\begin{equation}
\Gamma^{(n)}[\Phi](Z_1,\ldots,Z_n)
\equiv
\frac{\delta^n\Gamma[\Phi]}
{\delta\Phi(Z_1)\cdots\delta\Phi(Z_n)}.
\label{eq:effective_action_vertices}
\end{equation}
Setting the source to zero selects the physical background. Evaluated at
this background, \(\Gamma^{(2)}\) is the full inverse propagator.

To construct the loop expansion, we split the integration superfield into its
background and fluctuation parts,
$
\Psi
=
\Phi+\delta\Phi ,
$
where \(\delta\Phi\) is the fluctuating superfield. Expanding the action about the background gives
\begin{equation}
\begin{aligned}
S[\Phi+\delta\Phi]
&=
S[\Phi]
+
\int_1 S^{(1)}(1)\delta\Phi(1)
+
\frac{1}{2}\int_{1,2}\delta\Phi(1)S^{(2)}(1,2)\delta\Phi(2)
\\
&
+
\frac{1}{3!}\int_{1,2,3}
S^{(3)}(1,2,3)\delta\Phi(1)\delta\Phi(2)\delta\Phi(3)
\\
&+
\frac{1}{4!}\int_{1,2,3,4}
S^{(4)}(1,2,3,4)
\delta\Phi(1)\delta\Phi(2)\delta\Phi(3)\delta\Phi(4)
+\cdots ,
\label{eq:background_expansion}
\end{aligned}
\end{equation}
where \(1\equiv Z_1\), \(\int_1\equiv \int_{Z_1}\), and
\begin{equation}
S^{(n)}[\Phi](1,\ldots,n)
=
\frac{\delta^n S[\Phi]}
{\delta\Phi(1)\cdots\delta\Phi(n)} .
\label{eq:functional_derivatives}
\end{equation}
The tree-level inverse propagator in the background field is \(S^{(2)}[\Phi]\), and the
corresponding superpropagator is
\begin{equation}
\Delta_{\Phi}
=
\left(S^{(2)}[\Phi]\right)^{-1}
.
\label{eq:Delta_background}
\end{equation}
At vanishing background, this reduces to the free superpropagator
$\Delta_0\equiv\Delta_{\Phi=0}.
$

\subsection{Loop expansion}
\label{subsec31}
The effective action can be organized according to the number of loops:
\begin{equation}
\Gamma[\Phi]
=
S[\Phi]
+
\Gamma_{1 \ell}[\Phi]
+
\Gamma_{2 \ell}[\Phi]
+\cdots .
\label{eq:Gamma_loop_expansion}
\end{equation}
The one-loop term is
\begin{equation}
\Gamma_{1 \ell}[\Phi]
=
\frac{1}{2}\,{\rm Tr}\log S^{(2)}[\Phi],
\label{eq:one_loop_effective_action}
\end{equation}
where the trace includes the integration over spacetime and Grassmann
coordinates. The Grassmann integrations encode the ghost contributions and the
relative signs fixed by the MSRJD supersymmetry.

At two-loop order, the two possible 1PI topologies are the double-bubble and
the sunset diagrams \cite{Berges2004}.  In condensed notation their combined contributions are 
\begin{equation}
\begin{aligned}
    \Gamma_{2\ell}[\Phi]
    &=
    \frac{1}{8}
    \int_{1,2,3,4}
    S^{(4)}(1,2,3,4)
    \Delta_{\Phi}(1,2)\Delta_{\Phi}(3,4)
\\ &
-
    \frac{1}{12}
    \int_{1,\ldots,6}
    S^{(3)}(1,2,3)S^{(3)}(4,5,6)
    \Delta_{\Phi}(1,4)
    \Delta_{\Phi}(2,5)
    \Delta_{\Phi}(3,6).
    \label{eq:two_loop_effective_action}
\end{aligned}
\end{equation}
 For the local quartic interaction
 \begin{equation}
    S_{\mathrm{int}}[\Phi]
    =
    \frac{g}{4!}\int_Z \Phi^4(Z),
    \label{eq:quartic_superfield_interaction}
\end{equation}
the background-dependent inverse propagator is
\begin{equation}
    S^{(2)}[\Phi](1,2)
    =
    \Delta_{0}^{-1}(1,2)
    +\frac{g}{2}\Phi^2(1)\delta(1-2),
    \label{eq:quartic_background_hessian}
\end{equation}
while the interaction vertices are
\begin{align}
    S^{(3)}(1,2,3)
    &=
    g\,\Phi(1)\delta(1-2)\delta(1-3),
    \quad
    S^{(4)}(1,2,3,4)
    =
    g\,\delta(1-2)\delta(1-3)\delta(1-4).
    \label{eq:quartic_four_point_vertex}
\end{align}
Substituting these expressions into the general loop expansion (Eqs.~\eqref{eq:one_loop_effective_action} and
\eqref{eq:two_loop_effective_action}) gives
\begin{equation}
    \Gamma_{1 \ell}[\Phi] =\frac{1}{2} \Tr \log(\Delta_{ 0}^{-1} + \frac{g}{2}  \Phi^2 ),
\end{equation}
and 
\begin{equation}
\Gamma_{2 \ell}[\Phi]
=
\frac{g}{8}
\int_1
\Delta_\Phi(1,1)^2
-
\frac{g^2}{12}
\int_{1,2}
\Phi(1)\Phi(2)
\Delta_\Phi(1,2)^3 .
\label{eq:Gamma_two_loop_phi4}
\end{equation}
\subsection{Free superpropagator}
\label{subsec:superprop}
The loop expansion derived above is expressed in terms of the background-dependent
superpropagator \(\Delta_\Phi\). In perturbation theory this propagator is
expanded about the vanishing background $\Phi=0$. After the required functional derivatives
with respect to \(\Phi\) have been taken, the internal lines of the two-point
diagrams are therefore expressed in terms of \(\Delta_{ 0}\). We now construct
this free superpropagator explicitly by inverting the quadratic part of the
superspace action.

The dissipative and inertial terms act differently in Grassmann space. 
Using the covariant derivatives defined in
Eq.~\eqref{eq:superspace-covariant-derivatives}, 
the inertial term gives\footnote{The sign of the inertial term can be checked directly by defining
\(
I_{\mathrm{in}}[\Phi]\equiv
\int_Z(D\bar D\Phi)(\bar D D\Phi)
\).
\[
\delta I_{\mathrm{in}}
= \int_Z (D \bar D \delta\Phi)( \bar D D \Phi) + (D \bar D \Phi)( \bar D D \delta\Phi)  =-\int_Z\delta\Phi\,((D \bar D)^2+(\bar D  D)^2)\Phi
 =-\int_Z\delta\Phi\,[\bar D,D]^2\Phi
 =-\int_Z\delta\Phi\,\partial_t^2\Phi.
\]
We therefore obtain
\begin{align*}
    \frac{\delta^2 I_{\mathrm{in}}}
         {\delta\Phi(Y)\delta\Phi(X)}
    &=
    -\partial_{t_X}^2
    \frac{\delta\Phi(X)}{\delta\Phi(Y)}
    =
    -\partial_{t_X}^2\delta(X-Y).
\end{align*}}

 \begin{equation}
 \frac{\delta^2}{\delta\Phi(Z)\delta\Phi(Y)} \left[ -\frac{1}{c^2} \int_{Z'} (D\bar D\Phi)(\bar D D\Phi) \right] \notag = \frac{1}{c^2} [\bar D_Z,D_Z]^2\delta(Z-Y) = \frac{1}{c^2} \partial_t^2\delta(Z-Y). 
 \end{equation}
By contrast, the second variation of the dissipative
bilinear is
\begin{equation}
     \frac{\delta^2 }{\delta \Phi (Z)\delta \Phi (Y)} \left[X\int \widebar D\Phi(Z^\prime)  D  \Phi(Z^\prime)   \right]= - X[\widebar D_Z ,D_Z ] \delta(Z-Y) .
\end{equation}
This motivates introducing the normalized dissipative kernel
\begin{equation}
    K_X(Z,Y)
    \equiv
    \frac{1}{2}
    [\widebar D_Z,D_Z]\delta(Z-Y),
    \label{eq:KX-coordinate}
\end{equation}
and its explicit form is
\begin{equation}
    K_X(Z,Y)
    =
    \left[
        1+
        \frac{1}{2}\theta_{12}\bar\theta^{+}_{12}
        \partial_t
    \right]
    \delta(t-t')\delta^{(d)}(\bm{x}-\bm{y}).
    \label{eq:KX-explicit}
\end{equation}
Using the Fourier convention
\(f(\omega)=\int \dd t\,e^{\ii\omega t}f(t)\), this becomes\footnote{Note that \(K_X\)
acts as a time-translation kernel, 
\[
    \int \dd t_2\,
    K_X
    (t_1-t_2;\theta_{12},\bar\theta_{12}^{+})f(t_2)
    =
    f\left(
        t_1+\frac{1}{2}
        \theta_{12}\bar\theta_{12}^{+}
    \right) = f(\mathsf{t}_1) .
\]
The shifted argument is the supersymmetric time separation introduced in
Eq.~\eqref{eq:supersymmetric-time-interval}.}
\begin{equation}
    K_X(\omega;\theta_{12},\bar\theta_{12}^+)
    =
    1-\frac{  \ii \omega}{2}
    \theta_{12}\bar\theta_{12}^{+}.
    \label{eq:KX-fourier}
\end{equation} 
The
Grassmann-independent part of \(K_X\) represents the noise vertex, while
the term linear in the frequency represents the dissipative response.

Combining the temporal kernels with the spatial and mass terms, it is useful to introduce the propagating box operator
$$
\Box = -\frac{1}{c^2}  \partial_t^2 +\boldsymbol{\nabla}^2.
$$
The free superpropagator  satisfies the equations 
\begin{equation}
    \left(-X[\widebar D , D]    -\Box +m^2  \right) \Delta_{ 0}(Z,Y) = \delta(Z-Y) ,
\end{equation}
where all derivatives act on the first argument. 
The inversion is simplified by the identity
\begin{equation}
    [\widebar D, D]^2  =(\widebar D D -D \widebar D )^2   = \widebar D \{ D ,\widebar D \}D +  D \{\widebar D, D \}\widebar D 
    = \left\{D ,\widebar D\right\}^2 = \partial_t^2,
\end{equation}
which motivates viewing \(D\) and \(\widebar D\) as Grassmann square roots of the time derivative.
Consequently, we obtain the useful identity
\begin{equation}
    \left(-X[\widebar D, D ]    -\Box +m^2  \right)  \left(X[\widebar D, D ]    -\Box +m^2  \right)  =  -X^2[\widebar D, D ]^2 + (-\Box +m^2)^2 = - X^2\partial_t^2+
    (-\Box +m^2)^2 \,.
\end{equation}
Then, the free superpropagator can be written as
\begin{equation}
\left(- X^2\partial_t^2 +
    (-\Box +m^2)^2  \right) \Delta_{ 0}(Z,Y) = 
\left(X[\widebar D, D ]    -\Box +m^2  \right) 
\delta(Z-Y).  
\end{equation}
Since the operator on the left-hand side is diagonal in Grassmann space, the
remaining task is to apply the differential operator on the right-hand side to
the delta function. The propagator can therefore be written formally
as
\begin{equation}
    \Delta_{ 0}(Z,Y) = 
\frac{X[\widebar D, D ]    -\Box +m^2   }{- X^2\partial_t^2 +
    (-\Box +m^2)^2 }
\delta(Z-Y) .
	\label{eq:free_superpropagator_operator_form}
\end{equation}
The commutator $[\widebar D,D] $ gives
\begin{equation}
    [\widebar D ,D] = \partial_{\bar \theta} (\partial_\theta-\bar \theta \partial_t)- (\partial_\theta-\bar \theta \partial_t)\partial_{\bar \theta} = 2\frac{\partial^2}{\partial\bar\theta \partial \theta }  + 2\bar\theta\frac{\partial}{\partial \bar\theta \partial t}
    -\partial_t. 
\end{equation}
Therefore, its action on the delta function is
\begin{equation}
    \left( 2\frac{\partial^2}{\partial\bar\theta_1 \partial \theta_1 }  + 2\bar\theta_1\frac{\partial}{\partial \bar\theta_1 \partial t}
    -\partial_t  \right)\left( \theta_1- \theta_2 \right)
    \left( \bar\theta_1- \bar \theta_2 \right) = 2\left(1 +\frac12\theta_{12}\bar \theta^{+}_{12}\partial_t \right)
    \label{eq:commutatiorDelta}.
\end{equation}
We next express Eq.~\eqref{eq:free_superpropagator_operator_form} in its components. As shown in Eq.~\eqref{eq:Delta123-decomposition}, the
supersymmetric Ward identities imply the decomposition
\begin{align}
	\Delta_{ 0}(t_{12},\bm p; \theta_{12}, \bar \theta_{12}, \bar\theta^{+}_{12})
	=&
	\Delta_1(t_{12},\bm p)
	+\theta_{12}\bar \theta^+_{12}
	\Delta_2(t_{12},\bm p)+\theta_{12}\bar \theta_{12}
	\Delta_3(t_{12},\bm p).
	\label{eq:general_superpropagator_decomposition}
\end{align} 
Using \eqref{eq:commutatiorDelta}, we identify
\begin{equation}
    \Delta_1=\frac{2 X}{-X^2 \partial_t^2+(-\Box +m^2)^2},
\end{equation}
or, equivalently, in Fourier space,
\begin{equation}
    \Delta_1(\omega,\bm p)=\frac{2 X}{X^2\omega^2+(-\omega^2/c^2 +\bm p^2 +m^2)^2} . 
\end{equation}
This propagator coincides with the statistical propagator. In this formalism, it can be used as the fundamental block  to construct all the other components $\Delta_2$ and $\Delta_3$. 

\subsubsection{Two Gaussian scaling limits}
\label{subsec:Gaussian-scaling-limits}

The general free superpropagator contains both the dissipative coefficient
\(X\) and the inertial coefficient \(c^{-2}\). The two Gaussian theories
studied below are obtained by retaining only one of these temporal
structures. Throughout this subsection, we use
\begin{equation}
    \omega_{\bm p}^2\equiv \bm p^2+m^2.
\end{equation}

\paragraph{Overdamped limit.}

When the inertial term is discarded, the symmetric component of the
superpropagator becomes
\begin{equation}
    \Delta_{1,\mathrm{od}}(\omega,\bm p)
    =
    \frac{2X}
    {\omega_{\bm p}^4+X^2\omega^2}.
\end{equation}
Fourier transformation gives
\begin{equation}
    \Delta_{1,\mathrm{od}}(t,\bm p)
    =
    \frac{1}{\omega_{\bm p}^2}
    \exp\left(
        -\frac{\omega_{\bm p}^2}{X}|t|
    \right).
\end{equation}
The remaining components follow from the supersymmetry Ward identity and
causality, and lead to
\begin{align}
    \Delta_{2,\mathrm{od}}(t,\bm p)
    &=
    -\frac{\operatorname{sgn}(t)}{2X}
    \exp\left(
        -\frac{\omega_{\bm p}^2}{X}|t|
    \right),
    \quad
    \Delta_{3,\mathrm{od}}(t,\bm p)
    =
    \frac{1}{2X}
    \exp\left(
        -\frac{\omega_{\bm p}^2}{X}|t|
    \right).
\end{align}
At criticality, \(\omega_{\bm p}^2=\bm p^2\), and hence
\(\omega\sim\bm p^2\). The corresponding Gaussian dynamic exponent is
therefore \(z_{\mathrm{od},\mathrm{G}}=2\).
The corresponding retarded and advanced response functions are
\begin{align}
    G_{R,\mathrm{od}}(t,\bm p)
    &=
    \frac{\Theta(t)}{X}
    \exp\left(
        -\frac{\omega_{\bm p}^2}{X}t
    \right),
    \quad
    G_{A,\mathrm{od}}(t,\bm p)
    =
    \frac{\Theta(-t)}{X}
    \exp\left(
        -\frac{\omega_{\bm p}^2}{X}|t|
    \right).
\end{align}
In frequency space,
\begin{equation}
    G_{R,\mathrm{od}}(\omega,\bm p)
    =
    \frac{1}{\omega_{\bm p}^2-iX\omega},
    \qquad
    G_{A,\mathrm{od}}(\omega,\bm p)
    =
    \frac{1}{\omega_{\bm p}^2+iX\omega}.
\end{equation}
\paragraph{Dissipationless limit.}

In the absence of dissipation, the bulk noise vanishes. Equilibrium
fluctuations nevertheless remain because the initial field and velocity are
sampled from the Gibbs ensemble. The symmetric correlator is obtained by
taking the distributional limit \(X\to0^+\):
\begin{equation}
    \Delta_{1,\mathrm{prop}}(\omega,\bm p)
    =
    \frac{\pi}{\omega_{\bm p}^2}
    \left[
        \delta(\omega-c\omega_{\bm p})
        +
        \delta(\omega+c\omega_{\bm p})
    \right].
\end{equation}
Equivalently, in the time domain,
\begin{equation}
    \Delta_{1,\mathrm{prop}}(t,\bm p)
    =
    \frac{1}{\omega_{\bm p}^2}
    \cos(c\omega_{\bm p}t).
    \label{eq:propD1}
\end{equation}
The remaining components are
\begin{align}
    \Delta_{2,\mathrm{prop}}(t,\bm p)
    &=
    -\frac{c}{2\omega_{\bm p}}
    \sin(c\omega_{\bm p}t),
    \quad
    \Delta_{3,\mathrm{prop}}(t,\bm p)
    =
    \frac{c}{2\omega_{\bm p}}
    \sin(c\omega_{\bm p}|t|).
     \label{eq:propD23}
\end{align}
The two poles,
$
    \omega=\pm c\omega_{\bm p},
$
describe propagating modes. At criticality,
\(\omega_{\bm p}=|\bm p|\), so \(\omega\sim|\bm p|\), and the corresponding
Gaussian dynamic exponent is \(z_{\mathrm{prop},\mathrm{G}}=1\).
The corresponding causal response functions are
\begin{align}
    G_{R,\mathrm{prop}}(t,\bm p)
    &=
    \Theta(t)\frac{c}{\omega_{\bm p}}
    \sin(c\omega_{\bm p}|t|),
    \quad
    G_{A,\mathrm{prop}}(t,\bm p)
    =
    \Theta(-t)\frac{c}{\omega_{\bm p}}
    \sin(c\omega_{\bm p}|t|).
\end{align}
Equivalently,
\begin{equation}
    G_{R,\mathrm{prop}}(\omega,\bm p)
    =
    \frac{1}{
        \omega_{\bm p}^2-(\omega+i0^+)^2/c^2
    },
    \qquad
    G_{A,\mathrm{prop}}(\omega,\bm p)
    =
    \frac{1}{
        \omega_{\bm p}^2-(\omega-i0^+)^2/c^2
    }.
\end{equation}
The retarded propagator has poles at
\(\omega=\pm c\omega_{\bm p}-i0^+\), corresponding to propagating modes.

\subsection{Two-point vertex}
\label{sec:two_loop}

The loop expansion of the effective action and the free superpropagators
derived above provide the ingredients needed to construct the 1PI two-point
vertex. The contributing topologies are the same in the overdamped and
dissipationless theories, but their internal propagators and temporal
projections are different. 
At two-loop order,
\begin{equation}
    \Gamma^{(2)}(Z_1,Z_2)
    =
    \Gamma_{\mathrm{tree}}^{(2)}(Z_1,Z_2)
    +
    \Gamma_{1\ell}^{(2)}(Z_1,Z_2)
    +
    \Gamma_{2\ell}^{(2)}(Z_1,Z_2)
    +
    \mathcal{O}(g^3).
\end{equation}
The tree-level kernels are presented separately below. The tadpole, double-bubble, and sunset topologies are then derived in a form valid for both the propagating and overdamped theories. Their static projections will be compared in
Sec.~\ref{sec:static}, whereas their dynamical projections will be evaluated separately
in Secs.~\ref{sec:overdamped} and~\ref{sec:dissipationless}.

\subsubsection{Tree-level vertices}

\paragraph{Overdamped theory.}

For \(c^{-2}=0\), the tree-level two-point vertex is
\begin{equation}
    \Gamma_{\mathrm{tree},\mathrm{od}}^{(2)}(Z_1,Z_2)
    =
    \left(
        -X[\bar D_1,D_1]
        -\nabla_1^2
        +m^2
    \right)
    \delta(Z_1-Z_2).
\end{equation}
In frequency and momentum space,
\begin{equation}
\label{eq:tree_gamm2}
    \Gamma_{\mathrm{tree},\mathrm{od}}^{(2)}
    (\omega,\bm p;
    \theta_{12},\bar\theta_{12},\bar\theta_{12}^+)
    =
    -2X
    \left(
        1-\frac{ \ii \omega}{2}
        \theta_{12}\bar\theta_{12}^{+}
    \right)
    +
    \left(\bm p^2+m^2\right)
    \theta_{12}\bar\theta_{12}.
\end{equation}
The Grassmann-independent noise vertex and the term proportional to
\( \ii \omega\) form the equilibrium dissipative multiplet and carry the common
coefficient \(X\). The factor in parentheses is precisely the Fourier-space
kernel \(K_X\) defined in Eq.~\eqref{eq:KX-fourier}.

\paragraph{Dissipationless theory.}

For \(X=0\), the tree-level two-point vertex is
\begin{equation}
    \Gamma_{\mathrm{tree},\mathrm{prop}}^{(2)}(Z_1,Z_2)
    =
    \left(
        \frac{1}{c^2}\partial_{t_1}^2
        -\nabla_1^2
        +m^2
    \right)
    \delta(Z_1-Z_2).
\end{equation}
In frequency and momentum space,
\begin{align}
    &\Gamma_{\mathrm{tree},\mathrm{prop}}^{(2)}
    (\omega,\bm p;
    \theta_{12},\bar\theta_{12},\bar\theta_{12}^+)
    =
    \left(
        -\frac{\omega^2}{c^2}
        +\bm p^2+m^2
    \right)
    \theta_{12}\bar\theta_{12}.
\end{align}
There is no bulk noise and local dissipative-response vertex in this limit.

\subsubsection{One-loop tadpole}

The one-loop contribution to
the two-point vertex is the local superspace tadpole
\begin{equation}
    \Gamma_{1\ell}^{(2)}(Z_1,Z_2)
    =
    \raisebox{0.33cm}{$
    \mathord{\vcenter{\hbox{
    \begin{tikzpicture}[x=1.5cm,y=1.5cm]
        \draw[line width=0.85pt] (-0.95,0) -- (0.95,0);
        \draw[line width=0.85pt] (0,0.36) circle (0.36);
        \node[
            circle,
            fill=black,
            inner sep=0pt,
            minimum size=3.8pt
        ] at (0,0) {};
    \end{tikzpicture}
    }}}
    $}
    =
    \frac{g}{2}
    \Delta_{0}(Z_1,Z_1)
    \delta(Z_1-Z_2).
\end{equation}
At coincident superspace points, all terms containing a relative Grassmann
coordinate vanish. The loop therefore depends only on the equal-time
symmetric component,
\begin{equation}
    \Delta_{0}(Z,Z)
    =
    \int_{\bm q}
    \Delta_{1}(0,\bm q).
\end{equation}

\paragraph{Overdamped theory.}

Using the overdamped propagator derived in Subsec.~\ref{subsec:superprop},
\begin{equation}
    \Delta_{1,\mathrm{od}}(0,\bm q)
    =
    \frac{1}{\omega_{\bm q}^2},
\end{equation}
and hence
\begin{equation}
    T_{1,\mathrm{od}}
    \equiv
    \int_{\bm q}
    \Delta_{1,\mathrm{od}}(0,\bm q)
    =
    \int_{\bm q}
    \frac{1}{\omega_{\bm q}^2}.
        \label{eq:T1o}
\end{equation}

\paragraph{Dissipationless theory.}

The propagating equal-time correlator gives
\begin{equation}
    \Delta_{1,\mathrm{prop}}(0,\bm q)
    =
    \frac{1}{\omega_{\bm q}^2},
\end{equation}
so that
\begin{equation}
    T_{1,\mathrm{prop}}
    \equiv
    \int_{\bm q}
    \Delta_{1,\mathrm{prop}}(0,\bm q)
    =
    \int_{\bm q}
    \frac{1}{\omega_{\bm q}^2}.
\end{equation}

Although the two theories have different unequal-time propagators, their
coincident symmetric components are identical. We may therefore define the
common static tadpole
$
    T_1
    \equiv
    T_{1,\mathrm{od}}
    =
    T_{1,\mathrm{prop}}.
$
After Fourier transformation, the one-loop vertex is consequently
\begin{equation}
    \Gamma_{1\ell}^{(2)}
    (\omega,\bm p;\theta_{12},\bar\theta_{12})
    =
    \frac{g}{2}T_1\,
    \theta_{12}\bar\theta_{12}.
\end{equation}
Importantly, the tadpole is independent of the external frequency and momentum and, therefore, in both theories it renormalizes only the static mass.

\subsubsection{Two-loop contributions}
\label{sec:two_loop_db_sun}
At order \(g^2\), the two-point vertex receives contributions from the
double-bubble and sunset topologies:
\begin{equation}
    \Gamma_{2\ell}^{(2)}(Z_3,Z_4)
    =
    \Gamma_{2\ell,\mathrm{db}}^{(2)}(Z_3,Z_4)
    +
    \Gamma_{2\ell,\mathrm{sun}}^{(2)}(Z_3,Z_4).
\end{equation}
We consider the two contributions separately below.

\paragraph{Double-bubble.}

The double-bubble term in the background-field effective action is
\begin{equation}
    \Gamma_{2\ell,\mathrm{db}}[\Phi]
    =
    \frac{g}{8}
    \int_Z \Delta_{\Phi}(Z,Z)^2 .
\end{equation}
Although this expression contains no explicit power of \(\Phi\), it depends
on the background through $\Delta_{\Phi}$. Its inverse is
\begin{equation}
    \Delta_{\Phi}^{-1}(Z_1,Z_2)
    =
    \Delta_{0}^{-1}(Z_1,Z_2)
    +
    \frac{g}{2}\Phi^2(Z_1)\delta(Z_1-Z_2).
\end{equation}
Using
\begin{equation}
    \delta\Delta_{\Phi}
    =
    -\Delta_{\Phi}
    \left(\delta\Delta_{\Phi}^{-1}\right)
    \Delta_{\Phi},
\end{equation}
one finds
\begin{equation}
    \left.
    \frac{\delta\Delta_{\Phi}(A,B)}
         {\delta\Phi(Z_3)}
    \right|_{\Phi=0}
    =
    0,
\end{equation}
whereas
\begin{align}
    &\left.
    \frac{\delta^2\Delta_{\Phi}(A,B)}
         {\delta\Phi(Z_3)\delta\Phi(Z_4)}
    \right|_{\Phi=0}
    =
    -g\,\delta(Z_3-Z_4)
    \Delta_{0}(A,Z_3)
    \Delta_{0}(Z_3,B).
\end{align}
Differentiating the double-bubble effective action twice therefore gives
\begin{equation}
    \Gamma_{2\ell,\mathrm{db}}^{(2)}(Z_3,Z_4)
    =
    \raisebox{0.45cm}{$
    \mathord{\vcenter{\hbox{
    \begin{tikzpicture}[x=1.5cm,y=1.5cm]

        \draw[line width=0.85pt] (-0.65,0) -- (0.65,0);

        \draw[line width=0.85pt] (0,0.40) circle (0.40);

        \draw[line width=0.85pt] (0,1.20) circle (0.40);

        \node[
            circle,
            fill=black,
            inner sep=0pt,
            minimum size=3.8pt
        ] at (0,0) {};

        \node[
            circle,
            fill=black,
            inner sep=0pt,
            minimum size=3.8pt
        ] at (0,0.80) {};

    \end{tikzpicture}
    }}}
    $}
    =
    -\frac{g^2}{4}
    \delta(Z_3-Z_4)
    \int_Z
    \Delta_{0}(Z,Z)\,
    \Delta_{0}(Z,Z_3)^2.
\end{equation}
The tadpole contribution is a number 
\begin{equation}
    \Delta_{0}(Z,Z)
    =
    \int_{\bm q}\Delta_{1}(0,\bm q)
    =
    T_1.
\end{equation}
The coincident propagator gives the common tadpole \(T_1\) derived above.
We consequently define
\begin{equation}
    T_{2}
    \equiv
    \int_Z
    \Delta_{0}(Z,Z_3)^2,
\end{equation}
which is independent of \(Z_3\) by translation invariance. Performing the integral over the Grassmann variables, the resulting contribution is only in terms of the
symmetric and advanced components of the free propagator,
\begin{equation}
    T_{2}
    =
    2\int_{t,\bm p}
    \Delta_{1}(t,\bm p)
    G_{A}(t,\bm p).
\end{equation}
Causality restricts the advanced response to \(t<0\), where
\(G_{A}=\partial_t\Delta_{1}\). Hence
\begin{align}
    T_{2}
    &=
    2\int_{\bm p}\int_{-\infty}^{0} \dd t\,
    \Delta_{1}(t,\bm p)
    \partial_t\Delta_{1}(t,\bm p)
    =
    \int_{\bm p}
    \Delta_{1}(0,\bm p)^2.
\end{align}
 Using the
equal-time propagators derived in Subsec.~\ref{subsec:superprop}, one obtains separately
\begin{equation}
    T_{2,\mathrm{od}}
    =
    \int_{\bm p}
    \frac{1}{\omega_{\bm p}^4},
    \qquad
    T_{2,\mathrm{prop}}
    =
    \int_{\bm p}
    \frac{1}{\omega_{\bm p}^4}.
    \label{eq:T2o}
\end{equation}
We may therefore introduce the common static integral
\begin{equation}
    T_2
    \equiv
    T_{2,\mathrm{od}}
    =
    T_{2,\mathrm{prop}}.
\end{equation}
The double-bubble contribution is consequently
\begin{equation}
    \Gamma_{2\ell,\mathrm{db}}^{(2)}
    (\omega,\bm p;\theta_{12},\bar\theta_{12})
    =
    -\frac{g^2}{4}T_1T_2\,
    \theta_{12}\bar\theta_{12}.
\end{equation}
Like the one-loop tadpole, this contribution is independent of the external
frequency and momentum and therefore renormalizes only the static mass term
and does not contribute to any kinetic coefficient.
\paragraph{Sunset.}

The sunset term in the two-loop effective action contains two explicit
background fields. Differentiating twice and then setting the background to
zero gives
\begin{equation}
    \Gamma_{2\ell,\mathrm{sun}}^{(2)}(Z_1,Z_2)
    =
    \mathord{\vcenter{\hbox{
    \begin{tikzpicture}[x=1.5cm,y=1.5cm]
        \coordinate (vL) at (-0.62,0);
        \coordinate (vR) at ( 0.62,0);

        \draw[line width=0.85pt] (-1.12,0) -- (vL);
        \draw[line width=0.85pt]
            (vL) to[out=48,in=132,looseness=1.15] (vR);
        \draw[line width=0.85pt] (vL) -- (vR);
        \draw[line width=0.85pt]
            (vL) to[out=-48,in=-132,looseness=1.15] (vR);
        \draw[line width=0.85pt] (vR) -- (1.12,0);

        \node[
            circle,
            fill=black,
            inner sep=0pt,
            minimum size=3.8pt
        ] at (vL) {};

        \node[
            circle,
            fill=black,
            inner sep=0pt,
            minimum size=3.8pt
        ] at (vR) {};
    \end{tikzpicture}
    }}}
    =
    -\frac{g^2}{6}
    \Delta_{0}(Z_1,Z_2)^3.
\end{equation}
Unlike the tadpole and double bubble, this contribution depends
nontrivially on the external separation. It is therefore the only two-loop
topology capable of renormalizing the kinetic operators. To expose its superspace structure, introduce the relative coordinates
\begin{equation}
    t\equiv t_1-t_2,
    \qquad
    \bm{x}\equiv\bm{x}_1-\bm{x}_2.
\end{equation}
Using the Grassmann decomposition of the free propagator, its cube can be
written as
\begin{align}
    \Delta_{0}(Z_1,Z_2)^3
    &=
    \Sigma_{1}(t,\bm{x})
    +3\theta_{12}
    \left[
        \bar\theta_{12}^{+}
        \Sigma_{2}(t,\bm{x})
        +
        \bar\theta_{12}
        \Sigma_{3}(t,\bm{x})
    \right],
\end{align}
where
\begin{equation}
    \Sigma_{1}
    \equiv
    \Delta_{1}^3,
    \qquad
    \Sigma_{2}
    \equiv
    \Delta_{1}^2\Delta_{2},
    \qquad
    \Sigma_{3}
    \equiv
    \Delta_{1}^2\Delta_{3},
    \label{eq:sunset-component-definitions}
\end{equation}
and all functions on the right-hand side are evaluated at
\((t,\bm{x})\). The supersymmetry Ward identity and causality relation imply
\begin{align}
    \Sigma_{2}(t,\bm{x})
    &=
    \frac{1}{2}
    \Delta_{1}(t,\bm{x})^2
    \partial_t\Delta_{1}(t,\bm{x})
   =
    \frac{1}{6}
    \partial_t
    \Sigma_{1}(t,\bm{x}),
    \\
    \Sigma_{3}(t,\bm{x})
    &=
    -\frac{1}{2}
    \operatorname{sgn}(t)
    \Delta_{1}(t,\bm{x})^2
    \partial_t\Delta_{1}(t,\bm{x})
    =
    -\frac{1}{6}
    \operatorname{sgn}(t)
    \partial_t
    \Sigma_{1}(t,\bm{x}),
\end{align}
or equivalently,
\begin{equation}
    3\Sigma_{2}
    =
    \frac{1}{2}\partial_t \Sigma_{1},
    \qquad
    3\Sigma_{3}
    =
    -\frac{1}{2}\operatorname{sgn}(t)
    \partial_t \Sigma_{1}.
\end{equation}
Thus, for either Gaussian theory, supersymmetry reconstructs the complete
sunset superkernel from the single bosonic function
\( \Sigma_{1}\). 
To express the result in frequency space, define the positive-time
transform
\begin{equation}
    \widetilde{\Sigma}_{1}(\omega,\bm p)
    \equiv
    \int_0^\infty \dd t
    \int \dd^d x\,
    e^{\ii(\omega t-\bm p\cdot\bm x)}
    \Sigma_{1}(t,\bm x).
\end{equation}
Since \( \Sigma_{1}\) is real and even in time, its
negative-time contribution is
\(\widetilde{\Sigma}_{1}^{\,*}\). Integrating the
derivative terms by parts gives
\begin{align}
\label{eq:sunset}
    \Gamma_{2\ell,\mathrm{sun}}^{(2)}
    (\omega,\bm p;
    \theta_1,\bar\theta_1,\theta_2,\bar\theta_2)
    &=
    -\frac{g^2}{6}
    \left(
        1-\frac{ \ii \omega}{2}
        \theta_{12}\bar\theta_{12}^{+}
    \right)
    \left[
        \widetilde{\Sigma}_{1}(\omega,\bm p)
        +
        \widetilde{\Sigma}_{1}^{\,*}(\omega,\bm p)
    \right]
    \notag\\
    &\quad
    -\frac{g^2}{6}
    \left[
        \Sigma_{1}(0,\bm p)
        +
        \frac{ \ii \omega}{2}
        \left(
            \widetilde{\Sigma}_{1}(\omega,\bm p)
            -
            \widetilde{\Sigma}_{1}^{\,*}(\omega,\bm p)
        \right)
    \right]
    \theta_{12}\bar\theta_{12},
\end{align}
where we defined
\begin{equation}
    \Sigma_{1}(0,\bm p)
    \equiv
    \int \dd^d x\,
    e^{- \ii \bm p\cdot\bm x}
    \Sigma_{1}(0,\bm x).
\end{equation}
For the overdamped  theory, every internal line in
the diagrams refers to the overdamped propagator $\Delta_{1,\mathrm{od}}$. 
Its explicit
low-frequency expansion is evaluated in Sec.~\ref{sec:overdamped}.
For the propagating one, instead, every internal line  is the propagating correlator $\Delta_{1,\mathrm{prop}}$. 
These projections are
evaluated separately in Sec.~\ref{sec:dissipationless}.

\section{Static sector}
\label{sec:static}

In this section, we isolate the equilibrium static sector shared by the propagating and overdamped theories. In fact, the dynamics determines how equilibrium is approached, but not the equilibrium distribution itself. To see this explicitly, consider the propagating theory, whose equilibrium Gibbs distribution is defined in terms of the total mechanical energy introduced in Eq.~\eqref{eq:energy_functional}, \begin{equation*} P_{\mathrm{eq}}[\phi,\dot\phi] \propto e^{-\beta E_{\mathrm{tot}}[\phi,\dot\phi]}. 
\end{equation*} As the kinetic term is quadratic in \(\dot\phi\) and independent of \(\phi\),  integrating over the velocity gives \begin{equation*} P_{\mathrm{stat}}[\phi] \propto \int\mathcal D\dot\phi\, e^{-\beta E_{\mathrm{tot}}[\phi,\dot\phi]} \propto e^{-\beta H[\phi]}, \end{equation*} where the Gaussian integral over \(\dot\phi\) contributes only a field-independent normalization. Since the overdamped theory is governed by the same equilibrium weight \(e^{-\beta H[\phi]}\),  the propagating and overdamped formulations have different dynamics but the same static field distribution.

In the rest of this section, we show how this equivalence emerges directly at the level of the one-particle-irreducible vertices.
In Subsec.~\ref{subsec:staticloop}, we project the dynamical two- and four-point vertices onto their static components and recover the standard Euclidean $\phi^4$ loop corrections through two-loop order. In Subsec.~\ref{subsec:static-RG-flow}, we use the resulting renormalization factors to derive the static RG flow and identify the Gaussian and Wilson--Fisher fixed points that will subsequently be combined with the two dynamical  regimes in the following sections.

\subsection{Static loop corrections}
\label{subsec:staticloop}
The
static renormalization is determined by the two- and four-point vertices: the
former fixes the mass and field renormalizations, while the latter determines
the renormalization of the quartic coupling. In the dynamical theory, these vertices contain both frequency and Grassmann
structures. Their static parts are obtained by setting the external
frequencies to zero and projecting onto the appropriate local Grassmann
component. The equilibrium Ward identities then reduce the internal
frequency integrals to products of equal-time correlators. 
We demonstrate this reduction for the two-point vertex and then for the
four-point vertex, retaining the contributions required up to two-loop order.

\subsubsection{Two-point vertex}
The static inverse propagator is extracted from the two-point vertex by
setting the external frequency to zero and projecting onto its local
superspace structure. Since this structure is proportional to the superspace
delta function, its coefficient can be isolated by integrating over one pair
of Grassmann coordinates. We therefore define the static inverse propagator as
\begin{equation}
    \Gamma_{\rm stat}^{(2)}(\bm p) \equiv \int \mathrm{d} \bar \theta_1 \mathrm{d} \theta_1 \; \Gamma^{(2)}(\omega=0,\bm p ;\theta_1,\bar \theta_1, \theta_2,\bar \theta_2).
\end{equation}
Applying this projection to the two-point vertex derived in the preceding section at two loops gives
\begin{equation}
\Gamma_{\rm stat}^{(2)}(\bm p) 
=
\bm p^2+m^2
+
\frac{g}{2} T_1
-\frac{g^2}{4}T_1T_2
-\frac{g^2}{6} \Sigma_{1}(t=0,\bm p)
+
\mathcal O(g^3),
\label{eq:static}
\end{equation}
with $T_1$ and $T_2$ defined in \eqref{eq:T1o}
and \eqref{eq:T2o} respectively, while 
\begin{equation}
    \begin{aligned}
        \Sigma_{1}(t=0,\bm p) = \int_{\bm p_1, \bm p_2}
        \frac{1}{\omega_{\bm p_1}^2} 
        \frac{1}{\omega_{\bm p_2}^2} 
        \frac{1}{\omega_{\bm p-\bm p_1-\bm p_2}^2} .
    \end{aligned}
    \label{eq:Sigma_Sim}
\end{equation}
Equation~\eqref{eq:static} is precisely the static two-point function of
Euclidean \(\phi^4\) theory \cite{10.1093/oso/9780198834625.001.0001, Berges2004, Berges2015}.
In particular, no dependence on \(X\) or \(c\) remains after the equal-time
projection.

\subsubsection{Four-point vertex}
\label{subsec:static-four-point-vertex}

The same reduction applies to the four-point vertex, although one intermediate
step is important.  Before the Grassmann integrations are performed, the
dynamical loop is not simply a product of two symmetric correlators.
Superspace instead combines a symmetric component with a response component;
the equilibrium Ward identities then reduce their frequency integral to a
product of equal-time correlators.  We show this mechanism explicitly at one
loop before quoting the standard two-loop counterterm. The one-loop effective action contains
\begin{equation}
    \Gamma_{1\ell}[\Phi]
    =
    \frac12
    \operatorname{Tr}
    \log\left(
        \Delta_{ 0}^{-1}+\frac{g}{2}\Phi^2
    \right).
    \label{eq:static-one-loop-effective-action}
\end{equation}
Expanding the logarithm to second order in the background-dependent insertion
gives the term containing four external fields,
\begin{equation}
    \left.\Gamma_{1\ell}[\Phi]\right|_{\Phi^4}
    =
    -\frac{g^2}{16}
    \int_{ Z_1, Z_2}
    \Phi^2(Z_1)\,
    \Delta_{ 0}(Z_1,Z_2)\,
    \Phi^2(Z_2)\,
    \Delta_{0}(Z_1,Z_2).
    \label{eq:static-one-loop-Phi4-term}
\end{equation}
Taking four functional derivatives produces the three crossing channels, \begin{equation} \Gamma^{(4)}_{1\ell}(Z_1,Z_2,Z_3,Z_4) = -\frac{g^2}{2} \sum_{(ij)(kl)} \delta(Z_i-Z_j)\, \delta(Z_k-Z_l)\, \Delta_0(Z_i,Z_k)^2, \end{equation} where the sum runs over \( (ij)(kl)=(12)(34),(13)(24),(14)(23) \).
Transforming to frequency--momentum space, we define
\begin{equation}
\begin{aligned}
\mathcal{B}(\omega, \bm p;\theta_{12},\bar \theta_{12},\bar \theta_{12}^+)
&=
\int_{\nu, \bm q}\;
\Delta_0\!\big(\nu ,\bm q;\theta_{12},\bar \theta_{12},\bar \theta_{12}^+\big)\;
\Delta_0\!\big(\omega -\nu,\bm p -\bm q;\theta_{12},\bar \theta_{12},\bar \theta_{12}^+\big) \\
&=  \mathcal{B}_1(\omega, \bm p)+\mathcal{B}_2(\omega, \bm p)\,\theta_{12}\bar \theta^+_{12}+\mathcal{B}_3(\omega, \bm p)\,\theta_{12}\bar \theta_{12},
\label{eq:bubble-general-decomposition}
\end{aligned}
\end{equation}
with  
\begin{equation}
    \mathcal B_2(\omega,\bm p) = -\frac{\ii \omega }{2}  \mathcal B_1(\omega,\bm p),
\end{equation}
such that the bubble becomes the usual expansion 
\begin{equation}
\begin{aligned}
\mathcal{B}(\omega, \bm p;\theta_{12},\bar \theta_{12},\bar \theta_{12}^+)
&=  \left(1-\frac{ \ii \omega}{2}\,\theta_{12}\bar \theta^+_{12} \right)\mathcal{B}_1(\omega, \bm p)+\mathcal{B}_3(\omega, \bm p)\,\theta_{12}\bar \theta_{12}.
\label{eq:bubble-susy-decomposition}
\end{aligned}
\end{equation}
The four-point vertex can be expressed as
\begin{equation}
\begin{aligned}
\Gamma^{(4)}_{\rm 1loop}(\omega_1,\bm p_1,\theta_1 \bar \theta_1;\ldots;\omega_4,\bm p_4,\theta_4 \bar \theta_4)
= -\frac{g^2}{2}\Big[
&\theta_{12}\bar \theta_{12}\theta_{34}\bar \theta_{34}\,\;
\mathcal{B}(\omega_{12}, \bm p_{12};\theta_{13},\bar \theta_{13},\bar \theta_{13}^+)
\\
+&\theta_{13}\bar \theta_{13}\theta_{24}\bar \theta_{24}\;
\mathcal{B}(\omega_{13}, \bm p_{13};\theta_{12},\bar \theta_{12},\bar \theta_{12}^+)
\\
+&\theta_{14}\bar \theta_{14}\theta_{23}\bar \theta_{23}\;
\mathcal{B}(\omega_{14}, \bm p_{14};\theta_{12},\bar \theta_{12},\bar \theta_{12}^+)
\Big] .
\end{aligned}
\label{eq:Gamma4_1loop}
\end{equation}
The static part of the four-point function is extracted by setting all
external frequencies to zero and integrating over three pairs of Grassmann
coordinates:
\begin{equation}
   \Gamma^{(4)}_{\mathrm{stat},1\ell}(\{\bm p_i\})
   =
   \int\prod_{i=2}^{4}\dd\bar\theta_i\,\dd\theta_i\,
   \Gamma^{(4)}_{1\ell}
   (0,\bm p_1,\theta_1,\bar\theta_1;\ldots;
    0,\bm p_4,\theta_4,\bar\theta_4).
\end{equation}
It follows that
\begin{equation}
   \Gamma^{(4)}_{\mathrm{stat},1\ell}(\{\bm p_i\})
   =
   -\frac{g^2}{2}
   \left[
   \mathcal B_3(0,\bm p_{12})
   +\mathcal B_3(0,\bm p_{13})
   +\mathcal B_3(0,\bm p_{14})
   \right],
\end{equation}
corresponding to the usual \(s\)-, \(t\)-, and \(u\)-channel contributions.
The remaining superspace loop is fixed by the Grassmann projection, which at
zero external frequency selects
\begin{equation}
    \mathcal B_3(0,\bm p)
    =
    \int_{\bm q}\int\frac{\dd\omega}{2\pi}
    \left[
        \Delta_1(\omega,\bm q)\Delta_3 (\omega,\bm q+\bm p)
        +
        \Delta_1(\omega,\bm q+\bm p)
        \Delta_3 (\omega,\bm q)
    \right].
    \label{eq:static-superspace-bubble}
\end{equation}
The reduction can be established without choosing either Gaussian scaling
limit. 
Moreover, stationarity and equilibrium time-reversal invariance give
\(\Delta_1(-t,\bm p)=\Delta_1(t,\bm p)\). Introducing
\(\bm k\equiv \bm q+\bm p\) and using
\[
    \int\frac{\dd \omega}{2\pi}\,
    f(\omega)g(\omega)
    =
    \int_{-\infty}^{\infty}\dd t\,f(t)g(-t),
\]
Eq.~\eqref{eq:static-superspace-bubble} becomes
\begin{equation}
\begin{aligned}
    \mathcal B_3(0,\bm p)
   & =
    \int_{\bm q}\int_{-\infty}^{\infty}\dd t\,
    \bigl[
        \Delta_1(t,\bm q)\Delta_3 (-t,\bm k)
        +
        \Delta_3(t,\bm q)
        \Delta_1(-t,\bm k)
    \bigr]
    \\&=
    -\frac12 \int_{\bm  q}\int_0^\infty \dd t\,
    \bigl[
        \Delta_1(t,\bm  q)\partial_t\Delta_1(t,\bm  k)
        +\partial_t\Delta_1(t,\bm  q)\Delta_1(t,\bm  k)
    \bigr].
    \end{aligned}
\end{equation}
As in the evaluation of \(T_2\), the integrand is a total time derivative:
\begin{equation}
    \mathcal B_3(0,\bm p)
    =
    -\int_{\bm q}\int_0^\infty \dd t\,
    \partial_t\!
    \left[
        \Delta_1(t,\bm  q)\Delta_1(t,\bm  k)
    \right].
\end{equation}
The remaining boundary is fixed entirely by the equal-time equilibrium
correlator. Using the common equal-time correlator derived in
Subsec.~\ref{subsec:Gaussian-scaling-limits}, one obtains
\begin{equation}
    \mathcal B_3(0,\bm p)
    =
    \int_{\bm q}
    \Delta_1(0,\bm q)\Delta_1(0,\bm q+\bm p)
    =
    \int_{\bm q}
    \frac{1}{\omega_{\bm q}^2}
    \frac{1}{\omega_{\bm q +\bm p}^2}.
\end{equation}
This derivation depends only on the equilibrium Ward identity and causality,
and not on the relative magnitude of the inertial and dissipative
coefficients.
In dimensional regularization \cite{10.1093/oso/9780198834625.001.0001},
\begin{equation}
\Gamma_{\mathrm{stat},\,1\ell}^{(4)}
    =
    -\frac{3g^2}{(4\pi)^2(4-d)}
    +\text{finite}.
    \label{eq:static-four-point-one-loop-pole}
\end{equation}

\subsection{Static RG flow}
\label{subsec:static-RG-flow}
We can now obtain the static RG functions in the usual way \cite{10.1093/oso/9780198834625.001.0001}.
The divergence is cancelled by the leading coupling counterterm,
\begin{equation}
    \frac{Z_g}{Z_\phi^2}
    =
    1+\frac{3g_r}{(4\pi)^2 (4-d)}
    +\mathcal O(g_r^2),
    \label{eq:static-one-loop-coupling-renormalization}
\end{equation}
with the renormalized quartic coupling defined as 
\begin{equation}
    g=\mu^{4-d}\frac{Z_g}{Z_\phi^2}\,g_r.
\end{equation}
In the modified minimal-subtraction
\(\overline{\mathrm{MS}}\) scheme, \(Z_\phi=1\) at one-loop order, and
Eq.~\eqref{eq:static-one-loop-coupling-renormalization} agrees with the
corresponding result for the one-component Euclidean \(\phi^4\) theory. At two
loops, the reduction proceeds diagram by diagram in the same way. We therefore
quote the standard Euclidean counterterms from
Ref.~\cite{10.1093/oso/9780198834625.001.0001}, rather than reproducing the
intermediate momentum integrals.
To state the result, we introduce the dimensionless renormalized mass
parameter \(r\) and define the bare parameters by
\begin{equation}
    \Phi=Z_\phi^{1/2}\Phi_r,
    \qquad
    m^2=\mu^2\frac{Z_m}{Z_\phi}\,r,
    \qquad \epsilon=4-d.
    \label{eq:static-bare-renormalized-relations}
\end{equation}
Here the subscript \(r\) denotes renormalized quantities. The wavefunction
renormalization is determined by the inverse propagator,
\begin{equation}
    Z_\phi
    =
    1-\frac{g_r^2}{12(4\pi)^4(4-d)}
    +\mathcal O(g_r^3).
    \label{eq:static-field-renormalization}
\end{equation}
In the \(\overline{\mathrm{MS}}\) scheme, the complete renormalization factors
needed below are
\begin{align}
    \frac{Z_m}{Z_\phi}
    &=
    1+
    \frac{g_r}{(4\pi)^2\epsilon}
    +\frac{g_r^2}{(4\pi)^4}
    \left(
        \frac{2}{\epsilon^2}
        -\frac{5}{12\epsilon}
    \right)
    +\mathcal O(g_r^3),
    \label{eq:static-mass-renormalization}
    \\
    \frac{Z_g}{Z_\phi^2}
    &=
    1+
    \frac{3g_r}{(4\pi)^2\epsilon}
    +\frac{g_r^2}{(4\pi)^4}
    \left(
        \frac{9}{\epsilon^2}
        -\frac{17}{6\epsilon}
    \right)
    +\mathcal O(g_r^3).
    \label{eq:static-coupling-renormalization}
\end{align}
Together with \(Z_\phi\) in
Eq.~\eqref{eq:static-field-renormalization}, these relations contain all
static ultraviolet poles required below.  The double poles are fixed by
lower-order counterterm insertions, whereas the simple poles determine the
two-loop RG functions.

Acting with \(\mu\,\dd/\dd\mu\) at fixed bare parameters on
Eq.~\eqref{eq:static-bare-renormalized-relations}, and using the pole parts in
Eqs.~\eqref{eq:static-field-renormalization},
\eqref{eq:static-mass-renormalization}, and
\eqref{eq:static-coupling-renormalization}, gives
\begin{align}
    \beta_g
    &=
    -\epsilon g_r
    +\frac{3g_r^2}{(4\pi)^2}
    -\frac{17g_r^3}{3(4\pi)^4}
    +\mathcal O(g_r^4),
    \label{eq:static-beta-function}
    \\
    \gamma_\phi(g_r)
    &=
    \frac{g_r^2}{6(4\pi)^4}
    +\mathcal O(g_r^3),
    \label{eq:static-field-anomalous-dimension}
    \\
    \gamma_m(g_r)
    &=
    \frac{g_r}{(4\pi)^2}
    -\frac{5g_r^2}{6(4\pi)^4}
    +\mathcal O(g_r^3).
    \label{eq:static-mass-anomalous-dimension}
\end{align}
Here \(\gamma_m\) is defined through
\(\beta_r=(-2+\gamma_m)r\). These coincide with the standard two-loop
renormalization-group functions of the Euclidean \(\phi^4\) theory
\cite{WilsonFisher1972,Wilson:1973jj,10.1093/oso/9780198834625.001.0001}, confirming that the
supersymmetric dynamical formulation reproduces the known static
renormalization.  The trivial zero of
Eq.~\eqref{eq:static-beta-function} is the usual Gaussian fixed point $g_{\rm G}=0$, while the nontrivial zero is the Wilson--Fisher fixed point \cite{Wilson:1973jj},
\begin{equation}
    g_{\rm WF}
    =
    (4\pi)^2
    \left(
        \frac{\epsilon}{3}
        +\frac{17}{81}\epsilon^2
        +\mathcal O(\epsilon^3)
    \right)
    .
    \label{eq:Wilson-Fisher-fixed-point}
\end{equation}
Linearizing the beta function about either fixed point,
\(\beta_g\simeq y_\ast(g_r-g_\ast)\), gives
\begin{equation}
    y_{\rm G} = -\epsilon,
    \qquad
    y_{\rm WF} = \epsilon-\frac{17}{27}\epsilon^2.
\end{equation}
The corresponding static anomalous dimension is 
$
    \eta
    =
    \gamma_\phi(g_{\rm WF})
    =
   \epsilon^2/54
    +\mathcal O(\epsilon^3).
$
The Gaussian and Wilson--Fisher zeros of the static beta function can each be
combined with either of the two dynamical scaling regimes. In the following
sections we first keep the static coupling general, so that the Gaussian
results follow by setting \(g_r=0\), and then evaluate the interaction-induced
corrections at \(g_{\mathrm{WF}}\). This yields the four fixed points, which will be discussed
in detail in Sec.~\ref{sec:fixed-point-summary}.
\section{Overdamped dynamics: Model A}
\label{sec:overdamped}
We first consider the strictly overdamped theory, obtained by setting
\(c^{-2}=0\). At the Gaussian fixed point its dynamic exponent is
\(z_{\mathrm{od},\mathrm{G}}=2\), and the inertial operator is canonically
irrelevant. This is the Gaussian starting point of Model A. Interactions drive
the static coupling to the Wilson--Fisher fixed point and renormalize the
dissipative operator, producing the interacting exponent
\(z_{\mathrm{od},\mathrm{WF}}\). A detailed discussion  of Model A and the related dynamic exponent can be found in Refs. \cite{10.1093/oso/9780198834625.001.0001, Tauber2014, HohenbergHalperin1977, Canet:2011wf}. Having already fixed the static renormalizations in the previous section, we now focus on the genuinely
dynamical renormalization of the dissipative coefficient \(X\). In
Subsec.~\ref{subsec:two_loop_overdamped_vertex}, we isolate its superspace
structure and extract the ultraviolet pole. In Subsec.~\ref{sec:model-a-rg},
we then determine the interacting dynamic exponent
\(z_{\mathrm{od},\mathrm{WF}}\).

\subsection{Dissipative projection and two-loop counterterm}
\label{subsec:two_loop_overdamped_vertex}

The tree-level contribution to the dissipative term in $\Gamma^{(2)}$ is
\(\sim X K_X\) (see Eq.~\eqref{eq:KX-fourier}), so the renormalization of
\(X\) is obtained by projecting the self-energy onto the same Grassmann
structure as the tree-level term. At two-loop order, the only nonvanishing
correction arises from the sunset diagram: the one-loop tadpole and the
double-bubble contribution are independent of the external frequency and
momentum and therefore renormalize only the static mass term.
The complete superspace decomposition of the sunset was obtained in
Subsec.~\ref{sec:two_loop}.  We write the generic dissipative term as 
\begin{equation}
    \left.
    \Gamma_{2\ell,\mathrm{sun}}^{(2)}
    (\omega,\bm p;\theta_{12}, \bar \theta_{12}, \bar \theta_{12}^+)
    \right|_{X}
    =
    -\frac{g^2}{6}\,
    K_X(\omega;\theta_{12}, \bar \theta^+_{12})\Sigma_1(\omega,\bm p).
\label{eq:model-A-sunset-X-projection}
\end{equation} 
All Grassmann dependence is carried by \(K_X\). Consequently, the ultraviolet
coefficient can be obtained by setting the external frequency and momentum
of $\Sigma_1$ to zero:
\begin{equation}
      X\Sigma_X  \equiv\Sigma_1(0, \bm 0) =  \int_{-\infty}^{+\infty} \dd t\,\Sigma_{1} (t,\bm 0)  .
   \label{eq:definition_Sigma_X}
\end{equation}
Explicitly, using Eq.~\eqref{eq:sunset-component-definitions}, we obtain
\begin{equation}
\begin{aligned}
X\Sigma_X
&=
\int_{\bm q,\bm k}
\int_{-\infty}^{+\infty} \mathrm{d} t\,
\Delta_1(t, \bm q)\Delta_1(t,\bm k)\Delta_1(t,\bm q+\bm k)
.
\label{eq:Xsigma_time_integral}
\end{aligned}
\end{equation}
The time integration produces exactly one factor of $X$ that cancels the factor used in the definition of $\Sigma_X$, leaving a purely spatial, dimensionless two-loop integral:
\begin{equation}
\Sigma_X
=
2
\int_{\bm q,\bm k}
\frac{1}
{\omega^2_{\bm q}\omega^2_{\bm k}\omega^2_{\bm q+\bm k}
\left(
\omega^2_{\bm q}+\omega^2_{\bm k}+\omega^2_{\bm q+\bm k}
\right)} .
\label{eq:SigmaX_zero_final_integral}
\end{equation}
This shows explicitly
that the sunset renormalizes the existing dissipative operator.
To extract its ultraviolet divergence, we introduce Schwinger parameters. For instance, we write
\begin{equation}
\frac{1}{\omega^2_{\bm q}+\omega^2_{\bm k}+\omega^2_{\bm q+\bm k}}
=
\int_0^\infty \dd s\,
e^{-s(\omega^2_{\bm q}+\omega^2_{\bm k}+\omega^2_{\bm q+\bm k})} \, ,
\label{eq:schwinger_sum_denominator}
\end{equation}
and similar expressions for the other denominators. 
 We can rewrite Eq. \eqref{eq:SigmaX_zero_final_integral} as 
\begin{align}
\Sigma_X
&=
2
\int_0^\infty
\dd s\,\dd \alpha\,\dd \beta\,\dd \gamma
\int_{\bm q,\bm k}
\exp\left[
-(\alpha+s)\omega^2_{\bm q}
-(\beta+s)\omega^2_{\bm k}
-(\gamma+s)\omega^2_{\bm q+\bm k}
\right].
\label{eq:SigmaX_schwinger_before_momenta}
\end{align}
Since the momentum integral is Gaussian, we can perform the integrations over
\(\bm q\) and \(\bm k\), obtaining
\begin{align}
\Sigma_X
&=
\frac{2}{(4\pi)^d}
\int_0^\infty
\dd s\,\dd \alpha\,\dd \beta\,\dd \gamma\,
\frac{
\exp\left[
-m^2(\alpha+\beta+\gamma+3s)
\right]
}
{
\left[
(\alpha+s)(\beta+s)
+
(\alpha+s)(\gamma+s)
+
(\beta+s)(\gamma+s)
\right]^{d/2}
}.
\label{eq:SigmaX_schwinger_final}
\end{align}
In dimensional regularization, with \(d=4-\epsilon\), the pole part is\footnote{In order to show it explicitly, set
\(a=\alpha+s\), \(b=\beta+s\), and \(c=\gamma+s\).  For fixed
\(a,b,c\), the original integration domain becomes
\(0\leq s\leq\min(a,b,c)\).  After performing the \(s\) integral, use
permutation symmetry to choose \(a=\min(a,b,c)\), multiply by three, and set
\(b=ax\), \(c=ay\), with \(x,y\geq1\).  This gives
\(\Sigma_X=\frac{6}{(4\pi)^d}
\int_1^\infty\!\dd x\,\dd y\,
(x+y+xy)^{-d/2}
\int_0^\infty\!\dd a\,a^{3-d}
e^{-m^2a(1+x+y)}\).
For \(d=4-\epsilon\), the scale integral equals
\(\Gamma(\epsilon)[m^2(1+x+y)]^{-\epsilon}\), so its pole is
\(1/\epsilon\).  The remaining integral is
\(\int_1^\infty\!\dd x\,\dd y\,(x+y+xy)^{-2}
=\int_1^\infty\!\dd x\,[(x+1)(2x+1)]^{-1}
=\log(4/3)\).}
\begin{equation}
\left.\Sigma_X\right|_{\mathrm{pole}}
=
\frac{1}{(4\pi)^4}
\frac{6}{4-d}
\log\frac{4}{3}.
\label{eq:SigmaX_divergent_part}
\end{equation}
Thus, displaying only the ultraviolet pole, the dissipative term is
\begin{equation}
	\begin{split}
		\left.
		\Gamma^{(2)}
		(\omega,\bm p;
		\theta_{12}, \bar \theta_{12}, \bar \theta_{12}^+)
		\right|_{X}
		={}&
		-2X
		K_X
		\left[
		1+ \frac{g^2}{2(4\pi)^4 (4-d)}
\log\frac{4}{3}
		\right] +\text{finite}.
	\end{split}
	\label{eq:model-a-corrected-superspace-kernel}
\end{equation}

\subsection{RG flow and dynamic exponent}
\label{sec:model-a-rg}
We renormalize \(X\) by writing
\begin{equation}
    X
    =
    \mu^{2-z}\frac{Z_X}{Z_\phi}\,x.
    \label{eq:model-A-X-renormalization-definition}
\end{equation}
This definition separates the field renormalization \(Z_\phi\), already
fixed by the static sector, from the new dynamical factor \(Z_X\).  Indeed,
because \(\Gamma_r^{(2)}=Z_\phi\Gamma^{(2)}\), the factor \(Z_\phi\)
cancels the denominator in
Eq.~\eqref{eq:model-A-X-renormalization-definition}.  Finiteness of the
renormalized dissipative projection therefore requires
\begin{equation}
    Z_X
    =
    1-
    \frac{g_r^2}{2(4\pi)^4\epsilon}
    \log\frac{4}{3}
    +\mathcal O(g_r^3).
    \label{eq:model-A-ZX}
\end{equation}
This is the only genuinely dynamical counterterm required for Model A at
two-loop order.  To determine the
physical scaling of the complete frequency term, it must be combined with the
static field renormalization \(Z_\phi\) 
derived in Sec.~\ref{sec:static} as
\begin{equation}
\log\frac{Z_X}{Z_\phi}
=
-
\frac{g_r^{\,2}}{2(4\pi)^4\epsilon}
\left(
\log\frac{4}{3}-\frac{1}{6}
\right)
+
\mathcal O(g_r^{\,3}).
\label{eq:model-a-log-ZX-over-Z}
\end{equation}
Since this expression starts at order \( g_r^{\,2}\), only the
leading term
$
\beta_{ g}
=
-\epsilon g_r
+
\mathcal O( g_r^{\,2})
$
is required to determine \(\gamma_X\) at two-loop order. Substituting into
\begin{equation}
\gamma_X( g_r)
=
-\beta_{g}(g_r)
\frac{\partial}{\partial g_r}
\log\frac{Z_X}{Z_\phi},
\label{eq:model-a-gamma-X-definition}
\end{equation}
 one obtains
\begin{align}
\gamma_X( g_r)
=
-
 \frac{g_r^{\,2}}{(4 \pi)^4}
\left(
\log\frac{4}{3}-\frac{1}{6}
\right)
+
\mathcal O( g_r^{\,3}).
\label{eq:model-a-gamma-X}
\end{align}
We now use the RG equation for the dimensionless dissipative coefficient
\(x\). 
At an overdamped fixed point, the coefficient \(x\) is finite and nonzero.
The fixed-point condition \(\beta_x=0\) therefore gives
\(z=2-\gamma_X(g_r^*)\). At the Gaussian fixed point,
\(g_r^*=g_{\mathrm{G}}=0\), this immediately reproduces
\(z_{\mathrm{od},\mathrm{G}}=2\). At the Wilson--Fisher fixed point,
\(\gamma_X\) starts at order \(g_r^{\,2}\), so only the leading term
\(g_{\mathrm{WF}}=(4\pi)^2\epsilon/3+\mathcal O(\epsilon^2)\) is needed through
order \(\epsilon^2\). Equation~\eqref{eq:model-a-gamma-X} then gives
\begin{equation}
\gamma_X( g_{\rm WF})
=
-
\frac{\epsilon^2}{9}
\left(
\log\frac{4}{3}-\frac{1}{6}
\right)
+
\mathcal O(\epsilon^3),
\label{eq:model-a-gamma-X-fixed-point}
\end{equation}
and hence
\begin{equation}
z_{\mathrm{od},\mathrm{WF}}
=
2+
\frac{6\log(4/3)-1}{54}\,
\epsilon^2
+
\mathcal O(\epsilon^3).
\label{eq:model-a-z-equivalent}
\end{equation}
Using the static result,
$\eta
=
\epsilon^2/54
+
\mathcal O(\epsilon^3),
$ quoted in Sec.~\ref{sec:static}, the exponent can be written in the familiar form 
\begin{equation}
z_{\mathrm{od},\mathrm{WF}}=2+\left[6 \log \frac{4}{3}-1\right] \eta+\mathcal{O}\left(\epsilon^3\right). 
\end{equation}
This reproduces the known two-loop Model~A exponent, first obtained in
Refs.~\cite{HalperinHohenbergMa1972,HalperinHohenbergMa1974} and rederived
within the renormalized field theory of critical dynamics in
Ref.~\cite{BauschJanssenWagner1976}. Its $\epsilon$-expansion has since been
carried to five loops \cite{AdzhemyanEtAl2022}.

\section{Dissipationless dynamics}
\label{sec:dissipationless}
In the last section, we studied the overdamped limit, or Model A. We now restrict the theory to the opposite dissipationless limit, setting \(X=0\). Propagating critical scaling in relativistic scalar theories has been investigated both analytically and on the lattice \cite{BoyanovskyDeVega2002,SchweitzerSchlichtingVonSmekal2020,SchweitzerSchlichtingVonSmekal2022}.
Two distinct questions arise: first, in Subsec.~\ref{subsec:61},  we introduce a convenient representation of the propagators;  then, we calculate how
interactions renormalize the velocity coefficient \(c\) in Subsec.~\ref{subsec:hankel-prop}. Next, in Subsec. \ref{subsec:absenceofrenorm}, we determine whether fluctuations
generate a local dissipative operator when dissipation is absent in the microscopic theory.
Finally, in Subsec.~\ref{sec:propagating-RG-flow}, we obtain the Gaussian and Wilson--Fisher propagating exponents, \(z_{\mathrm{prop},\mathrm{G}}\) and \(z_{\mathrm{prop},\mathrm{WF}}\).

\subsection{Hankel representation }
\label{subsec:61}
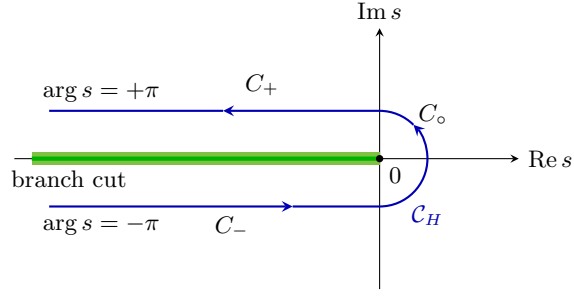
\begin{figure}
\centering
\begin{tikzpicture}[
    scale=1.15,
    >=stealth,
    contour/.style={thick, blue!70!black},
    axis/.style={->, black},
    every node/.style={font=\small}
]

\draw[axis] (-4.2,0) -- (1.6,0) node[right, black] {$\Re s$};
\draw[axis] (0,-1.5) -- (0,1.5) node[above, black] {$\Im s$};

\def\rad{0.55}

\fill[yellow!45!green] (-4.0,-0.07) rectangle (0,0.07);
\draw[ultra thick, green!70!black] (-4.0,0) -- (0,0);
\node[below left, black] at (-2.8,-0.03) {branch cut};

\draw[contour,->]
    (-3.8,-\rad) -- (-1,-\rad)
    node[near end, below , black] {$C_{-}$};
\draw[contour]
    (-1,-\rad) -- (0,-\rad);

\draw[contour,->]
    (0,-\rad) arc[start angle=-90,end angle=45,radius=\rad];

\draw[contour]
    ({\rad*cos(45)},{\rad*sin(45)})
    arc[start angle=45,end angle=90,radius=\rad]
    node[midway, right=4pt, black] {$C_{\circ}$};

\draw[contour,->]
    (0,\rad) -- (-1.8,\rad)
    node[near end, above=2pt, black] {$C_{+}$};
\draw[contour]
    (-1.8,\rad) -- (-3.8,\rad);

\fill[black] (0,0) circle (1.2pt);
\node[below right, black] at (0,0) {$0$};

\node[blue!70!black] at (0.55,-0.65) {$\mathcal C_H$};
\node[above, black] at (-3.2,\rad) {$\arg s=+\pi$};
\node[below, black] at (-3.2,-\rad) {$\arg s=-\pi$};

\end{tikzpicture}
\caption{Hankel contour $\mathcal C_H$ around the negative real axis, decomposed into the lower branch $C_{-}$, the small circular arc $C_{\circ}$, and the upper branch $C_{+}$.}
\label{fig:hankel-contour}
\end{figure}

As already shown in Sec.~\ref{sec:effective_action}, the one-loop tadpole and the two-loop double-bubble diagrams are independent of the external
frequency and momentum, and consequently, they contribute only to the static sector and \textit{cannot} renormalize $c$. The
first nonvanishing frequency-dependent correction is therefore due to the two-loop sunset diagram.
Its complete superspace
decomposition was derived in Subsec.~\ref{sec:two_loop}, and we report it here
\begin{equation}
\begin{aligned}
            \Gamma_{2\ell, \rm sun }^{(2)} (\omega ,\bm p;\theta_1,\bar \theta_1,\theta_2,\bar \theta_2 )  = &
    -\frac{g^2}{6} \left\{\left[1 -\frac{ \ii \omega}{2}\theta_{12}\bar \theta^+_{12}\right]
\Sigma_1(\omega,\bm p)
  +
   3\Sigma_3(\omega,\bm  p)
   \theta_{12}
   \bar \theta_{12}\right\},
\end{aligned}
\end{equation}
where 
\begin{equation}
\begin{aligned}
 \Sigma_1(\omega, \bm p)  
   &
   =\int_{-\infty}^{\infty} \dd t \dd^d x e^{ \ii \omega t - \ii \bm p \cdot \bm x } \Delta^3_1(t, \bm x),
   \\
   \Sigma_3(\omega,\bm p) &
  =  \int_{-\infty}^{\infty} \dd t \dd^d x e^{ \ii \omega t - \ii \bm p \cdot \bm x } \Delta^2_1(t,x)\Delta_3(t,\bm x).
   \end{aligned}
\end{equation}
In terms of the propagators [Eqs. \eqref{eq:propD1} and \eqref{eq:propD23}], the two integrals become
\begin{equation}
\begin{aligned}
  \Sigma_1(\omega, \bm p)  
   &=\int_{-\infty}^{\infty} \dd t  e^{ \ii \omega t }\int_{\bm p_1 \bm p_2 }
  \frac{\cos(c \omega_{\bm p_1} |t|)}{\omega^2_{\bm p_1}} \frac{\cos(c \omega_{\bm p_2} |t|)}{\omega^2_{\bm p_2}}  \frac{\cos(c \omega_{\bm p - \bm p_1-\bm p_2} |t|)}{\omega^2_{\bm p-\bm p_1-\bm p_2}} ,
  \\
  \Sigma_3(\omega,\bm p) &= 
  \int_{-\infty}^{\infty} \dd t  e^{ \ii \omega t }\int_{\bm p_1 \bm p_2 }
  \frac{\cos(c \omega_{\bm p_1} |t|)}{\omega^2_{\bm p_1}} \frac{\cos(c \omega_{\bm p_2} |t|)}{\omega^2_{\bm p_2}}  \frac{c \sin(c \omega_{\bm p - \bm p_1-\bm p_2} |t|)}{2\omega_{\bm p-\bm p_1-\bm p_2}}  . 
   \end{aligned}
   \label{eq:Sigmacandsym}
\end{equation}
The integrals in \eqref{eq:Sigmacandsym} differ from their overdamped
counterparts in Sec.~\ref{sec:overdamped}. In the overdamped theory, the time-dependent part of the
statistical propagator is
\begin{equation}
    \Delta_{1, \mathrm{od}}(t,\bm p)
    =
    \frac{1}{\omega_{\bm p}^{2}}
    \exp\left(
        -\frac{\omega_{\bm p}^{2}}{X}|t|
    \right) = \sum_{n=0}^{\infty} \frac{|t|^n}{X^{n}}\frac{(-1)^n(\omega_{\bm p}^2)^{n-1}}{n!}.
    \label{eq:overdamped-time-dependence}
\end{equation}
The momentum dependence therefore enters directly through an exponential
quadratic in \(\bm p\).  Now, by contrast, the statistical and response propagators
contain the oscillatory functions
$
    \cos(c\omega_{\bm p}t),$ and $
   \sin(c\omega_{\bm p}t)/\omega_{\bm p}$. At criticality, these depend on \(c|\bm p|t\). Consequently, exponentiating the
explicit denominators does not by itself make the momentum integrations
Gaussian, and the square-root dispersion remains inside the trigonometric
functions.
This square-root dependence is nevertheless only apparent. Indeed,
\begin{equation}
\begin{aligned}
    \cos(c\omega_{\bm p}t)
    &=
    \sum_{n=0}^{\infty}
    \frac{(-1)^{n}}{(2n)!}
    \left(c^{2}t^{2}\omega_{\bm p}^{2}\right)^{n},
    \\
    \frac{\sin(c\omega_{\bm p}t)}{\omega_{\bm p}}
    &=
    ct\sum_{n=0}^{\infty}
    \frac{(-1)^{n}}{(2n+1)!}
    \left(c^{2}t^{2}\omega_{\bm p}^{2}\right)^{n}.
    \label{eq:sine-analytic-expansion}
\end{aligned}
\end{equation}
Both functions are therefore analytic in
\(\omega_{\bm p}^{2}\), even though they are naturally written
in terms of \(\omega_{\bm p}\).
 A direct term-by-term use of these series is not convenient inside the loop
integral, because it obscures the large-time behavior and does not provide a
uniform representation from which the ultraviolet pole can be extracted. We
instead resum the series through a Hankel-contour representation \cite[\S6.2]{Watson1944}. We introduce
the function\footnote{We thank D.~Seminara for pointing out this representation.}
\begin{equation}
    \mathcal H_n(x)
    =
    \frac{\sqrt{\pi}}{2\pi \ii}
    \int_{\mathcal C_H} \dd s\,
    s^{-n-\frac12}
    \exp\left(s-\frac{x}{4s}\right).
    \label{eq:Hankel-representation}
\end{equation}
The contour $\mathcal{C}_H$ is shown in Fig.~\ref{fig:hankel-contour} and
encircles the negative real axis counterclockwise. It can be deformed into
three pieces: the two branches $C_{\pm}$, running just above and below the
negative real axis, and the small circular arc $C_{\circ}$ around the origin.
Setting $
    x=c^{2}t^{2}\omega_{\bm p}^{2} $, in particular, one has
\begin{equation}
    \mathcal H_0(x)=\cos\sqrt{x},
    \qquad
    \mathcal H_1(x)
    =
    2\frac{\sin\sqrt{x}}{\sqrt{x}}.
    \label{eq:Hankel-trigonometric-functions}
\end{equation}
For fixed contour parameter \(s\), the dependence on the loop momentum is now
Gaussian. The Hankel representation therefore makes it possible to extract
the ultraviolet divergence using standard dimensional-regularization
techniques. In the remainder of this section, we introduce Schwinger parameters for the explicit
propagator denominators and perform the two Gaussian momentum integrations.
The explicit evaluation of the dimensionless contour integral is explained in detail in
Appendix~\ref{app:hankel-integrals}.
\subsection{Renormalization of the velocity sector}
\label{subsec:hankel-prop}

We now apply the representation introduced above to $\Sigma_3$. Substituting
Eqs.~\eqref{eq:Hankel-representation} and
\eqref{eq:Hankel-trigonometric-functions} into the second line of
Eq.~\eqref{eq:Sigmacandsym}, we obtain
\begin{align}
\Sigma_{3}(t,\bm  q)
={}&
\frac{c^{2}|t|\pi^{3/2}}{4}
\int_{\bm  p_{1},\bm  p_{2}}
\frac{1}
{\omega_{\bm  p_{1}}^{2}\omega_{\bm p_{2}}^{2}}
\prod_{i=1}^{3}
\left[
\int_{\mathcal C_{H}}
\frac{\dd s_{i}}{2\pi \ii}\,e^{s_{i}}
\right]
s_{1}^{-1/2}s_{2}^{-1/2}s_{3}^{-3/2}
\exp\left[
-\rho\sum_{i=1}^{3}
\frac{\omega_{\bm  p_{i}}^{2}}{s_{i}}
\right],
\label{eq:Sigmac-Hankel}
\end{align}
where we defined
$
\rho\equiv c^{2}t^{2}/4.
$
It remains to exponentiate the two explicit propagator denominators. We use
\begin{equation}
\frac{1}
{\omega_{\bm  p_{1}}^{2}\omega_{\bm  p_{2}}^{2}}
=
\rho^{2}
\int_{0}^{\infty} \dd x_{1}  \dd x_{2}\,
\exp\left[
-\rho
\left(
x_{1}\omega_{\bm  p_{1}}^{2}
+x_{2}\omega_{\bm  p_{2}}^{2}
\right)
\right].
\label{eq:Sigmac-Schwinger}
\end{equation}
Combining Eqs.~\eqref{eq:Sigmac-Hankel} and
\eqref{eq:Sigmac-Schwinger}, the complete momentum dependence takes the
quadratic form
\begin{equation}
	\exp\left[
	-\rho\left(
	S_1\omega_{\bm p_1}^{2}
	+S_2\omega_{\bm p_2}^{2}
	+S_3\omega_{\bm p_3}^{2}
	\right)
	\right],
	\label{eq:sigmac-quadratic-form}
\end{equation}
with
\begin{equation}
S_{1}=x_{1}+\frac{1}{s_{1}},
\qquad
S_{2}=x_{2}+\frac{1}{s_{2}},
\qquad
S_{3}= \frac{1}{s_{3}} .
\label{eq:Sigmac-Si}
\end{equation}
Thus, for fixed Schwinger and contour parameters, both loop-momentum
integrations are Gaussian. Defining
\begin{equation}
    \mathcal{D} =
S_1 S_2
+S_1 S_3
+S_2 S_3  ,
\label{eq:Sigmac-D-A}
\end{equation}
and
\begin{equation}
    A_{\bm q}
=
S_{1}+S_{2}+S_{3}
+
\frac{S_{1}S_{2}S_{3}}{\mathcal D}\frac{\bm q^{2}}{m^{2}},
\label{eq:A_q}
\end{equation}
the two Gaussian momentum integrals then give
\begin{align}
&\int_{\bm  p_{1},\bm  p_{2}}
\exp\left[
-\rho
\left(
S_{1}\omega_{\bm  p_{1}}^{2}
+S_{2}\omega_{\bm  p_{2}}^{2}
+S_{3}\omega_{\bm  p_{3}}^{2}
\right)
\right]
=
\frac{1}{(4\pi)^{d}}
\rho^{-d} \mathcal D^{-d/2}
\exp\left[-m^{2}\rho A_{\bm  q}\right].
\label{eq:Sigmac-momentum-integral}
\end{align}
Consequently,
\begin{align}
\Sigma_{3}(t,\bm  q)
={}&
\frac{c^{2}|t|\pi^{3/2}}
     {4(4\pi)^{d}}
\int_{0}^{\infty} \dd x_{1} \dd x_{2}
\prod_{i=1}^{3}
\left[
\int_{\mathcal C_{H}}
\frac{\dd s_{i}}{2\pi \ii}\,e^{s_{i}}
\right]
s_{1}^{-1/2}s_{2}^{-1/2}s_{3}^{-3/2}
\mathcal D^{-d/2}
\rho^{\,2-d}
e^{-m^{2}\rho A_{\bm  q}}.
\label{eq:Sigmac-time-parametric}
\end{align}
The above equation is the central simplification
provided by the Hankel representation: all dependence on the external time is
contained in the single scale \(\rho\), while the remaining integrations are
dimensionless.  Since \(\Sigma_3(t,\bm q)\) is even in time, its Fourier transform can be
written as
\begin{equation}
    \Sigma_3(\omega,\bm q)
    =
    2\int_{0}^{\infty} \dd t\,
    \cos(\omega t)\Sigma_3(t,\bm q),
    \label{eq:sigmac-fourier-transform}
\end{equation}
leading to
\begin{align}
    \Sigma_3(\omega,\bm q)
    &=
    \frac{\pi^{3/2}}{(4\pi)^{d}}
    \int_{0}^{\infty}\dd x_1\,\dd x_2
    \left[
        \prod_{i=1}^{3}
        \int_{\mathcal C_H}\frac{\dd s_i}{2\pi \ii}
    \right]
    e^{s_1+s_2+s_3}
    s_1^{-1/2}s_2^{-1/2}s_3^{-3/2}
    \mathcal D^{-d/2}
    \nonumber\\
    &\qquad\times
    \int_{0}^{\infty}\dd\rho\,
    \rho^{2-d}
    e^{-m^{2}A_{\bm q}\rho}
    \cos\left(
        \frac{2\omega}{c}\sqrt{\rho}
    \right).
    \label{eq:sigmac-rho-integral}
\end{align}
At this stage the dependence on the external frequency is entirely contained
in the last one-dimensional integral. 
The scale integral can be performed exactly using the confluent hypergeometric
function ${}_1F_1$ \cite{AbramowitzStegun1964},
\begin{equation}
    \int_{0}^{\infty}\dd\rho\,
    \rho^{a-1}e^{-\beta\rho}
    \cos(2\gamma\sqrt{\rho})
    =
    \Gamma(a)\beta^{-a}
    {}_1F_1\left(
        a;\frac{1}{2};-\frac{\gamma^{2}}{\beta}
    \right).
    \label{eq:rho-integral-identity}
\end{equation}
Taking
$
    a=3-d,
    \beta=m^{2}A_{\bm q},
    \gamma=\omega/c,
$
gives
\begin{align}
    \Sigma_3(\omega,\bm q)
    &=
    \frac{\pi^{3/2}}{(4\pi)^{d}}
    \Gamma(3-d)
    \int_{0}^{\infty}\dd x_1\,\dd x_2
    \left[
        \prod_{i=1}^{3}
        \int_{\mathcal C_H}\frac{\dd s_i}{2\pi \ii}
    \right]
    e^{s_1+s_2+s_3}
    s_1^{-1/2}s_2^{-1/2}s_3^{-3/2}
    \mathcal D^{-d/2}
    \nonumber\\
    &\qquad\times
    \left(m^{2}A_{\bm q}\right)^{d-3}
    {}_1F_1\left(
        3-d;\frac{1}{2};
        -\frac{\omega^{2}}{c^{2}m^{2}A_{\bm q}}
    \right).
    \label{eq:sigmac-hypergeometric}
\end{align}
We are interested in the coefficient \(\sim \omega^{2}\) at zero external
momentum.
For fixed \(m>0\) and small external frequency,
\begin{equation}
    {}_1F_1\left(
        3-d;\frac{1}{2};-x
    \right)
    =
    1-2(3-d)x+\mathcal O(x^{2}).
    \label{eq:hypergeometric-small-omega}
\end{equation}
Substitution into Eq.~\eqref{eq:sigmac-hypergeometric} yields
\begin{equation}
    \Sigma_3(\omega,\bm 0)
    =
    \Sigma_3(0,\bm 0)
    -
    \frac{\omega^{2}}{c^{2}}
    \frac{2\pi^{3/2}}{(4\pi)^{d}}
    \Gamma(4-d)
    (m^{2})^{d-4}
    I_3(d)
    +\mathcal O(\omega^{4}),
    \label{eq:sigmac-low-frequency-Ic}
\end{equation}
where
\begin{align}
    I_3(d)
    &\equiv
    \int_{0}^{\infty}\dd x_1\,\dd x_2
    \left[
        \prod_{i=1}^{3}
        \int_{\mathcal C_H}\frac{\dd s_i}{2\pi \ii}
    \right]
    e^{s_1+s_2+s_3}
    s_1^{-1/2}s_2^{-1/2}s_3^{-3/2}
    \mathcal D^{-d/2}
    A_{\bm 0}^{d-4}.
    \label{eq:Ic-d-definition}
\end{align}
The first term in Eq.~\eqref{eq:sigmac-low-frequency-Ic} is independent of
the external frequency and belongs to the static sunset contribution analyzed
in Sec.~\ref{sec:static}. It therefore does not renormalize
the coefficient $c^{-2}$.
Near \(d=4-\epsilon\), the entire ultraviolet pole is carried by
\begin{equation}
    \Gamma(4-d)
    =\frac{1}{4-d}+\mathcal O(1).
    \label{eq:Gamma-uv-pole}
\end{equation}
The remaining dimensionless integral is regular at \(d=4\), so that only
\(I_3(4)\) is needed for the pole coefficient. Its explicit evaluation,
including the deformation of the Hankel contours and the remaining
Schwinger-parameter integrations, is given in
Appendix~\ref{app:hankel-inertial}. The result is
\begin{equation}
    I_3(4)
    =
    -\frac{2(3\log 2-2)}{3\pi^{3/2}}.
    \label{eq:Ic-four-dimensional-result}
\end{equation}
It is convenient to define
\begin{equation}
    \sigma_3\equiv 4(3\log 2-2),
    \label{eq:sigma-c-definition}
\end{equation}
thus
\begin{equation}
    \left.
    \Sigma_3(\omega,\bm 0)
    \right|_{\omega^{2},\mathrm{div}}
    =
    \frac{\omega^{2}}{c^{2}}
    \frac{\sigma_3}{3(4\pi)^{4}(4-d)}.
    \label{eq:sigmac-single-divergence}
\end{equation}
Combining this result with the tree-level inertial kernel gives
\begin{equation}
    \left.
    \Gamma^{(2)}(\omega,\bm 0,\theta_{12}\bar \theta_{12})
    \right|_{\omega^{2}}
    =
    -\frac{\omega^{2}}{c^{2}}
    \left[
        1+
        \frac{g^{2}\sigma_3}{6(4\pi)^{4}(4-d)}
    \right]
    \theta_{12}\bar \theta_{12}
    +\text{finite}.
    \label{eq:complete-inertial-kernel}
\end{equation}
The ultraviolet pole is removed in minimal subtraction by
\begin{equation}
    Z_c
    =
    1-
    \frac{g_r^{2}}{6(4\pi)^{4}\epsilon}\sigma_3
    +\mathcal O(g_r^{3}).
    \label{eq:Zc-propagating}
\end{equation}
\subsection{Absence of a generated local dissipative term}
\label{subsec:absenceofrenorm}
We next consider the diagram $\Sigma_1$
in Eq.~\eqref{eq:Sigmacandsym}.
According to the sunset decomposition in Eq.~\eqref{eq:sunset},
its contribution to the dissipative part of the two-point function is proportional to
$
 \ii \omega\,\Sigma_1(\omega,\bm q).
$
It can therefore renormalize the local coefficient $X$ only if
$\Sigma_1(\omega,\bm q)$ approaches a finite, nonzero constant as
$\omega\to0$. We now show that this does not occur. This establishes the
invariance of the surface $X=0$; its transverse stability is analyzed in
Sec.~\ref{sec:X-perturbation}.
Repeating the steps used for $\Sigma_3$, we obtain
\begin{equation}
\begin{aligned}
    \Sigma_1(\omega, \bm q) = \frac{2\pi^{\frac32} }{(4\pi)^dc  }
    \prod_{i=1}^3\left[
    \int \dd x_i \frac{\dd s_i}{2\pi \ii } e^{s_i} s_i^{-\frac{1}{2}}
    \right] 
    \mathcal{D}^{-d/2} 
(m^2A_{\bm q})^{d-\frac72} \Gamma\left(d-\frac{7}{2}\right) \tensor[_1]{F}{_1}\left( \frac{7}{2} -d ,\frac{1}{2}, -\frac{\omega^2}{ c^2 m^2  A_{\bm q}}\right),
\end{aligned}
\end{equation}
where we used some shorthand notation defined in Eq.~\eqref{eq:A_q}. With a slight abuse of notation, for this integral $S_3= x_3 + 1/s_3$. 
This difference from Eq.~\eqref{eq:Sigmac-time-parametric} comes from the fact that now there is one more integral over the Schwinger parameter $x_3$. 
Naively, possible singularities are signaled by $\Gamma(d-7/2)$, which has
poles at $d=\frac72, \frac52, \frac32, \frac12$.
Around $d=4$ this Gamma function is finite, and one must instead check
whether the remaining parameter integrations develop endpoint singularities.
As this task is rather involved in general, we focus on two physically motivated limits, which we discuss next. 
\subsubsection{Finite mass} 
For \(m>0\), it is possible to avoid evaluating the full Schwinger--Hankel
representation in order to determine whether
\(\Sigma_1\) contains a constant term. This question can be
answered directly from the spectral support of the integral. Writing the three
cosines as sums of phases gives
\begin{equation}
   \prod_{i=1}^{3}\cos\!\big(c\,\omega_{\bm p_i}|t|\big)
   =\frac{1}{8}\sum_{\{s_i=\pm 1\}} e^{\,\ii c\,\Omega_{\bm s}|t|},
   \qquad
   \Omega_{\bm s}\equiv s_1\omega_{\bm p_1}+s_2\omega_{\bm p_2}+s_3\omega_{\bm p_3} ,
\end{equation}
where $\bm p_3\equiv \bm p-\bm p_1-\bm p_2$.
The Fourier transforms lead to
\begin{equation}
   \Sigma_1(\omega,\bm p)
   =\frac{\pi}{4}\int_{\bm p_1\bm p_2}
   \frac{1}{\omega^2_{\bm p_1}\,\omega^2_{\bm p_2}\,\omega^2_{\bm p_3}}
   \sum_{\{s_i=\pm 1\}}\delta \big(\omega-c\,\Omega_{\bm s}\big).
\end{equation}
We now set $\omega=0$. The channel in which all three energies enter
with the same sign has no support because every
\(\omega_{\bm p_i}\) is positive.
Any mixed-sign channel would instead require one energy to equal the sum of
the other two. For the massive dispersion
$
	\omega_{\bm p}=\sqrt{\bm p^{2}+m^{2}},
$
such an equality cannot be satisfied. Indeed, interpreting
\(\omega_{\bm p}\) as the Euclidean norm of the \((d+1)\)-dimensional vector
\((\bm p,m)\), the triangle inequality gives
\begin{align}
	\omega_{\bm p_i}+\omega_{\bm p_j}
	>
	\omega_{\bm p_i+\bm p_j}.
	\label{eq:massive-triangle-inequality}
\end{align}
Thus none of the energy-conserving delta functions has support at zero
external frequency and momentum. It follows that no local dissipative term is generated in the massive theory.
For every finite mass, the statistical correlator instead develops threshold
behavior at nonzero frequency.
The massless limit needs special care, and we discuss it next.  

\subsubsection{Massless limit at fixed external frequency} Instead of sending
\(\omega\to0\) at fixed \(m>0\) as before, we now take \(m\to0\) while keeping the
external frequency finite.
Under this assumption, the argument used above is not valid.
The confluent hypergeometric function is controlled by the dimensionless
ratio
\begin{equation}
	x_{\bm q}
	\equiv
	\frac{\omega^{2}}{c^{2}m^{2}A_{\bm q}}.
	\label{eq:mass-frequency-ratio}
\end{equation}
The massless limit we are considering
corresponds to \(x_{\bm q}\to+\infty\). Then, we can expand the hypergeometric function as 
\begin{equation}
    {}_1F_1\left(
        \frac72-d;\frac12;-x_{\bm q}
    \right)
    =
    \frac{\Gamma\left(\frac12\right)}
         {\Gamma(d-3)}
    x_{\bm q}^{-\left(\frac72-d\right)}
    \left[1+\mathcal O(x_{\bm q}^{-1})\right],
    \qquad x_{\bm q}\to\infty.
    \label{eq:hypergeometric-large-x}
\end{equation}
Substituting the expression in the self-energy gives
\begin{align}
    \Sigma_1(\omega,\bm q)
    &=
    \frac{2\pi^2}{(4\pi)^d c}
    \frac{
        \Gamma\left(\frac72-d\right)
    }{
        \Gamma(d-3)
    }
    \left(
        \frac{\omega^2}{c^2}
    \right)^{d-\frac72}
    \prod_{i=1}^3
    \left[
        \int \dd x_i
        \int_{\mathcal C_H}\frac{\dd s_i}{2\pi \ii}\,
        e^{s_i}s_i^{-1/2}
    \right]
    \mathcal D^{-d/2},
    \label{eq:symmetric-massless-general-d}
\end{align}
since the factor \(x_{\bm q}^{-a}\) exactly cancels the explicit mass and momentum dependence:
\begin{align}
	\left(m^{2}A_{\bm q}\right)^{d-\frac72}
	x_{\bm q}^{-\left(\frac72-d\right)}
	&=
	\left(
	\frac{\omega^{2}}{c^{2}}
	\right)^{d-\frac72}.
	\label{eq:mass-cancellation}
\end{align}
The massless limit is therefore finite but nonanalytic in the external
frequency and independent of the external momentum $\bm q$.
\begin{equation}
	\Sigma_1(\omega,\bm q)
	\propto
	\left(
	\frac{\omega^{2}}{c^{2}}
	\right)^{d-\frac72}.
	\label{eq:symmetric-general-scaling}
\end{equation}
In particular, at \(d=4\),
\begin{equation}
\begin{aligned}
    \Sigma_1(\omega, \bm q) = \frac{2\pi^{2} }{(4\pi)^4 c  }
    \Gamma\left(-\frac{1}{2}\right)
\frac{|\omega|}{ c  }
    \prod_{i=1}^3\left[
    \int \dd x_i \int \frac{\dd s_i}{2\pi \ii } e^{s_i} s_i^{-\frac{1}{2}}
    \right] 
    \mathcal{D}^{-2} 
    = \frac{2\pi^{2} }{(4\pi)^4 c  }
    \Gamma\left(-\frac{1}{2}\right) \frac{|\omega|}{ c  }I_1,
    \label{eq:Sigma_sym}
\end{aligned}
\end{equation}
where 
\begin{equation}
    I_1 \equiv \prod_{i=1}^3\left[
    \int_{\mathcal C_H} \frac{\dd s_i}{2\pi \ii } e^{s_i} s_i^{-\frac{1}{2}}
    \right] 
    \int_0^{+\infty} \dd x_1\dd x_2\dd x_3
    \frac{1}{\mathcal{D}^{2}}, 
    \label{eq:Isym-def}
\end{equation}
The integral over the Hankel function is computed in Appendix \ref{app:hankel-symmetric}. 
The final result is 
\begin{equation}
   \Sigma_1(\omega,\bm q) 
   = 
\frac{1 }{(4\pi)^3  }
\frac{|\omega|}{ c^2  } 
(6\log 2-4  ). 
\label{eq:Sigma_sym_final}
\end{equation}
Since the contribution of the symmetric self-energy to the dissipative sector is
proportional to $ \ii \omega\Sigma_1(\omega,\bm 0)$, Eq.~\eqref{eq:Sigma_sym}
implies
$
     \ii \omega\Sigma_1(\omega,\bm 0)
    \propto  \ii \omega|\omega| .
$
This finite contribution is nonanalytic and vanishes faster than
$\ii\omega$ as $\omega\to0$. It therefore cannot be absorbed into a
renormalization of the local dissipative coefficient $X$.

Together with the finite-mass result, we conclude that no local
dissipative term is generated on the surface \(X=0\). This establishes that
the surface is invariant under the perturbative RG flow; its stability against
a pre-existing dissipative perturbation is addressed separately in
Sec.~\ref{sec:X-perturbation}.
We close this section by determining the propagating dynamic exponent.

\subsection{RG flow on the dissipationless surface}
\label{sec:propagating-RG-flow}

Having shown that no local dissipative term is generated, we conclude
that \(X=0\) defines an invariant surface of the RG flow. We may therefore
consistently determine the dynamical scaling within this surface by
following the running of the velocity coefficient \(c_r\). In accordance with
our conventions in Appendix~\ref{app:RG_conventions}, we write
\begin{equation}
    c^{-2}
    =
    \mu^{2-2z}\frac{Z_c}{Z_\phi}\,c_r^{-2} .
    \label{eq:cr-renormalization}
\end{equation}
The factor \(Z_\phi\) normalizes the spatial momentum term, whereas
\(Z_c\) contains the additional renormalization of the relativistic temporal
kinetic operator.
Using \eqref{eq:sigma-c-definition}
together with the static field renormalization, we find
\begin{equation}
    \frac{Z_c}{Z_\phi}
    =
    1+
    \frac{g_r^2}{(4\pi)^4\epsilon}
    \left(
        \frac{1}{12}-\frac{\sigma_3}{6}
    \right)
    +\mathcal O(g_r^3).
    \label{eq:Zc_over_Zphi}
\end{equation}
The anomalous contribution associated with the inertial coefficient is
defined by
\begin{equation}
    \gamma_c(g_r)
    \equiv
    -\beta_g(g_r)
    \frac{\partial}{\partial g_r}
    \log\frac{Z_c}{Z_\phi}.
    \label{eq:gamma_c_definition}
\end{equation}
Since the ratio \(Z_c/Z_\phi\) starts at order \(g_r^2\), only
\(\beta_g(g_r)=-\epsilon g_r+\mathcal O(g_r^2)\) is required at this order.
Hence
\begin{align}
    \gamma_c(g_r)
    &=
    \frac{g_r^2}{(4\pi)^4}
    \left(
        \frac{1}{6}-\frac{\sigma_3}{3}
    \right)
    +\mathcal O(g_r^3)
    =
    \frac{g_r^2}{(4\pi)^4}
    \left(
        \frac{17}{6}-4\log2
    \right)
    +\mathcal O(g_r^3).
    \label{eq:gamma_c}
\end{align}
The beta function of the renormalized velocity is
\begin{equation}
    \beta_{c_r}
    \equiv
    \left.\mu\frac{\mathrm d c_r}{\mathrm d\mu}\right|_{\mathrm{bare}}
    =
    \left(1-z-\frac{1}{2}\gamma_c\right)c_r.
    \label{eq:beta-cr}
\end{equation}
At a propagating fixed point, \(c_r\) is finite and nonzero, so
\(\beta_{c_r}=0\). At the Gaussian fixed point, where \(g_r^*=0\) and
\(\gamma_c=0\), this gives \(z_{\mathrm{prop},\mathrm{G}}=1\). At the
Wilson--Fisher fixed point it gives
\begin{equation}
    z_{\mathrm{prop},\mathrm{WF}}
    =
    1-\frac{1}{2}\gamma_c(g_{\mathrm{WF}}).
    \label{eq:zprop_definition}
\end{equation}
Using the Wilson--Fisher fixed point value
we obtain
\begin{equation}
    z_{\mathrm{prop},\mathrm{WF}}
    =
    1-
    \frac{\epsilon^2}{9}
    \left(
        \frac{17}{12}-2\log 2
    \right)
    +\mathcal O(\epsilon^3)
    \simeq
    1-0.00337\,\epsilon^2 .
\end{equation}
The above equation implies that the characteristic critical time
scale behaves as
\begin{equation}
    \tau(\xi)\sim\xi^{z_{\mathrm{prop},\mathrm{WF}}}, \qquad
    \omega\sim k^{z_{\mathrm{prop},\mathrm{WF}}}.
\end{equation}
We conclude that, within the two-loop $\epsilon$-expansion, interactions produce
a small negative correction to the ballistic value $z=1$, so that the
critical time scale grows slightly more slowly than the correlation
length. The correction is numerically very small: setting $\epsilon=1$
gives \(z_{\mathrm{prop},\mathrm{WF}}\simeq0.9966\). This exponent describes the RG flow tangent to the invariant surface
$X=0$. Whether that surface is stable against a nonzero dissipative
perturbation is a separate question, addressed in the next Sec.~\ref{sec:X-perturbation}.

\section{Renormalization of the dissipative perturbation}
\label{sec:X-perturbation}
We now test the stability of the propagating theory discussed in the previous section against a small local
dissipative perturbation. From the canonical dimensions derived in
Subsec.~\ref{eq:Subsection canonical dim}, the dissipative coupling has
$[X]=2-z$ and is therefore relevant at tree level near the propagating fixed
point. Interactions modify this scaling through the renormalization of the
corresponding composite operator. In Subsec.~\ref{subsec:composite_field}, we
define its insertion using a local source. In
Subsec.~\ref{subsec:homogeneous-X-projection}, we project onto a homogeneous
coupling and evaluate the loop corrections. Finally,
Subsec.~\ref{subsec:RG_X} extracts the anomalous contribution and the
associated RG eigenvalue.

\subsection{Composite-operator insertion} \label{subsec:composite_field}

The dissipative perturbation is generated by the microscopic superspace
operator
\begin{equation}
    \mathcal{O}_X[\Psi](Z) = \widebar D \Psi(Z) \cdot D \Psi(Z) \,.
\end{equation}
To define insertions of this operator, we introduce a local source $X(Z)$ and consider the deformed action
\begin{equation}
   S[\Psi;X]=S_{\rm prop}[\Psi]+\int_Z X(Z)\mathcal O_X[\Psi](Z),
\end{equation}
where \(S_{\mathrm{prop}}\) denotes the interacting dissipationless theory (evaluated at
\(X=0\)). 
The generating functional in the presence of the ordinary source
\(J(Z)\) and the composite source \(X(Z)\) is
\begin{equation}
\mathcal Z[J;X]
=
\int\mathcal D\Psi\,
\exp\left[
-S[\Psi;X]
+
\int_Z J(Z)\Psi(Z)
\right].
\label{eq:Z-with-composite-source}
\end{equation}
We define
$
W[J;X]
=
\log\mathcal Z[J;X].
$
Using the Legendre transform
\(\Gamma[\Phi;X]=J\cdot\Phi-W[J;X]\), and differentiating at fixed
background field \(\Phi\), one gets
\begin{equation}
\left\langle\mathcal O_X(Z)\right\rangle_{\Phi,X} =\left.
\frac{\delta\Gamma[\Phi;X]}{\delta X(Z)}
\right|_\Phi
=
-
\left.
\frac{\delta W[J;X]}{\delta X(Z)}
\right|_J
.
\end{equation}
Note that setting \(X=0\) evaluates the insertion in the unperturbed theory. We consequently define
\begin{equation}
\Gamma_{\mathcal O_X}[Z;\Phi]
\equiv
\left.
\frac{\delta\Gamma[\Phi;X]}
     {\delta X(Z)}
\right|_{X=0}.
\label{eq:Gamma-OX-functional}
\end{equation}
This is a functional of the background field \(\Phi\). Its expansion
starts as
\begin{align}
\Gamma_{\mathcal O_X}[Z;\Phi]
=&
\Gamma_{\mathcal O_X}^{(1;0)}(Z)+
\frac{1}{2}
\int_{Z_1,Z_2}
\Phi(Z_1)\Phi(Z_2)
\Gamma_{\mathcal O_X}^{(1;2)}
(Z;Z_1,Z_2)
+\mathcal O(\Phi^4).
\label{eq:Gamma-OX-background-expansion}
\end{align}
The 1PI vertex with one insertion of \(\mathcal O_X\) and two external
superfields is then
\begin{equation}
\Gamma_{\mathcal O_X}^{(1;2)}
(Z;Z_1,Z_2)
=
\left.
\frac{
\delta^3\Gamma[\Phi;X]
}{
\delta X(Z)\,
\delta\Phi(Z_1)\,
\delta\Phi(Z_2)
}
\right|_{\Phi=0,\,X=0}.
\label{eq:OX_composite_vertex}
\end{equation}
The first superscript counts insertions of \(\mathcal O_X\), while the
second counts external superfield legs.

\subsection{Projection onto a homogeneous dissipative coupling}
\label{subsec:homogeneous-X-projection}
 
To determine the anomalous scaling of $X$, it is sufficient to integrate the
position of the composite insertion and consider the vertex
\begin{equation}
\overline\Gamma_{\mathcal O_X}^{(1;2)}
(Z_1,Z_2)
\equiv
\int_Z
\Gamma_{\mathcal O_X}^{(1;2)}
(Z;Z_1,Z_2).
\label{eq:integrated-composite-vertex}
\end{equation} 
This projection sets the momentum carried by the insertion to zero; it does
not set the momenta of the external superfields to zero. Equivalently, it is
obtained by differentiating with respect to a homogeneous coupling $X_{\rm c}$:
\begin{align}
\overline{\Gamma}_{\mathcal O_X}^{(1;n)}
(Z_1,\ldots,Z_n)
&\equiv
\left.
\int_Z
\frac{\delta^{n+1}\Gamma[\Phi;X]}
{\delta X(Z)\,
 \delta\Phi(Z_1)\cdots\delta\Phi(Z_n)}
\right|_{\Phi=0,\,X=0}=
\left.
\frac{\partial}{\partial X_{\mathrm c}}
\frac{\delta^n
\Gamma[\Phi;X(Z)=X_{\mathrm c}]}
{\delta\Phi(Z_1)\cdots\delta\Phi(Z_n)}
\right|_{\Phi=0,\,X_{\mathrm c}=0},
\label{eq:79}
\end{align}
where in the last expression the action is evaluated at constant coupling.
At tree level, the insertion vertex is
\begin{equation}
\mathcal{V}_X(Z,Z_1,Z_2) =  \frac{\delta \mathcal{ O}_X[\Phi](Z)}{ \delta \Phi(Z_1)\delta \Phi(Z_2)},
   \label{eq:dissipative_vertex_free}
\end{equation}
and its integral is
\begin{align}
\int_Z
\mathcal{V}_X(Z,Z_1,Z_2)
=
\left[
D_{Z_1},
\widebar D_{Z_1}
\right]
\delta(Z_1-Z_2)
=-2K_X(Z_1,Z_2).
\label{eq:integrated-tree-insertion}
\end{align}
Here we used the normalization
$K_X=\frac12[\widebar D,D]\delta$ introduced in
Eq.~\eqref{eq:KX-coordinate}. The renormalization constant $Z_X$ is determined by the
ultraviolet divergence multiplying this same kernel.
More explicitly, the short-distance expansion of the integrated two-point vertex
may contain several local structures,
\begin{align}
\overline{\Gamma}_{\mathcal O_X}^{(1;2)}
={}&
C_X\, K_X(Z_1, Z_2)
+
C_0\,\delta(Z_1-Z_2)
+\cdots .
\label{eq:local-operator-decomposition}
\end{align}
 Only the coefficient \(C_X\) contributes to the
multiplicative renormalization of the dissipative coupling\footnote{In
particular, a term proportional to \(\delta(Z_1-Z_2)\), although local,
does not by itself renormalize \(X\).}.
The second term $\sim C_0$ is a distinct
static contact structure.
Up to two-loop order, the integrated composite vertex can be organized as
\begin{align}
\overline{\Gamma}_{\mathcal O_X}^{(1;2)}
={}&
-2K_X
+
\overline{\Gamma}_{\mathcal O_X,1\ell}^{(1;2)}
+
\overline{\Gamma}_{\mathcal O_X,\mathrm{db}}^{(1;2)}
+
\overline{\Gamma}_{\mathcal O_X,\mathrm{sun}}^{(1;2)}
+
\mathcal O(g^3).
\label{eq:projected-vertex-loop-decomposition}
\end{align}
The different terms have distinct roles, which we discuss in turn. The
perturbative series follows from Eq.~\eqref{eq:79} and the loop expansion of
the effective action in Eq.~\eqref{eq:two_loop_effective_action}. Before
differentiating, the theory is evaluated at finite homogeneous $X_{\rm c}$.
Since $\mathcal O_X$ is quadratic in the field, the topologies are the same as
at $X=0$ and the derivative acts only on the internal propagators. Thus
\begin{equation}
    \overline{\Gamma}_{\mathcal O_X,i}^{(1;2)}
    (Z_1,Z_2)
    =
    \left.
    \frac{\partial}{\partial X_{\rm c}}
    \Gamma_{i}^{(2)}
    (Z_1,Z_2;X_{\rm c})
    \right|_{X_{\rm c}\to0^+}.
    \label{eq:insertion-derivative-general}
\end{equation}
To keep the signs explicit, we define the inserted propagator by
\begin{align}
\Delta_{X}(Z_1,Z_2)
&\equiv
-\left.
\frac{\partial\Delta_0(Z_1,Z_2;X_{\rm c})}{\partial X_{\rm c}}
\right|_{X_{\rm c}\to0^+}
=
\int_{Z, Z_3,Z_4}
\Delta_0(Z_1,Z_3)\,
\mathcal V_X(Z;Z_3,Z_4)\,
\Delta_0(Z_4,Z_2).
\label{eq:local-inserted-propagator}
\end{align}
where the second equality follows from
$\partial_X\Delta_0=-\Delta_0(\partial_X\Delta_0^{-1})\Delta_0$ and
$\partial_X\Delta_0^{-1}=\mathcal V_X$. Consequently, differentiating a loop
diagram replaces each differentiated internal line according to
$\partial_X\Delta_0=-\Delta_X$.
Inserting the tree composite vertex
in Eq. \eqref{eq:dissipative_vertex_free}
we get
\begin{align}
\Delta_{X}(Z_1,Z_2)
= \int_Z \Delta_0(Z_1,Z) [D_Z,\widebar D_Z]\Delta_0(Z,Z_2).
\label{eq:local-inserted-propagator-explicit}
\end{align}
Using the results of Appendix \ref{app:superspace-convolution}, 
we get
\begin{equation}
    \Delta_X = \Delta_{1X} +\theta_{12}\left[\bar \theta_{12}^+ \Delta_{2X}+\bar\theta_{12}\Delta_{3X}\right],
\end{equation}
with
\begin{equation}
 [D_{Z_1},\widebar D_{Z_1}] \Delta_0(Z_1,Z_2)  = \Delta_1^{\prime}(t_{12}) - 2 \left[
   \Delta_2(t_{12}) 
   + \Delta_3(t_{12}) \right] -\theta_{12}
   \left[
   \bar\theta_{12} 
   \Delta_2^{\prime}(t_{12}) +
   \bar \theta_{12}^+
   \Delta_3^{\prime}(t_{12})
   \right],
\end{equation}
where the prime denotes differentiation with respect to $t_{12}$. Explicitly\footnote{
The first component can equivalently be written as
$
    \Delta_{1X} = -\frac{c^2t^2}{2\omega_{\bm p}}j_{1}(c\omega_{\bm p}|t|)\, 
$
with $j_1$ the spherical Bessel function. 
},
\begin{equation}
\begin{aligned}
\Delta_{1X}(t,\bm p)
&=
\frac{c}{2\omega_{\bm p}^3}
\left[
c\omega_{\bm p}|t| \cos(c\omega_{\bm p} |t|)
-
\sin(c\omega_{\bm p} |t|)
\right],
\\
\Delta_{2X}(t,\bm p)
&=
-\,
\frac{c^3}{4\omega_{\bm p}}\,
t\sin(c\omega_{\bm p} |t|),
\\
\Delta_{3X}(t,\bm p)
&=
\frac{c^3}{4\omega_{\bm p}}\,
|t| \sin(c\omega_{\bm p} |t|),
\end{aligned}
\label{eq:Delta123X}
\end{equation}
These components satisfy the equilibrium Ward identities by construction:
\begin{equation}
    \Delta_{2X} = \frac12 \partial_t \Delta_{1X},  \qquad
    \Delta_{3X} = -\frac12 \operatorname{sgn}(t)\partial_t \Delta_{1X}.
\end{equation}

\subsubsection{One-loop}
After integration over the insertion point, the one-loop term is
proportional to the inserted propagator evaluated at coincident
superspace points:
\begin{equation}
\overline{\Gamma}_{\mathcal O_X,1\ell}^{(1;2)}
(Z_1,Z_2)
\propto
g\,
\delta(Z_1-Z_2)\,
\Delta_X(Z_1,Z_1).
\label{eq:one-loop-insertion-coincident}
\end{equation}
At coincident Grassmann coordinates, only the scalar component \(\Delta_{1X}\)
of the inserted propagator survives. Using Eq. \eqref{eq:Delta123X},
one immediately finds
$
\Delta_{1X}(0,\bm  p)=0.
$
Consequently,
\begin{equation}
\Delta_X(Z,Z)
=
\int_{\bm  p}\Delta_{1X}(0,\bm  p)
=
0,
\label{eq:coincident-inserted-propagator-zero}
\end{equation}
and hence
$
\overline{\Gamma}_{\mathcal O_X,1\ell}^{(1;2)}
=0.
$
There is therefore no one-loop counterterm for the dissipative operator.
\subsubsection{Double-bubble }

We next consider the double-bubble topology.  The double-bubble contribution to the ordinary
two-point vertex has the form
\begin{equation}
    \Gamma_{\mathrm{db}}^{(2)}
    (Z_1,Z_2;X_{\mathrm c})
    =
    -\frac{g^2}{4}\,
    \delta(Z_1-Z_2)\,
    T_1(X_{\mathrm c})T_2(X_{\mathrm c}),
    \label{eq:db_finite_X}
\end{equation}
where
\begin{equation}
    T_1(X_{\mathrm c})
    =
    \int_{\bm p} \Delta_{1}(0,\bm p;X_c),
    \qquad
    T_2(X_{\mathrm c})
    =
    \int_Y
    \left[\Delta_{1}(Y,Z;X_c)\right]^2 .
    \label{eq:T1_T2_finite_X}
\end{equation}
The propagator $\Delta_{1}(Y,Z;X_{\rm c})$ is evaluated in the theory with
finite homogeneous dissipative coupling $X_{\rm c}$.
The vertex with one homogeneous insertion of $\mathcal O_X$ is therefore
\begin{equation}
    \overline{\Gamma}_{\mathcal O_X,\mathrm{db}}^{(1;2)}
    (Z_1,Z_2)
    =
    \left.
    \frac{\partial}{\partial X_{\mathrm c}}
    \Gamma_{\mathrm{db}}^{(2)}
    (Z_1,Z_2;X_{\mathrm c})
    \right|_{X_{\mathrm c}\to0^+}.
    \label{eq:db-insertion-derivative}
\end{equation}
 The Ward identity and the causal decomposition can therefore be applied and
give 
\begin{equation}
    T_2(X_{\mathrm c})
    =
    \int_{\bm p}
    \left[
        \Delta_1(0,\bm p; X_c)
    \right]^2 .
    \label{eq:T2_equal_time_finite_X}
\end{equation}
Equilibrium fixes the equal-time correlator entirely through the static
quadratic kernel:
\begin{equation}
    \Delta_{1}(0,\bm p;X_{\rm c})
    =
    \frac{1}{\omega_{\bm p}^2},
    \label{eq:equal_time_independent_X}
\end{equation}
independently of $X_{\mathrm c}$. It follows that
\begin{equation}
    T_1(X_{\mathrm c})
    =
    \int_{\bm p}\frac{1}{\omega_{\bm p}^2},
    \qquad
    T_2(X_{\mathrm c})
    =
    \int_{\bm p}\frac{1}{\omega_{\bm p}^4},
\end{equation}
and therefore
\begin{equation}
    \frac{\partial T_1}{\partial X_{\mathrm c}}
    =
    \frac{\partial T_2}{\partial X_{\mathrm c}}
    =
    0.
\end{equation}
Substituting these results into
Eq.~\eqref{eq:db-insertion-derivative}, we obtain
\begin{equation}
\overline{\Gamma}_{\mathcal O_X,\mathrm{db}}^{(1;2)}
    (Z_1,Z_2)=0
    .
\end{equation}
Thus the double-bubble topology generates neither the static contact
structure proportional to $C_0$ nor the dissipative derivative structure
proportional to $C_X$.
The same result can also be verified using the Hankel representation.

\subsubsection{Sunset}
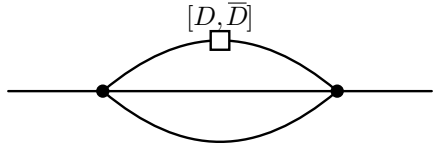
\begin{figure}[t]
    \centering
    \begin{tikzpicture}[
        baseline=(current bounding box.center),
        propagator/.style={
            draw=black,
            line width=0.9pt,
            line cap=round
        },
        vertex/.style={
            circle,
            fill=black,
            inner sep=0pt,
            minimum size=5pt
        },
        insertion/.style={
            rectangle,
            draw=black,
            fill=white,
            line width=0.9pt,
            inner sep=0pt,
            minimum size=7pt
        }
    ]
        \coordinate (vL) at (-1.55,0);
        \coordinate (vR) at ( 1.55,0);

        \draw[propagator] (-2.8,0) -- (vL);
        \draw[propagator] (vR) -- (2.8,0);

        \draw[propagator]
            (vL)
            to[out=40,in=140,looseness=1.15]
            coordinate[pos=0.5] (ins)
            (vR);

        \draw[propagator] (vL) -- (vR);

        \draw[propagator]
            (vL)
            to[out=-40,in=-140,looseness=1.15]
            (vR);

        \node[insertion] at (ins) {};
        \node at ([yshift=9pt]ins) {$[D,\widebar D]$};

        \node[vertex] at (vL) {};
        \node[vertex] at (vR) {};
    \end{tikzpicture}

    \caption{
        Representative nonlocal two-loop sunset contribution with one
        dissipative insertion, denoted by the square.
    }
    \label{fig:sunset-X}
\end{figure}

The first non-vanishing correction to the operator
\(\mathcal O_X\) is generated by the two-loop sunset diagram. 
 The corresponding 1PI vertex with one homogeneous insertion and two external superfields
is 
\begin{equation}
\overline{\Gamma}_{\mathcal O_X,\mathrm{sun}}^{(1;2)}
(Z_1,Z_2)
=
\frac{g^2}{2}\,
\Delta_{0}(Z_1,Z_2)^2
\Delta_X(Z_1,Z_2)
.
\label{eq:projected-sunset-insertion}
\end{equation}
A representative insertion is shown in Fig. \ref{fig:sunset-X}.
Unlike the one-loop and double-bubble terms, this contribution is nonlocal in 
\(Z_1-Z_2\). Its short-distance expansion nevertheless contains a local
ultraviolet divergence proportional to the tree-level dissipative
kernel. 
Following Appendix~\ref{app:superspace-convolution}, we introduce the three-component kernel
\begin{equation}
	\Pi_X(Z_1,Z_2)
	\equiv
	\Delta_0(Z_1,Z_2)^2\Delta_X(Z_1,Z_2)
	=
	\Pi_{1X}
	+
	\theta_{12}
	\left[
	\bar \theta_{12}\Pi_{3X}
	+\bar \theta^+_{12}\Pi_{2X}
	\right].
	\label{eq:PiX-superkernel}
\end{equation}
Because every Grassmann-dependent term contains the common factor
\(\theta_{12}\), the pointwise-product rule of Appendix~\ref{app:superspace-convolution} gives
\begin{align}
	\Pi_{1X}
	&=
	\Delta_1^2\Delta_{1X},
	\nonumber\\
	\Pi_{2X}
	&=
	2\Delta_1\Delta_2\Delta_{1X}
	+\Delta_1^2\Delta_{2X},
	\label{eq:PiX-components}\\
	\Pi_{3X}
	&=
	2\Delta_1\Delta_3\Delta_{1X}
	+\Delta_1^2\Delta_{3X}.
	\nonumber
\end{align}
Thus the complete superspace kernel \(\Pi_X\) is fixed once
\(\Pi_{1X}\) is known. In spatial Fourier space, its first component is
\begin{align}
	\Pi_{1X}(t,\bm q)
	=
	\int_{\bm p_1,\bm p_2}
	\Delta_1(t,\bm p_1)\Delta_1(t,\bm p_2)
	\Delta_{1X}(t,\bm q-\bm p_1-\bm p_2).
	\label{eq:Pi1X-kernel}
\end{align}
Equation~\eqref{eq:PiX-superkernel} ensures that a local pole in \(\Pi_{1X}\)
reconstructs the complete local dissipative kernel.  It is therefore
sufficient to evaluate \(\Pi_{1X}(\omega=0,\bm q= \bm 0)\).
For the propagating theory, the two component functions entering
Eq.~\eqref{eq:Pi1X-kernel} are $\Delta_1(t, \bm p)$ and 
\begin{align}
	\Delta_{1X}(t,\bm p)
	&=
	\frac{c}{2\omega_{\bm p}^3}
	\left[
	c\omega_{\bm p}|t|\cos(c\omega_{\bm p}|t|)
	-\sin(c\omega_{\bm p}|t|)
	\right]
	=
	-\frac{c^4|t|^3}{8}\,
	\mathcal H_2(c^2\omega_{\bm p}^2t^2).
	\label{eq:Delta1-and-Delta1X-Hankel}
\end{align}
Thus the two ordinary lines supply two \(\mathcal H_0\) factors, whereas
the inserted line supplies one \(\mathcal H_2\).  Explicitly,
\begin{equation}
  \Pi_{1X}(t,\bm p_1,\bm p_2,\bm p_3)
  =
  -\frac{c^4|t|^3}
  {8\omega_{\bm p_1}^2\omega_{\bm p_2}^2}
  \mathcal H_0(c^2\omega_{\bm p_1}^2t^2)
  \mathcal H_0(c^2\omega_{\bm p_2}^2t^2)
  \mathcal H_2(c^2\omega_{\bm p_3}^2t^2),
  \label{eq:Pi1X-Hankel-product}
\end{equation}
with
\(\bm p_3=\bm q-\bm p_1-\bm p_2\).
The remaining reduction parallels the symmetric-sunset calculation in
Subsec.~\ref{subsec:absenceofrenorm}, but we spell out the intermediate quantities needed below:
\begin{align}
  \Pi_{1X}(t,\bm q)
  &=
  -\frac{c^4|t|^3\pi^{3/2}}{16(4\pi)^d}
  \int_0^\infty\!\dd x_1\dd x_2
  \prod_{i=1}^3
  \left[
    \int_{\mathcal C_H}\frac{\dd s_i}{2\pi \ii}
  \right]
  e^{s_1+s_2+s_3}
  s_1^{-1/2}s_2^{-1/2}s_3^{-5/2}
  \mathcal D^{-d/2}\rho^{2-d}
  e^{-m^2\rho A_{\bm q}}.
  \label{eq:Pi1X-after-momenta}
\end{align}
This form makes the origin of the ultraviolet pole transparent.  Since the
kernel is even in \(t\), its Fourier transform is
\begin{equation}
  \Pi_{1X}(\omega,\bm q)
  =
  2\int_0^\infty\!\dd t\,
  \cos(\omega t)\Pi_{1X}(t,\bm q)=
  -\frac{\pi^{3/2}}{(4\pi)^{4-\epsilon}}\,
  \Gamma(4-d)\,
  I_X(d;\omega,\bm q).
  \label{eq:Pi1X-epsilon-form}
\end{equation}
Here the dimensionless parameter integral is defined by
\begin{align}
  I_X(d;\omega,\bm q)
  &\equiv
  \int_0^\infty\!\dd x_1\dd x_2
  \prod_{i=1}^3
  \left[
    \int_{\mathcal C_H}\frac{\dd s_i}{2\pi \ii}
  \right]
  e^{s_1+s_2+s_3}
  s_1^{-1/2}s_2^{-1/2}s_3^{-5/2}
  \mathcal D^{-d/2}
  \nonumber\\
  &\quad\times
  (m^2A_{\bm q})^{-(4-d)}
  {}_1F_1\!\left(
    4-d;\frac12;
    -\frac{\omega^2}{c^2m^2A_{\bm q}}
  \right).
  \label{eq:IX-full-definition}
\end{align}
Thus all dependence on the external frequency and momentum is isolated in
the last line of Eq.~\eqref{eq:IX-full-definition}.  The remaining
parameter integrations are then regular at \(d=4\), while
\begin{equation}
  (m^2A_{\bm q})^{-4+d}=1+\mathcal O(4-d),
  \qquad
  {}_1F_1\!\left(
    4-d;\frac12;
    -\frac{\omega^2}{c^2m^2A_{\bm q}}
  \right)
  =
  1+\mathcal O(4-d).
  \label{eq:IX-external-independence}
\end{equation}
It follows that
\begin{equation}
  I_X(d;\omega,\bm q)
  =
  I_X(4)+\mathcal O(4-d),
  \label{eq:IX-expansion}
\end{equation}
where \(I_X(4)\) is independent of \(m\), \(\omega\), and \(\bm q\).  Hence the
pole is local:
\begin{equation}
\Pi_{1X}(\omega,\bm q)
  =
  -\frac{\pi^{3/2}I_X(4)}{(4\pi)^4(4-d)}+\mathcal{O}(1).
  \label{eq:Pi1X-local-pole}
\end{equation}
The remaining dimensionless integral is evaluated in Appendix~\ref{app:hankel-dissipative}, with
the result
\begin{equation}
  I_X(4)
  =
  \frac{2(2\log 2-1)}{\pi^{3/2}}.
  \label{eq:IX-result-main}
\end{equation}
Defining
$
  \widetilde{\sigma}_X
  \equiv
  2(2\log 2-1),
$
we obtain
\begin{equation}
\Pi_{1X}(0,\bm 0)
  =
  -\frac{\widetilde{\sigma}_X}{(4\pi)^4(4-d)} +\mathcal{O}(1).
  \label{eq:Pi1X-final-pole}
\end{equation}

\subsection{Renormalization-group flow of the dissipative perturbation}
\label{subsec:RG_X} 
The preceding calculation determines the ultraviolet renormalization of the
dissipative operator. Since the tree-level insertion is $-2K_X$, while the
sunset contributes $(g_r^2/2)\Pi_X$, pole cancellation requires
\begin{equation}
    Z_X
    =
    1-
    \frac{g_r^2}{4(4\pi)^4\epsilon}\,
    \widetilde{\sigma}_X
    +\mathcal O(g_r^3).
    \label{eq:ZX-propagating}
\end{equation}
Combining this result with the field renormalization
$
    Z_\phi
$
gives
\begin{equation}
    \frac{Z_X}{Z_\phi}
    =
    1+
    \frac{g_r^2}{(4\pi)^4\epsilon}
    \left(
        \frac{1}{12}-\frac{\widetilde{\sigma}_X}{4}
    \right)
    +\mathcal O(g_r^3).
\end{equation}
Using the minimal-subtraction identity collected in
Appendix~\ref{app:RG_conventions}, the anomalous contribution associated with
the dissipative perturbation is therefore
\begin{align}
    \gamma_X(g_r)
    &=
    \frac{g_r^2}{(4\pi)^4}
    \left(
        \frac{1}{6}-\frac{\widetilde{\sigma}_X}{2}
    \right)
    +\mathcal O(g_r^3)
    =
    -\frac{g_r^2}{(4\pi)^4}
    \left(
        2\log2-\frac{7}{6}
    \right)
    +\mathcal O(g_r^3).
\end{align}
Notice that \(\gamma_X(g_r)<0\). Evaluating this expression at the
Wilson--Fisher fixed point gives
\begin{equation}
    \gamma_X(g_{\rm WF} )
    =
    -\frac{\epsilon^2}{9}
    \left(
                2\log 2-\frac{7}{6}
    \right)
    +\mathcal O(\epsilon^3).
    \label{eq:gammaX_prop_fixed_point}
\end{equation}

\section{Fixed-point structure and local stability}
\label{sec:fixed-point-summary}
\begin{figure}
\centering
\begin{tikzpicture}[
    x=1.25cm,
    y=1cm,
    scale=0.96,
    transform shape,
    font=\small,
    >={Stealth[length=2.0mm,width=1.35mm]},
    axis/.style={
        ->,
        draw=axiscol,
        line width=0.75pt
    },
    guide/.style={
        draw=guidecol,
        line width=0.65pt
    },
    staticflow/.style={
        ->,
        draw=axiscol,
        line width=0.95pt
    },
    propflow/.style={
        ->,
        draw=propcol,
        line width=0.95pt
    },
    odflow/.style={
        ->,
        draw=odcol,
        line width=0.95pt
    },
    fp/.style={
        circle,
        inner sep=0pt,
        minimum size=5.6pt
    }
]

\coordinate (GP) at (1.80,1.40);
\coordinate (WP) at (6.80,1.40);
\coordinate (GO) at (1.80,5.00);
\coordinate (WO) at (6.80,5.00);

\draw[axis]
    (GP) -- (8.05,1.40)
    node[right=1pt] {\(g_r\)};

\draw[axis]
    (GP) -- (1.80,6.05)
    node[above=1pt] {\(\rho_r\)};

\draw[guide] (GP) -- (WP);
\draw[guide] (GO) -- (WO);
\draw[guide] (WP) -- (WO);

\node[fp,fill=propcol] at (GP) {};
\node[fp,fill=propcol] at (WP) {};
\node[fp,fill=odcol]   at (GO) {};
\node[fp,fill=odcol]   at (WO) {};

\node[
    anchor=north,
    align=center
] at (1.80,1.12)
    {Gaussian\\[-1pt]
     \scriptsize static repulsive};

\node[
    anchor=north,
    align=center
] at (6.80,1.12)
    {Wilson--Fisher\\[-1pt]
     \scriptsize static attractive};

\node[
    anchor=east,
    align=right,
    text=propcol
] at (1.55,1.40)
    {Propagating $X=0$\\[-1pt]
     \scriptsize dynamic repulsive};

\node[
    anchor=east,
    align=right,
    text=odcol
] at (1.55,5.00)
    {Overdamped $c=\infty$\\[-1pt]
     \scriptsize dynamic attractive};

\node[
    anchor=south west,
    text=propcol
] at (2.00,1.52)
    {\(z_{\mathrm{prop},\mathrm{G}}=1\)};

\node[
    anchor=south east,
    text=propcol
] at (6.62,1.52)
    {\(z_{\mathrm{prop},\mathrm{WF}}
      \simeq1-0.00337\,\epsilon^2\)};

\node[
    anchor=south west,
    text=odcol
] at (2.00,5.12)
    {\(z_{\mathrm{od},\mathrm{G}}=2\)};

\node[
    anchor=south east,
    text=odcol
] at (6.62,5.12)
    {\(z_{\mathrm{od},\mathrm{WF}}
      \simeq2+0.01345\,\epsilon^2\)};


\draw[staticflow]
    (2.05,1.40) -- (2.90,1.40);

\draw[staticflow]
    (5.70,1.40) -- (6.55,1.40);

\draw[staticflow]
    (2.05,5.00) -- (2.90,5.00);

\draw[staticflow]
    (5.70,5.00) -- (6.55,5.00);


\draw[propflow]
    (1.80,1.68) -- (1.80,2.48);

\draw[propflow]
    (6.80,1.68) -- (6.80,2.48);

\draw[odflow]
    (1.80,3.92) -- (1.80,4.72);

\draw[odflow]
    (6.80,3.92) -- (6.80,4.72);

\end{tikzpicture}
\caption{Local stability of the four fixed points on the mass-tuned critical
surface. The propagating and overdamped boundaries lie
at $\rho_r=0$ and $\rho_r=+\infty$, respectively. In the static direction, the Gaussian
fixed points are repulsive and the Wilson--Fisher fixed points are attractive.
In the dynamical direction, the propagating fixed points are repulsive and the
overdamped fixed points are attractive. Consequently, the propagating
Wilson--Fisher and overdamped Gaussian fixed points are saddles, while the
overdamped Wilson--Fisher fixed point is the only fixed point attractive in
both directions.}
\label{fig:dynamical-fixed-points}
\end{figure}
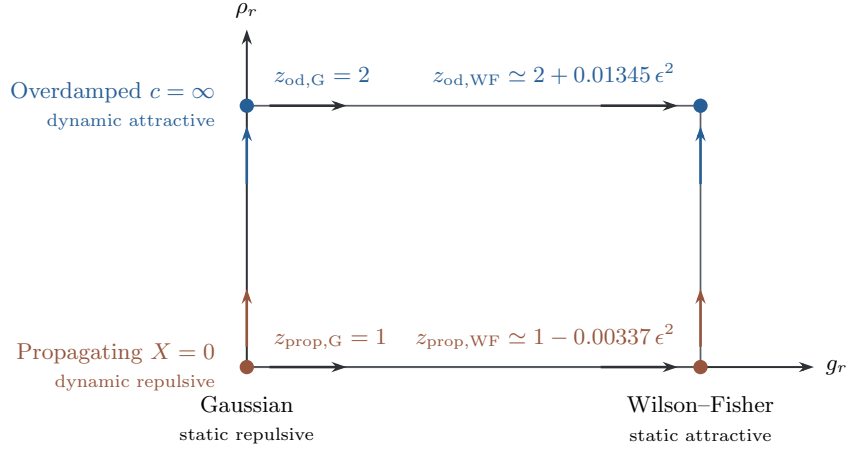
We now combine the two dynamical endpoints with the Gaussian and
Wilson--Fisher static fixed points, following the standard field-theoretic
classification of dynamical universality
\cite{HohenbergHalperin1977,FolkMoser2006}.
Figure~\ref{fig:dynamical-fixed-points}
summarizes their local stability; it does not assume a global RG trajectory
between the two dynamical limits.
The two dynamical endpoints can be combined with either the Gaussian or the
Wilson--Fisher static fixed point. A natural dimensionless coordinate measuring
the relative strength of the dissipative and propagating temporal operators is
\begin{equation}
    \rho_r\equiv x c_r .
    \label{eq:rho-definition}
\end{equation}
Indeed, Eq.~\eqref{eq:app-bare-renormalized} gives
\begin{equation}
    Xc
    =
    \mu\,
    \frac{Z_X}{\sqrt{Z_\phi Z_c}}\,\rho_r ,
    \label{eq:Xc-rho}
\end{equation}
so \(Xc\) has canonical mass dimension one independently of the choice of
the dynamic exponent. A dimensionless ratio of this kind, measuring the
relative strength of two competing dynamical operators, plays the same role
here as the time-scale ratio whose flow selects the stable dynamical fixed
point in Model~C \cite{FolkMoser2004,MesterhazyEtAl2013}.

\subsection{Propagating limit}

The surface \(\rho_r=0\), equivalently \(X=0\), is invariant but infrared
unstable. At the two static fixed points, the propagating exponents are
\begin{equation}
    z_{\mathrm{prop},\mathrm{G}}=1,
    \qquad
    z_{\mathrm{prop},\mathrm{WF}}
    =
    1-\frac{\epsilon^2}{9}
    \left(\frac{17}{12}-2\log2\right)
    +\mathcal O(\epsilon^3).
    \label{eq:zprop-endpoints}
\end{equation}
The beta function of the dissipative coupling remains proportional to the
coupling itself, so the exactly dissipationless theory remains on \(\rho_r=0\).

The transverse stability of this surface follows from
\begin{align}
    \beta_{\rho_r}
    &\equiv
    \left.\mu\frac{\mathrm d\rho_r}{\mathrm d\mu}\right|_{\mathrm{bare}}
    =
    \rho_r\left(\frac{\beta_x}{x}+\frac{\beta_{c_r}}{c_r}\right)
    =
    \left(-1+\gamma_X-\frac{1}{2}\gamma_c\right)\rho_r
    \equiv -y_X\rho_r .
    \label{eq:beta-rho}
\end{align}
At the propagating Gaussian fixed point, \(\gamma_X=\gamma_c=0\), and hence
\(y_{X,\mathrm{G}}=1\). At the propagating Wilson--Fisher fixed point,
\begin{equation}
    y_{X,\mathrm{WF}}
    =
    2-z_{\mathrm{prop},\mathrm{WF}}-\gamma_X(g_{\mathrm{WF}})
    =
    1+\frac{\epsilon^2}{36}
    +\mathcal O(\epsilon^3)>0.
    \label{eq:yX-WF}
\end{equation}
Thus any nonzero ratio of dissipative to propagating temporal couplings grows toward the infrared at
both propagating fixed points, whereas the surface \(\rho_r=0\) itself remains
invariant.

\subsection{Overdamped limit}

The opposite boundary is the strictly overdamped theory, for which
\(c_r\to\infty\) and hence \(\rho_r=x c_r\to\infty\). Near this boundary it is
convenient to introduce
\begin{equation}
    \kappa_r\equiv\frac{1}{\rho_r}=\frac{1}{x c_r}.
    \label{eq:kappa-definition}
\end{equation}
The overdamped fixed points are located at \(\kappa_r^*=0\). Their dynamic
exponents are
\begin{equation}
    z_{\mathrm{od},\mathrm{G}}=2,
    \qquad
    z_{\mathrm{od},\mathrm{WF}}
    =
    2+\frac{6\log(4/3)-1}{54}\epsilon^2
    +\mathcal O(\epsilon^3).
    \label{eq:zod-endpoints}
\end{equation}
In the Model-A scaling regime, the \(\omega^2\) projection is ultraviolet
finite at the order considered, so \(\gamma_c^{\mathrm{od}}=0\). The inverse
coordinate therefore obeys
\begin{align}
    \beta_{\kappa_r}
    &\equiv
    \left.\mu\frac{\mathrm d\kappa_r}{\mathrm d\mu}\right|_{\mathrm{bare}}
    =
    -\kappa_r\left(\frac{\beta_x}{x}+\frac{\beta_{c_r}}{c_r}\right)
    =
    \left(1-\gamma_X^{\mathrm{od}}\right)\kappa_r .
    \label{eq:beta-kappa}
\end{align}
At the Gaussian and Wilson--Fisher endpoints, respectively, the final factor
is \(z_{\mathrm{od},\mathrm{G}}-1\) and
\(z_{\mathrm{od},\mathrm{WF}}-1\), both positive. Hence \(\kappa_r\to0\) as
\(\mu\to0\), and both overdamped fixed points are dynamically attractive.

Combining static and dynamical stability gives four fixed points on the
mass-tuned critical surface. The propagating Gaussian point is repulsive in
both directions; the overdamped Gaussian and propagating Wilson--Fisher points
are saddles; and the overdamped Wilson--Fisher point is attractive in both
directions. These endpoint analyses do not determine the beta function at
finite \(\rho_r\), and therefore do not by themselves establish a global RG
trajectory connecting the propagating and overdamped limits.

\section{Conclusions}
\label{sec:conclusions}

In this work, we have investigated the competition between propagating and
relaxational critical dynamics for a non-conserved scalar order parameter in
thermal equilibrium. Starting from a stochastic equation containing both a
relativistic kinetic term and a local friction–noise sector, we formulated the
theory in superspace. We carried out its renormalization near four spatial
dimensions to two-loop order. The relativistic and dissipative contributions
remain distinct temporal operators, while the BRST and equilibrium KMS
symmetries constrain their renormalization and encode the
fluctuation--dissipation relation. The propagating and overdamped limits
therefore share the same static renormalization, including the Gaussian and
Wilson--Fisher fixed points, but obey different dynamical scaling laws.

At the Gaussian fixed points, the two limits have the exponents
$z_{\mathrm{od},\mathrm{G}}=2$ and
$z_{\mathrm{prop},\mathrm{G}}=1$. In the overdamped limit, interactions lead
to the standard Model-A result \cite{HalperinHohenbergMa1972,HalperinHohenbergMa1974}
\begin{equation}
z_{\mathrm{od},\mathrm{WF}}
2+\frac{6\log(4/3)-1}{54}\epsilon^2
+\mathcal{O}(\epsilon^3).
\end{equation}
The relativistic temporal kinetic operator is irrelevant in this regime,
making the overdamped Wilson–Fisher fixed point locally attractive in the
dynamical direction. On the dissipationless surface, by contrast, the critical modes remain
propagating. Renormalization of the temporal and spatial kinetic operators
produces the interacting dynamical exponent
\begin{equation}
z_{\mathrm{prop},\mathrm{WF}}
1-
\frac{\epsilon^2}{9}
\left(
\frac{17}{12}-2\log 2
\right)
+\mathcal O(\epsilon^3).
\label{eq:z-prop-conclusion}
\end{equation}
The correction to the Gaussian value $z_{\mathrm{prop},\mathrm{G}}=1$ is small, showing that interactions only weakly modify the ballistic scaling of the propagating theory.

Our main result is the distinction between invariance and stability of the
dissipationless surface. When $X=0$, the two-loop self-energy contains no
ultraviolet divergence linear in frequency, so coarse graining does not
generate local friction; the induced nonanalytic term
$\propto\ii\omega|\omega|$ vanishes faster than $\ii\omega$ as $\omega\to0$
and does not renormalize $X$. The surface $X=0$ is therefore invariant at
this order. It is not, however, infrared attractive: an infinitesimal
dissipative perturbation is relevant, in agreement with earlier RG analyses
of the $O(N)$ model with the canonical momentum retained as a dynamical
variable \cite{OhnishiKunihiro2006}, here obtained within a controlled
$\epsilon$-expansion. Combining the two static with the two dynamical
endpoints, the propagating Gaussian fixed point is repulsive in both
directions, the overdamped Gaussian and propagating Wilson–Fisher points are
saddles, and the overdamped Wilson–Fisher point is the only one locally
attractive in both. Relativistic propagation may thus persist at microscopic
or intermediate scales and become scale invariant on the fine-tuned surface
$X=0$, while Model~A governs the generic infrared.  The analysis is local in the dynamical phase diagram: it controls the
neighborhoods of the two endpoints, not the flow at finite relative strength
of the temporal operators, and therefore does not establish that a single
trajectory connects them.
Demonstrating the crossover requires a nonperturbative
framework like
the functional renormalization--group formulation \cite{Dupuis:2020fhh} formulated on the Schwinger–Keldysh contour \cite{Batini:2023nan}.

Two technical
ingredients may be useful more broadly: the Hankel-contour representation,
which reduces the oscillatory real-time two-loop integrals and isolates their
ultraviolet singularities, and the general solution of the supersymmetric
Ward identities for $n$-point functions given in
Appendix~\ref{app:n-pointkms}, which parametrizes the admissible correlators
in terms of differential forms on the space of relative times.

Several extensions follow. The most direct one concerns relativistic
hydrodynamics itself.
We demonstrate that a relativistic theory reduces to a nonrelativistic one in the infrared. It would be interesting to see if this is the case for Israel--Stewart type \cite{Israel:1979wp} theory that ends in the nonrelativistic Navier Stokes \cite{Kovtun:2012rj}.  Including the
conserved slow modes of QCD critical dynamics moves the problem to Model~G,
where propagating pion fluctuations coexist with diffusive charge modes
\cite{FlorioGrossiSolovievTeaney2022,Rajagopal:1992qz}, or to Model~H at the
critical endpoint \cite{Son:2004iv}, both directly relevant to heavy-ion
phenomenology.
Furthermore, relaxing temporal
locality by coupling to a non-Markovian environment generates memory kernels
$\sim|\omega|^\alpha$ \cite{Bonart:2012}, which may compete with both local
relaxation and propagation.  Finally, moving away from equilibrium, where
friction and noise are no longer tied by the fluctuation–dissipation
relation and may renormalize independently, would test how much of this
structure is specific to equilibrium dynamics.

\acknowledgments

We thank S. De Curtis, S. Fl\"orchinger, and N. Pinzani for useful discussions. We are particularly grateful to D. Seminara for many insightful discussions and for pointing out the elegant Hankel integral representation that played a crucial role in making this work possible.
L.B. acknowledges support from NCCR SwissMAP.

\appendix

\section{Renormalization-group conventions}
\label{app:RG_conventions}

In this appendix, we summarize the renormalization-group conventions used throughout the main text.  We define the renormalized superfield by
\begin{equation}
	\Phi = Z_\phi^{1/2}\Phi_r .
	\label{eq:app_field_renormalization}
\end{equation}
The remaining bare parameters are expressed in terms of their renormalized
counterparts as
\begin{align}
 g=
\mu^{4-d}
\frac{Z_g}{Z_\phi^2} g_r,\qquad
m^2=
\mu^2
\frac{Z_m}{Z_\phi}r,
\qquad
X
=
\mu^{2-z}
\frac{Z_X}{Z_\phi}x,
\qquad
c^{-2}
=
\mu^{2-2z}
\frac{Z_c}{Z_\phi}c_r^{-2}.
\label{eq:app-bare-renormalized}
\end{align}
Thus, $\Phi,g,m^2,X,$ and $c^{-2}$ denote bare quantities, whereas
$\Phi_r,g_r,r,x,$ and $c_r$ are their renormalized counterparts. At this stage, the dynamic exponent \(z\) is
left arbitrary.
It is convenient to formulate the RG flow directly at the level of the
renormalized vertex functions, which are related to the bare ones by
\begin{equation}
\Gamma_{r}^{(n)}
=
Z_\phi^{n/2}\Gamma^{(n)}.
\end{equation}
Because the bare theory is independent of the arbitrary
renormalization scale \(\mu\),  differentiation at fixed bare parameters yields the
Callan--Symanzik equation
\begin{equation}
	\left[
	\mu\frac{\partial}{\partial\mu}
	+\beta_{ g}
	\frac{\partial}{\partial g_r}
	+\beta_r\frac{\partial}{\partial r}
	+\beta_x\frac{\partial}{\partial x}
	+\beta_{c_r}\frac{\partial}{\partial c_r}
	-\frac{n}{2}\gamma_\phi
	\right]
	\Gamma_r^{(n)}
	=0.
	\label{eq:app-CS}
\end{equation}
The beta functions appearing here describe the variation of the
renormalized parameters with \(\mu\) at fixed bare theory:
\begin{equation}
	\beta_{ g}
	\equiv
	\left.
	\mu\frac{\mathrm{d} g_r}{\mathrm{d}\mu}
	\right|_{\rm bare},
	\qquad
	\beta_r
	\equiv
	\left.
	\mu\frac{\mathrm{d} r}{\mathrm{d}\mu}
	\right|_{\rm bare},
	\qquad
	\beta_x
	\equiv
	\left.
	\mu\frac{\mathrm{d}x}{\mathrm{d}\mu}
	\right|_{\rm bare},
	\qquad
	\beta_{c_r}
	\equiv
	\left.
	\mu\frac{\mathrm{d}c_r}{\mathrm{d}\mu}
	\right|_{\rm bare}.
	\label{eq:app-beta-definitions}
\end{equation}
Similarly, the field anomalous dimension is
\begin{equation}
	\gamma_\phi
	\equiv
	\left.
	\mu\frac{\mathrm{d}}{\mathrm{d}\mu}\log Z_\phi
	\right|_{\rm bare}.
	\label{eq:app-field-anomalous}
\end{equation}
Taking the logarithm of Eq.~\eqref{eq:app-bare-renormalized} and
differentiating with respect to \(\mu\) at fixed bare parameters gives the
following identities. In minimal subtraction, the renormalization factors
depend on \(\mu\) only through \(g_r\):
\begin{equation}
\begin{aligned}
0
&=
\beta_{ g}
\left(
 g_r
\frac{\mathrm d}{\mathrm d g_r}\log Z_g
-
2 g_r
\frac{\mathrm d}{\mathrm d g_r}\log Z_\phi
+1
\right)
+
(4-d) g_r,
\\
0
&=
r\beta_{ g}
\frac{\mathrm d}{\mathrm d g_r}\log Z_m
-
r\beta_{ g}
\frac{\mathrm d}{\mathrm d g_r}\log Z_\phi
+
\beta_r
+
2r,
\\
0
&=
x\beta_{ g}
\frac{\mathrm d}{\mathrm d g_r}\log Z_X
-
x\beta_{ g}
\frac{\mathrm d}{\mathrm d g_r}\log Z_\phi
+
\beta_x
+
(2-z)x,
\\
0
&=
\beta_{ g}
\frac{\mathrm d}{\mathrm d g_r}
\log\left(\frac{Z_c}{Z_\phi}\right)
-2\frac{\beta_{c_r}}{c_r}
+(2-2z).
\end{aligned}
\label{eq:RG-bare-derivatives}
\end{equation}
Therefore,
\begin{equation}
\begin{aligned}
\beta_{ g}
&=
-\frac{(4-d) g_r}{
1+
 g_r
\dfrac{\mathrm d}{\mathrm d g_r}
\log\left(Z_g/Z_\phi^2\right)
},
\\
\beta_r
&=
\left[-2
-
\beta_{ g}
\frac{\mathrm d}{\mathrm d g_r}
\log\left(\frac{Z_m}{Z_\phi}\right)\right]r,
\\
\beta_x
&=
\left[(z-2)
-
\beta_{ g}
\frac{\mathrm d}{\mathrm d g_r}
\log\left(\frac{Z_X}{Z_\phi}\right)\right]x,
\\
\beta_{c_r}
&=
\left[(1-z)
+\frac{1}{2}\beta_{ g}
\frac{\mathrm d}{\mathrm d g_r}
\log\left(\frac{Z_c}{Z_\phi}\right)\right]c_r.
\end{aligned}
\label{eq:RG-beta-functions-exact}
\end{equation}
To separate the canonical scaling from the contributions generated by
fluctuations, we introduce the anomalous dimensions
\begin{align}
\gamma_m
&\equiv
-\beta_{ g}
\frac{\mathrm d}{\mathrm d g_r}
\log\left(\frac{Z_m}{Z_\phi}\right), \quad 
\gamma_X
\equiv
-\beta_{ g}
\frac{\mathrm d}{\mathrm d g_r}
\log\left(\frac{Z_X}{Z_\phi}\right), \quad
\gamma_c
\equiv
-\beta_{ g}
\frac{\mathrm d}{\mathrm d g_r}
\log\left(\frac{Z_c}{Z_\phi}\right).
\label{eq:app-anomalous-dimensions}
\end{align}
The beta functions in Eq.~\eqref{eq:RG-beta-functions-exact}
can then be written in  compact form as
\begin{equation}
\begin{aligned}
\beta_r
&=
(-2+\gamma_m)r,
\quad
\beta_x
=
(z-2+\gamma_X)x,
\quad
\beta_{c_r}
=
\left(1-z-\frac{1}{2}\gamma_c\right)c_r.
\end{aligned}
\label{eq:app-beta-functions-gamma}
\end{equation}

To evaluate the anomalous dimensions in the minimal-subtraction scheme, we
write the renormalization factors as Laurent expansions in
\begin{equation}
Z_i( g_r,\epsilon)
=
1+
\sum_{n=1}^{\infty}
\frac{Z_i^{[n]}( g_r)}{\epsilon^n}.
\label{eq:app-Z-Laurent}
\end{equation}
Here, $Z_i^{[1]}$ denotes the residue of the simple pole. For each ratio
\(Z_i/Z_\phi\) entering the anomalous dimensions, with
\(i\in\{m,X,c\}\), the corresponding expansion reads
\begin{equation}
\log\left(\frac{Z_i}{Z_\phi}\right)
=
\frac{
Z_i^{[1]}-Z_\phi^{[1]}
}{\epsilon}
+
\sum_{n=2}^{\infty}
\frac{L_i^{[n]}}{\epsilon^n}.
\label{eq:app-log-ratio-Laurent}
\end{equation}
The simple-pole coefficient of $\log Z_i$ coincides with that of
$Z_i$, while the higher-pole coefficients $L_i^{[n]}$ are fixed by
the usual pole-consistency relations. Near four dimensions, the beta function begins with its canonical
contribution,
\begin{equation}
    \beta_g=-(4-d)g_r+\mathcal O(g_r^2).
\end{equation}
Substituting the Laurent expansions into the definitions in Eq.~\eqref{eq:app-anomalous-dimensions} and
requiring the anomalous dimensions to remain finite as \(d\to4\), the
higher-pole contributions cancel by virtue of the pole-consistency
relations. The finite remainder is therefore determined entirely by the
simple-pole residues:
\begin{align}
\gamma_m
=
 g_r
\frac{\mathrm d}{\mathrm d g_r}
\left(
Z_m^{[1]}-Z_\phi^{[1]}
\right),
\quad
\gamma_X
=
 g_r
\frac{\mathrm d}{\mathrm d g_r}
\left(
Z_X^{[1]}-Z_\phi^{[1]}
\right),
\quad
\gamma_c
=
 g_r
\frac{\mathrm d}{\mathrm d g_r}
\left(
Z_c^{[1]}-Z_\phi^{[1]}
\right).
\label{eq:app-gamma-simple-poles}
\end{align}
The same argument applied to the field-renormalization factor gives
\begin{equation}
\gamma_\phi
=
- g_r
\frac{\mathrm d Z_\phi^{[1]}}
     {\mathrm d g_r},
\label{eq:app-field-simple-pole}
\end{equation}
and the beta function of the quartic coupling becomes
\begin{equation}
\beta_{ g}
=
-\epsilon g_r
+
 g_r^{\,2}
\frac{\mathrm d}{\mathrm d g_r}
\left(
Z_g^{[1]}-2Z_\phi^{[1]}
\right).
\label{eq:app-beta-g-simple-pole}
\end{equation}
Together with Eq.~\eqref{eq:app-gamma-simple-poles}, these relations show that all the RG functions
needed in the main text are determined by the simple-pole residues\footnote{ The
higher poles are nevertheless required to ensure the consistency of
renormalization, although they do not contribute directly to the critical
exponents.}.

\section{Hankel-contour integrals}
\label{app:hankel-integrals}

In this appendix, we evaluate the three Hankel-contour integrals entering the
two-loop dynamical analysis. In the order presented below, $I_3$, $I_1$, and
$I_X$ determine the renormalization of the relativistic temporal coefficient
in Subsec.~\ref{subsec:hankel-prop}, the nonanalytic massless contribution in
Subsec.~\ref{subsec:absenceofrenorm}, and the local dissipative insertion in
Subsec.~\ref{subsec:homogeneous-X-projection}. Although they have different
physical interpretations, their evaluation follows a common strategy:
\begin{enumerate}
    \item perform the integrations over the Schwinger parameters \(x_i\),
    leaving only the Hankel-contour integrals over \(s_i\);
    \item deform each Hankel contour onto the two sides of the negative
    real axis;
    \item combine the contributions from the two sides of the branch cut.
    The imaginary parts cancel, leaving an ordinary real integral over
    the variables \(r_i\);
    \item recognize the resulting integration measures as normalized
    Gamma distributions and evaluate their logarithmic moments using
    digamma functions.
\end{enumerate}

To implement the second step, we first fix our contour and branch
conventions. We use the Hankel contour \(\mathcal C_H\) shown in Fig.~\ref{fig:hankel-contour},
with the orientation and decomposition introduced in Subsec.~\ref{subsec:61}. We adopt
the principal branch \(-\pi<\arg s<\pi\), so that the negative real axis
is the branch cut. Its two straight-line branches are parametrized as
\begin{equation}
    s=-r+\mathrm{i}0\,\sigma,
    \qquad
    r>0,
    \qquad
    \sigma=\pm1,
    \label{eq:r}
\end{equation}
where \(\sigma=+1\) and \(\sigma=-1\) denote the upper and lower contour,
respectively. With these conventions, the two contour identities needed
below are
\begin{align}
    \int_{\mathcal C_H}\frac{\mathrm ds}{2\pi \ii}\,
    e^s s^{-1/2}f(s)
    &=
    \frac{1}{2\pi}
    \int_0^\infty \mathrm dr\,
    e^{-r}r^{-1/2}
    \bigl[f(-r+i0)+f(-r-i0)\bigr],
    \label{eq:app-hankel-minus-half}
    \\
    \int_{\mathcal C_H}\frac{\mathrm ds}{2\pi \ii}\,
    e^s s^{1/2}f(s)
    &=
    -\frac{1}{2\pi}
    \int_0^\infty \mathrm dr\,
    e^{-r}r^{1/2}
    \bigl[f(-r+i0)+f(-r-i0)\bigr].
    \label{eq:app-hankel-plus-half}
\end{align}
The relative minus sign in Eq.~\eqref{eq:app-hankel-plus-half} follows from
the phase of $s^{1/2}$ and the orientations of the two contours. As a check, setting \(f=1\) reproduces the standard Hankel
representation of the reciprocal Gamma function
\begin{equation}
    \frac{1}{\Gamma(1/2)}=\frac{1}{\sqrt{\pi}},
    \qquad
    \frac{1}{\Gamma(-1/2)}=-\frac{1}{2\sqrt{\pi}}.
    \label{eq:app-hankel-check}
\end{equation}

\subsection{$I_3(4)$ contribution }
\label{app:hankel-inertial}

We first evaluate the ultraviolet pole of $\Sigma_3$, computed in Eq. \eqref{eq:sigmac-low-frequency-Ic}
\begin{equation}
    I_3(4) =I_3
    =
    \int_0^\infty\mathrm \dd x_1\,\mathrm \dd x_2
    \left[
        \prod_{i=1}^{3}
        \int_{\mathcal C_H}\frac{\mathrm \dd s_i}{2\pi \ii}\,e^{s_i}
    \right]
    s_1^{-1/2}s_2^{-1/2}s_3^{-3/2}\mathcal D^{-2}.
    \label{eq:app-Ic-definition}
\end{equation}
Here
\begin{equation}
    S_1=x_1+\frac{1}{s_1},
    \qquad
    S_2=x_2+\frac{1}{s_2},
    \qquad
    S_3=\frac{1}{s_3},
    \qquad
    \mathcal D=S_1S_2+S_1S_3+S_2S_3.
    \label{eq:app-inertial-definitions}
\end{equation}
\subsubsection{Step 1: Schwinger-parameter integrations}
First, we set
\begin{equation}
    a_i=\frac{1}{s_i},
    \qquad
    y_1=x_1+a_1,
    \qquad
    y_2=x_2+a_2.
\end{equation}
Since $S_3=a_3$, the denominator becomes
$
   \mathcal  D=y_1y_2+a_3(y_1+y_2).
$ For fixed $y_1$, the first integration gives
\begin{equation}
    \int_{a_2}^{\infty}
    \frac{\mathrm dy_2}{[y_1y_2+a_3(y_1+y_2)]^2}
    =
    \frac{1}{(y_1+a_3)[(a_2+a_3)y_1+a_2a_3]}.
    \label{eq:app-inertial-y2-integral}
\end{equation}
The remaining integral is
\begin{align}
    &\int_{a_1}^{\infty}
    \frac{\mathrm dy_1}{(y_1+a_3)[(a_2+a_3)y_1+a_2a_3]}
   =
    \frac{1}{a_3^2}
    \log\left[
        \frac{(a_1+a_3)(a_2+a_3)}
        {a_1a_2+a_1a_3+a_2a_3}
    \right].
    \label{eq:app-inertial-y1-integral}
\end{align}
The remaining integral in $s_i$ variables is 
\begin{equation}
    \int_0^\infty\mathrm \dd x_1\,\mathrm \dd x_2\,\mathcal D^{-2}
    =
    s_3^2
    \log\left[
        \frac{(s_1+s_3)(s_2+s_3)}
        {s_3(s_1+s_2+s_3)}
    \right].
    \label{eq:app-inertial-feynman-result}
\end{equation}
It is convenient to define
\begin{equation}
    L_3(s_1,s_2,s_3)
    \equiv
    \log\left[
        \frac{(s_1+s_3)(s_2+s_3)}
        {s_3(s_1+s_2+s_3)}
    \right].
    \label{eq:app-L-definition}
\end{equation}
The factor $s_3^2$ in Eq.~\eqref{eq:app-inertial-feynman-result} converts
the original factor $s_3^{-3/2}$ into $s_3^{1/2}$. Therefore,
\begin{equation}
    I_3
    =
    \left[
        \prod_{i=1}^{3}
        \int_{\mathcal C_H}\frac{\mathrm \dd s_i}{2\pi \ii}\,e^{s_i}
    \right]
    s_1^{-1/2}s_2^{-1/2}s_3^{1/2}L_3(s_1,s_2,s_3).
    \label{eq:app-Ic-reduced}
\end{equation}

\subsubsection{Step 2: From the Hankel contours to a real integral}
Before collapsing the contours onto the branch cut, we verify that the
small circles around the origin give no contribution.  Indeed, $L_3=\mathcal O(s_1)$ as $s_1\to0$
and similarly for $s_2\to0$, while
\begin{equation}
    L_3(s_1,s_2,s_3)
    =
    -\log s_3
    +\log\left(\frac{s_1s_2}{s_1+s_2}\right)
    +\mathcal O(s_3)
\end{equation}
as $s_3\to0$. Because the integrand contains $s_3^{1/2}$, the contribution
from a circle $|s_3|=\delta$ is
$\mathcal O(\delta^{3/2}|\log\delta|)$.
On the $ C_{\pm}$ branches, we write
$s_i=-r_i+\ii\sigma_i\eta_i$, where $\eta_i>0$ are kept distinct until the
sum over branch assignments has been performed. This prescription avoids an
ambiguity when two signs are opposite and the imaginary part of a sum
$s_i+s_j$ would otherwise be written as zero prematurely. The integral becomes
\begin{equation}
    I_3
    =
    -\sum_{\sigma_{1},\sigma_{2},\sigma_{3}=\pm}\left[
        \prod_{i=1}^{3}
        \int_{0}^{\infty}\frac{\mathrm dr_i}{2\pi}\,e^{-r_i}
    \right]
    r_1^{-1/2}r_2^{-1/2}r_3^{1/2}L_3(s_1,s_2,s_3),
    \label{eq:app-I3-branch-sum}
\end{equation}
where Eqs.~\eqref{eq:app-hankel-minus-half} and
\eqref{eq:app-hankel-plus-half} have been applied to the three contours. The
factor $1/(2\pi)^3$ is essential: the subsequent sum over the eight branch
assignments converts it into the factor $1/\pi^3$ below.
On the branch cut, the real part of $L_3$ is
\begin{equation}
    \Re L_3(r_1,r_2,r_3)
    =
    \log\left[
        \frac{(r_1+r_3)(r_2+r_3)}
        {r_3(r_1+r_2+r_3)}
    \right],
    \label{eq:app-L-real}
\end{equation}
The branch assignments occur in complex-conjugate pairs related by
$\sigma_i\mapsto-\sigma_i$. Their imaginary parts therefore cancel before the
regulators $\eta_i$ are sent to zero. Since the real part is the same for all
eight assignments, the integral reduces to
\begin{align}
    I_3
    ={}&
    -\frac{1}{\pi^3}
    \int_0^\infty
    \mathrm dr_1\,\mathrm dr_2\,\mathrm dr_3\,
    e^{-(r_1+r_2+r_3)}
r_1^{-1/2}r_2^{-1/2}r_3^{1/2}
    L_3(r_1,r_2,r_3).
    \label{eq:app-Ic-real}
\end{align}
The overall minus sign comes solely from the $s_3^{1/2}$ contour identity
in Eq.~\eqref{eq:app-hankel-plus-half}.

\subsubsection{Step 3: Evaluation as a Gamma-distribution average}
Recall the defining integral of the Gamma function,
\begin{equation}
    \Gamma(\alpha)
    =
    \int_0^\infty \mathrm dr\,e^{-r}r^{\alpha-1}.
    \label{eq:app-gamma-definition}
\end{equation}
It follows that
\begin{equation}
    p_\alpha(r)
    \equiv
    \frac{e^{-r}r^{\alpha-1}}{\Gamma(\alpha)},
    \qquad r>0,
    \label{eq:app-gamma-density}
\end{equation}
is normalized to unity. It is the Gamma density with shape parameter
$\alpha$ and unit scale. We suppress the scale parameter below because it is
always equal to one.
For clarity, we denote a one-dimensional average with the density
$p_\alpha$ by
\begin{equation}
    \langle f(r)\rangle_\alpha
    \equiv
    \int_0^\infty\mathrm dr\,p_\alpha(r)f(r).
    \label{eq:app-one-dimensional-average}
\end{equation}
The loop integral depends on 
two elementary Gamma-function identities. 
First, differentiating Eq.~\eqref{eq:app-gamma-definition} with respect to
$\alpha$ gives
\begin{equation}
    \Gamma'(\alpha) = \int_0^\infty
    \mathrm dr\,e^{-r}r^{\alpha-1}\log r
    .
\end{equation}
After division by $\Gamma(\alpha)$, the differentiated Gamma integral gives
\begin{equation}
    \langle\log r\rangle_\alpha
    =
    \frac{\Gamma'(\alpha)}{\Gamma(\alpha)}
    \equiv\psi(\alpha),
    \label{eq:app-gamma-log-moment}
\end{equation}
where $\psi$ is the digamma function \cite{AbramowitzStegun1964}. 
Replacing $\alpha$ by $\alpha+1$
in the same calculation gives
\begin{equation}
    \langle r\log r\rangle_\alpha
    =
    \alpha\psi(\alpha+1).
    \label{eq:app-gamma-rlogr-moment}
\end{equation}

Second, the sum of independent Gamma variables with the same scale is again
Gamma distributed, with the shape parameters added:
\begin{equation}
    r_a\sim\Gamma(\alpha),\quad
    r_b\sim\Gamma(\beta)
    \quad\Longrightarrow\quad
    r_a+r_b\sim\Gamma(\alpha+\beta),
    \label{eq:app-gamma-addition}
\end{equation}
where $\sim$ means that it is also Gamma-distributed
\footnote{ 
This fact follows directly by convolving the two densities:
$$
    p_{\alpha+\beta}(R)
    =
    \int_0^R\mathrm dr\,p_\alpha(r)p_\beta(R-r)
    =
    \frac{e^{-R}}{\Gamma(\alpha)\Gamma(\beta)}
    \int_0^R\mathrm dr\,
    r^{\alpha-1}(R-r)^{\beta-1}
    =
    \frac{e^{-R}R^{\alpha+\beta-1}}{\Gamma(\alpha+\beta)}.   
$$
In the last step we used the Euler beta integral
$$
B(\alpha,\beta)=\frac{\Gamma(\alpha)\Gamma(\beta)}{
\Gamma(\alpha+\beta)}
$$}.

In particular, in Eq.~\eqref{eq:app-Ic-real},
\begin{equation}
\begin{aligned}
    e^{-r_i}r_i^{-\frac12 }\,\mathrm dr_i
   & =
    \Gamma\left(\frac12\right)
    p_{\frac12}(r_i)\,\mathrm dr_i, 
    \\
    e^{-r_i}r_i^{\frac12}\,\mathrm dr_i
    &=
    \Gamma\left(\frac32\right)
    p_{\frac32}(r_i)\,\mathrm dr_i.
    \label{eq:app-half-weight-normalization}
    \end{aligned}
\end{equation}

The real measure of Eq. \eqref{eq:app-Ic-real} factorizes as
\begin{align}
    &e^{-(r_1+r_2+r_3)}r_1^{-\frac12}r_2^{-\frac12}r_3^{\frac12}
    \prod_{i=1}^{3}\mathrm dr_i
=
    \Gamma\left(\frac12\right)^2
    \Gamma\left(\frac32\right)
    p_{\frac12}(r_1)p_{\frac12}(r_2)p_{\frac32}(r_3)
    \prod_{i=1}^{3}\mathrm dr_i.
    \label{eq:app-Ic-normalization}
\end{align}
For any function $F(r_1,r_2,r_3)$, we define the brackets
\begin{equation}
\begin{aligned}
    \langle F\rangle_{\alpha\beta \gamma }
    \equiv
    \int_{0}^{+\infty}
    \mathrm \dd r_1\, \dd r_2\, \dd r_3 p_{\alpha}(r_1) p_{\beta }(r_2)  \,p_{\gamma}(r_3)
    F(r_1,r_2,r_3).
\end{aligned}
\label{eq:app-expectation-definition}
\end{equation}
The brackets are therefore shorthand for the normalized triple integral, and
\begin{equation}
    I_3
    =
    -\frac{1}{2\pi^{3/2}}
    \langle L_3(r_1,r_2,r_3)\rangle_{\frac12 \frac{1}{2} \frac{3}{2}},
    \label{eq:app-Ic-expectation}
\end{equation}
where we have used 
\begin{equation}
    \Gamma\left(\frac12\right)^2
    \Gamma\left(\frac32\right)
    =\frac{\pi^{3/2}}{2},
\end{equation}

To evaluate this expectation, we first expand the definition of $L_3$:
\begin{align}
    \langle L_3\rangle_{\frac12 \frac{1}{2} \frac{3}{2}}
    ={}&
    \langle\log(r_1+r_3)\rangle_{\frac12  \frac{3}{2}}
    +\langle\log(r_2+r_3)\rangle_{\frac12 \frac{3}{2}}
    -
    \langle\log r_3\rangle_{\frac{3}{2}}
    -\langle\log(r_1+r_2+r_3)\rangle_{\frac12 \frac{1}{2} \frac{3}{2}}.
    \label{eq:app-Ic-expectation-expanded}
\end{align}
Setting
\begin{equation}
    r_{ij}=r_i+r_j, \quad r=r_1+r_2+r_3,
    \label{eq:app-sym-R-definitions}
\end{equation}
the four arguments are distributed as
\begin{equation}
    r_{13},\;r_{23}\sim\Gamma(2),
    \qquad
    r_3\sim\Gamma\left(\frac32\right),
\quad 
    r
    \sim\Gamma\left(\frac52\right).
\end{equation}
Using Eq.~\eqref{eq:app-gamma-log-moment},  each term gives
\begin{align}
    \langle\log(r_{13})\rangle_{\frac12 \frac{1}{2} \frac{3}{2}}
    =
\langle\log(r_{23})\rangle_{\frac12 \frac{1}{2} \frac{3}{2}}=\psi(2),
    \quad
    \langle\log r_3\rangle_{\frac12 \frac{1}{2} \frac{3}{2}}
    =\psi\left(\frac32\right),
    \quad
\langle\log(r)\rangle_{\frac12 \frac{1}{2} \frac{3}{2}}
=\psi\left(\frac52\right).
    \label{eq:app-Ic-four-log-moments}
\end{align}
Substituting into Eq.~\eqref{eq:app-Ic-expectation-expanded} yields
\begin{equation}
    \langle L_3\rangle_{\frac12 \frac{1}{2} \frac{3}{2}}
    =
    2\psi(2)
    -\psi\left(\frac32\right)
    -\psi\left(\frac52\right),
    \label{eq:app-Ic-log-average}
\end{equation}
and therefore,
\begin{equation}
    I_3
    =
    -\frac{1}{2\pi^{3/2}}
    \left[
        2\psi(2)
        -\psi\left(\frac32\right)
        -\psi\left(\frac52\right)
    \right].
\end{equation}
Finally, using
\begin{equation}
    \psi(2)=1-\gamma_E,
    \qquad
    \psi\left(\frac32\right)=2-\gamma_E-2\log2,
    \qquad
    \psi\left(\frac52\right)=\frac83-\gamma_E-2\log2,
\end{equation}
where $\gamma_E$ is the Euler constant, one obtains
\begin{equation}
    I_3
    =
    -\frac{2(3\log2-2)}{3\pi^{3/2}}
.
    \label{eq:app-Ic-result}
\end{equation}
\subsection{$I_1(4)$ contribution}
\label{app:hankel-symmetric}

We next consider the dimensionless integral \(I_1\) appearing in the massless
symmetric self-energy in \(d=4\), (Eq.~\eqref{eq:Sigma_sym})
\begin{equation}
    I_1(4)=I_1
    =
    \left[
        \prod_{i=1}^{3}
        \int_{\mathcal{C}_H}\frac{\mathrm \dd s_i}{2\pi \ii}\,
        e^{s_i}s_i^{-1/2}
    \right]
    \int_0^\infty
    \frac{\mathrm \dd x_1\,\mathrm \dd x_2\,\mathrm dx_3}{\mathcal D^2},
    \label{eq:app-Isym-definition}
\end{equation}
with
\begin{equation}
    S_i=x_i+\frac{1}{s_i},
    \qquad
   \mathcal  D=S_1S_2+S_1S_3+S_2S_3.
    \label{eq:app-sym-denominator}
\end{equation}

\subsubsection{Step 1: Schwinger-parameter integrations}

At fixed contour variables \(s_i\), we define
\begin{equation}
    a_i=\frac{1}{s_i},
    \qquad
    y_i=x_i+a_i.
    \label{eq:app-sym-shift}
\end{equation}
We first perform the integrations in the domain \(a_i>0\), where the
integration bounds below have their usual real meaning. The resulting
analytic expression is then continued to the complex values
\(a_i=s_i^{-1}\) along the Hankel contours.
In terms of the shifted variables,
the denominator becomes $
    \mathcal D=y_1y_2+y_1y_3+y_2y_3.
$
Integrating first over \(y_3\) gives
\begin{equation}
    \int_{a_3}^{\infty}
    \frac{\mathrm dy_3}{[y_1y_2+y_3(y_1+y_2)]^2}
    =
    \frac{1}{(y_1+y_2)[y_1y_2+a_3(y_1+y_2)]}.
    \label{eq:app-sym-y3-integral}
\end{equation}
For the remaining two integrations, we set
\begin{equation}
    y_1=Ru,
    \qquad
    y_2=R(1-u),
    \qquad
    0<u<1.
    \label{eq:app-sym-radial-change}
\end{equation}
The Jacobian is $\mathrm dy_1\mathrm dy_2=R\,\mathrm dR\mathrm du$, and the
lower bounds $y_1\geq a_1$ and $y_2\geq a_2$ imply
\begin{equation}
    R\geq R_{\min}(u)
    \equiv
    \max\left(\frac{a_1}{u},\frac{a_2}{1-u}\right).
    \label{eq:app-sym-Rmin}
\end{equation}
At fixed \(u\), the radial integration can be performed directly
\begin{equation}
    \int_{R_{\min}(u)}^\infty
    \frac{\mathrm dR}{R[Ru(1-u)+a_3]}
    =
    \frac{1}{a_3}
    \log\left[1+
        \frac{a_3}{R_{\min}(u)u(1-u)}
    \right].
    \label{eq:app-sym-radial-integral}
\end{equation}
The two branches of $R_{\min}$ meet at
$
    u_*=a_1/(a_1+a_2).
$
We therefore split the $u$ integration at $u_*$. Performing the
remaining integrations gives
\begin{equation}
\begin{aligned}
    \int_0^\infty\frac{\mathrm d^3x}{\mathcal D^2}
    &=
    \frac{a_1+a_3}{a_1a_3}\log(a_1+a_3)
    +\frac{a_2+a_3}{a_2a_3}\log(a_2+a_3)
   \\
    &+
    \frac{a_1+a_2}{a_1a_2}\log(a_1+a_2)
    -
    \frac{a_1a_2+a_1a_3+a_2a_3}{a_1a_2a_3}
    \log(a_1a_2+a_1a_3+a_2a_3).
    \label{eq:app-sym-a-result}
    \end{aligned}
\end{equation}
Substituting $a_i=s_i^{-1}$ and collecting the logarithms produces the
manifestly symmetric kernel
\begin{align}
    L_1(s_1,s_2,s_3)
    ={}&
    \sum_{i<j}(s_i+s_j)\log(s_i+s_j)
    -
    \left(\sum_{i=1}^{3}s_i\right)
    \log\left(\sum_{i=1}^{3}s_i\right)
    -\sum_{i=1}^{3}s_i\log s_i.
    \label{eq:app-sym-kernel}
\end{align}
Thus, the original integral reduces to
\begin{equation}
    I_1
    =
    \left[
        \prod_{i=1}^{3}
        \int_{\mathcal C_H}\frac{\mathrm \dd s_i}{2\pi \ii}\,
        e^{s_i}s_i^{-1/2}
    \right]
    L_1(s_1,s_2,s_3).
    \label{eq:app-Isym-kernel}
\end{equation}

\subsubsection{Step 2: From the Hankel contours to a real integral}

Before deforming the contours, we verify that the small circular arcs around
the origin do not contribute. For example, at fixed $s_1,s_2$,
\begin{equation}
    L_1(s_1,s_2,s_3)
    =
    -s_3\log s_3
    +s_3\left[1+\log\left(\frac{s_1s_2}{s_1+s_2}\right)\right]
    +\mathcal O(s_3^2).
    \label{eq:app-sym-small-s}
\end{equation}
On a circle $|s_3|=\delta$, the corresponding contour contribution is
therefore $\mathcal O(\delta^{3/2}|\log\delta|)$ and vanishes as
$\delta\to0$. The same argument applies to $s_1$ and $s_2$.

We may consequently use the contour parametrization, obtaining for the real part
\begin{equation}
    \operatorname{Re}L_1
    =
    -\sum_{i<j}r_{ij}\log r_{ij}
    +r\log r
    +\sum_{i=1}^{3}r_i\log r_i.
    \label{eq:app-sym-real-kernel}
\end{equation}
The imaginary part is odd under the simultaneous reversal
$(\sigma_1,\sigma_2,\sigma_3)
    \longmapsto
    (-\sigma_1,-\sigma_2,-\sigma_3),
$
whereas the real part is unchanged. The imaginary contributions therefore
cancel pairwise in the sum over the eight contour assignments.

Applying Eq.~\eqref{eq:app-hankel-minus-half} to each contour then yields
the real integral
\begin{align}
    I_1
    ={}&
    \frac{1}{\pi^3}
    \int_0^\infty
    \prod_{i=1}^{3}
    \left[\mathrm dr_i\,e^{-r_i}r_i^{-1/2}\right]
    \left[
        -\sum_{i<j}r_{ij}\log r_{ij}
        +r\log r
        +\sum_{i=1}^{3}r_i\log r_i
    \right].
    \label{eq:app-Isym-real}
\end{align}

\subsubsection{Step 3: Evaluation as a Gamma-distribution average}

We now explain explicitly how the real integral is converted into the
Gamma-distribution notation used below. No physical randomness is being
introduced. The notation is only a convenient way of writing a normalized
integral. Applying this identity to the three integration variables in
Eq.~\eqref{eq:app-Isym-real}, the full measure factorizes as
\begin{align}
    &e^{-(r_1+r_2+r_3)}
    r_1^{-\frac12 }r_2^{-\frac12}r_3^{-\frac12}
    \prod_{i=1}^{3}\mathrm dr_i
  =
    \Gamma\left(\frac12\right)^3
    \prod_{i=1}^{3}
    \left[p_{\frac12}(r_i)\,\mathrm dr_i\right].
    \label{eq:app-Isym-normalization}
\end{align}
Equivalently, one may say that the auxiliary variables $r_i$ are independent
and distributed as
$r_i\sim\operatorname{Gamma}\left(\frac12\right)$.
Here the first argument is the shape parameter and the second is the scale
parameter. Again, this notation describes the integration measure; it does
not assign any stochastic meaning to the original field theory. Since $\Gamma(1/2)^3=\pi^{3/2}$, Eq.~\eqref{eq:app-Isym-real} becomes
\begin{align}
    I_1
    =\frac{1}{\pi^{3/2}}
    \left\langle
        -\sum_{i<j}r_{ij}\log r_{ij}
        +r\log r
        +\sum_{i=1}^{3}r_i\log r_i
    \right\rangle_{\frac{1}{2}\frac{1}{2}\frac{1}{2}}.
    \label{eq:app-Isym-expectation}
\end{align}
We can now evaluate the three groups of terms in
Eq.~\eqref{eq:app-Isym-expectation} separately:
\begin{align}
    \langle r_i\log r_i\rangle_{\frac12}
=\frac12\psi\left(\frac32\right),
    \quad
    \langle r_{ij}\log r_{ij}\rangle_{\frac12\frac12}
    =\psi(2),
    \quad
    \langle r \log r \rangle_{\frac12\frac12\frac12}
    =\frac32\psi\left(\frac52\right).
    \label{eq:app-Isym-three-moments}
\end{align}
There are three single-variable terms, three pair terms, and one total-sum
term. Substitution therefore gives
\begin{equation}
    I_1
    =
    \frac{1}{\pi^{3/2}}
    \left[
        -3\psi(2)
        +\frac32\psi\left(\frac52\right)
        +\frac32\psi\left(\frac32\right)
    \right].
    \label{eq:app-Isym-digamma}
\end{equation}
With
\begin{equation}
    \psi\left(\frac12\right)=-\gamma_E-2\log2,
    \qquad
    \psi(z+1)=\psi(z)+\frac{1}{z},
    \label{eq:app-digamma-identities}
\end{equation}
the Euler constants cancel, leaving
\begin{equation}
    I_1
    =
    \frac{4-6\log2}{\pi^{3/2}}.
    \label{eq:app-Isym-result}
\end{equation}
\subsection{$I_X(4)$ contribution}
\label{app:hankel-dissipative}

We finally evaluate the ultraviolet pole of the dissipative composite
insertion: 
\begin{equation}
    I_{X}(4)=I_X
    =
    \int_0^\infty\mathrm \dd x_1\,\mathrm \dd x_2
    \left[
        \prod_{i=1}^{3}
        \int_{\mathcal C_H}\frac{\mathrm \dd s_i}{2\pi \ii}\,e^{s_i}
    \right]
    s_1^{-1/2}s_2^{-1/2}s_3^{-5/2}\mathcal D^{-2}.
    \label{eq:app-IX-definition}
\end{equation}
The denominator $\mathcal D$ is the same as in
Eq.~\eqref{eq:app-inertial-definitions}.

\subsubsection{Step 1: Schwinger-parameter integrations}

The integrations over $x_1$ and $x_2$ are identical to those in Eq. \eqref{eq:app-Ic-definition}. Equation~\eqref{eq:app-inertial-feynman-result}
therefore gives
\begin{equation}
    \int_0^\infty\mathrm \dd x_1\,\mathrm \dd x_2\,\mathcal D^{-2}
    =s_3^2L_3(s_1,s_2,s_3),
\end{equation}
with $L_3$ defined in Eq.~\eqref{eq:app-L-definition}. The difference from
the relativistic temporal case lies entirely in the original power of $s_3$: the factor
$s_3^2$ now converts $s_3^{-5/2}$ into $s_3^{-1/2}$. Hence
\begin{equation}
    I_X
    =
    \left[
        \prod_{i=1}^{3}
        \int_{\mathcal C_H}\frac{\mathrm \dd s_i}{2\pi \ii}\,
        e^{s_i}s_i^{-1/2}
    \right]
    L_3(s_1,s_2,s_3).
    \label{eq:app-IX-reduced}
\end{equation}

\subsubsection{Step 2: From the Hankel contours to a real integral.}

The small circular arcs vanish also in this case. For $s_1\to0$ or
$s_2\to0$, the kernel $L_3$ is linear in the corresponding variable. For
$s_3\to0$, the kernel grows only logarithmically. Since the integrand now
contains $s_3^{-1/2}$, the contribution from a circle
$|s_3|=\delta$ is
$\mathcal O(\delta^{1/2}|\log\delta|)$ and therefore still vanishes. As in the $I_3$ calculation, the imaginary part cancels between
complex-conjugate contour assignments. The surviving real kernel is therefore the same
$L_3$ given in Eq.~\eqref{eq:app-L-real}. Applying
Eq.~\eqref{eq:app-hankel-minus-half} to all three contours gives
\begin{align}
    I_X
    =&
    \frac{1}{\pi^3}
    \int_0^\infty
    \mathrm dr_1\,\mathrm dr_2\,\mathrm dr_3\,
    e^{-(r_1+r_2+r_3)}
    r_1^{-\frac12}r_2^{-\frac12}r_3^{-\frac12}
    L_3(r_1,r_2,r_3).
    \label{eq:app-IX-real}
\end{align}
In contrast with Eq.~\eqref{eq:app-Ic-real}, there is no overall minus sign:
all three contour integrals now contain the power $s_i^{-\frac12}$.

\subsubsection{Step 3: Evaluation as a Gamma-distribution average.}

The measure factorizes as
\begin{align}
    &e^{-(r_1+r_2+r_3)}r_1^{-\frac12}r_2^{-\frac12}r_3^{-\frac12}
    \prod_{i=1}^{3}\mathrm dr_i
   =
    \Gamma\left(\frac12\right)^3
    \prod_{i=1}^{3}
    \left[p_{\frac12}(r_i)\,\mathrm dr_i\right].
    \label{eq:app-IX-normalization}
\end{align}
Since $\Gamma(1/2)^3=\pi^{3/2}$, Eq.~\eqref{eq:app-IX-real} becomes
\begin{equation}
    I_X
    =
    \frac{1}{\pi^{3/2}}
    \langle L_3(r_1,r_2,r_3)\rangle_{\frac12\frac12\frac12}.
    \label{eq:app-IX-expectation}
\end{equation}
Expanding the kernel inside the brackets gives
\begin{align}
    \langle L_3\rangle_{\frac12\frac12\frac{1}{2}}
    =&
    \langle\log(r_{13})\rangle_{\frac12\frac12}
    +\langle\log(r_{23})\rangle_{\frac12\frac12}
    -
    \langle\log r_3\rangle_{\frac12}
    -\langle\log(r)\rangle_{\frac12\frac12\frac12}.
    \label{eq:app-IX-expectation-expanded}
\end{align}
Equation~\eqref{eq:app-gamma-log-moment} then gives the four contributions
separately:
\begin{align}
\langle\log(r_{13})\rangle_{\frac12\frac12}
    = \langle\log(r_{23})\rangle_{\frac12\frac12}=\psi(1),
    \quad
    \langle\log r_3\rangle_{\frac12}
=\psi\left(\frac12\right),
    \quad \langle\log(r)\rangle_{\frac12\frac12\frac{1}{2}}
    &=\psi\left(\frac32\right).
    \label{eq:app-IX-four-log-moments}
\end{align}
Substitution into Eq.~\eqref{eq:app-IX-expectation-expanded} yields
\begin{equation}
    \langle L_3\rangle_{\frac12\frac12\frac12}
    =
    2\psi(1)
    -\psi\left(\frac12\right)
    -\psi\left(\frac32\right).
    \label{eq:app-IX-log-average}
\end{equation}
Therefore,
\begin{equation}
    I_X
    =
    \frac{1}{\pi^{3/2}}
    \left[
        2\psi(1)
        -\psi\left(\frac12\right)
        -\psi\left(\frac32\right)
    \right].
\end{equation}
Using
\begin{equation}
    \psi(1)=-\gamma_E,
    \qquad
    \psi\left(\frac12\right)=-\gamma_E-2\log2,
    \qquad
    \psi\left(\frac32\right)=2-\gamma_E-2\log2,
\end{equation}
the Euler constants again cancel, yielding
\begin{equation}
    I_X
    =
    \frac{2(2\log2-1)}{\pi^{3/2}}.
    \label{eq:app-IX-result}
\end{equation}

\subsubsection{Summary}
For reference, the three dimensionless integrals evaluated in this appendix
are
\begin{equation}
    \begin{aligned}
        I_1
        &=\frac{4-6\log2}{\pi^\frac{3}{2}} = -\frac{2}{\pi^{\frac{3}{2}}} (3\log 2 -2)\sim -0.0285,
        \\
        I_3
        &=-\frac{2(3\log2-2)}{3\pi^\frac{3}{2}}=-\frac{2}{3\pi^{\frac{3}{2}}} (3\log 2 -2) \sim -0.00951,
        \\
        I_X
        &=\frac{2(2\log2-1)}{\pi^\frac{3}{2}}\sim0.139.
    \end{aligned}
    \label{eq:app-hankel-summary}
\end{equation}
\section{Convolution of superspace kernels}
\label{app:superspace-convolution}
This appendix derives the convolution algebra for the class of
three-component superspace kernels used throughout this work.  The result
is purely algebraic and does not depend on the physical origin of the
kernel: it applies to any bilocal kernel with the Grassmann decomposition
given below.  To
keep the notation close to that of the main text, we write a generic kernel
as
\begin{equation}
	\begin{aligned}
		&K\!\left(t_{12};
		\theta_{12},\bar \theta_{12}, \bar \theta_{12}^+
		\right)
	 =
		K_1(t_{12})
		+\theta_{12}
		\left[
        \bar \theta^+_{12}K_2(t_{12})
		+\bar \theta_{12}K_3(t_{12})
		\right].
	\end{aligned}
	\label{eq:three-component-kernel}
\end{equation}
Here \(K_1,K_2,K_3\) denote the three bosonic components.  Their arguments
will be suppressed below. 
For the ordinary propagator one simply has
$
	(K_1,K_2,K_3)=(\Delta_1,\Delta_2,\Delta_3)$, and Eq.~\eqref{eq:three-component-kernel} reduces
precisely to Eq.~\eqref{eq:F123-decomposition}.

Next, let us consider two kernels of the form \eqref{eq:three-component-kernel}.  Their
superspace convolution is\footnote{Our Berezin convention is
$\int\!\dd\bar\theta\,\dd\theta\,\theta\bar\theta=1.$}
\begin{equation}
	\begin{aligned}
		&(K\star L)\!\left(t_{13};
		\theta_{13},\bar \theta_{13}, \bar \theta_{13}^+
		\right)
		 \equiv
		\int \dd t_2\,\dd^d x_2\,\dd\bar\theta_2\,\dd\theta_2\,
		K\!\left(
		t_{12},
		\theta_{12};\bar \theta_{12}, \bar \theta_{12}^+
		\right)
		L\!\left(
		t_{23};
		\theta_{23},\bar \theta_{23}, \bar \theta_{23}^+
		\right).
	\end{aligned}
	\label{eq:superspace-convolution}
\end{equation}
We use \(\circ\) for the remaining convolution over the bosonic coordinates,
\begin{equation}
	(f\circ g)(t_1, \bm x_1,t_3, \bm  x_3)
	\equiv
	\int \dd t_2\,\dd^d x_2\,
    f(t_1,\bm x_1,t_2,\bm x_2)g(t_2,\bm x_2,t_3,\bm x_3).
	\label{eq:bosonic-convolution}
\end{equation}
Performing the Grassmann integration over $\theta_2, \bar \theta_2$ gives
\begin{equation}
	\begin{aligned}
		&(K\star L)\!\left(
		t_{13};
		\theta_{13},\bar \theta_{13}, \bar \theta_{13}^+
		\right)
		 =
		(K\star L)_1
		+\theta_{13}
		\left[
        \bar\theta^+_{13}(K\star L)_2
        +
		\bar\theta_{13}(K\star L)_3
		\right].
	\end{aligned}
	\label{eq:convolved-kernel-form}
\end{equation}
The three bosonic components appearing here are
\begin{align}
	(K\star L)_1
	&=
	K_1\circ(L_3+L_2)
	+(K_3-K_2)\circ L_1,
	\nonumber\\
	(K\star L)_2
	&=
	K_3\circ L_2+K_2\circ L_3,
	\label{eq:convolution-components}\\
	(K \star L)_3
	&=
	K_3\circ L_3+K_2\circ L_2.
	\nonumber
\end{align}
Thus the convolution generates no additional Grassmann structure: it only
changes the three bosonic components.

To verify Eq.~\eqref{eq:convolution-components}, one expands the two kernels
in \(\theta_2,\bar\theta_2\) and retains only the terms proportional to
\(\theta_2\bar\theta_2\).  For example, the crossed terms acquire a minus
sign when the integration variables are brought into the chosen order,
\begin{equation}
	\int\!\dd\bar\theta_2\,\dd\theta_2\,
	(\theta_1\bar\theta_2)(\theta_2\bar\theta_3)
	=
	-\theta_1\bar\theta_3.
	\label{eq:crossed-berezin-signs}
\end{equation}
Collecting the surviving terms reproduces
Eq.~\eqref{eq:convolution-components}.  

\section{Supersymmetric Ward identities for \(n\)-point functions}
\label{app:n-pointkms}
In Subsec.~\ref{subsec:24}, we derived the supersymmetric Ward identity for a
two-point function. Here we generalize that construction to an arbitrary
$n$-point function.

Identities of this type are known: they follow from the supersymmetry of the
MSRJD generating functional
\cite{Parisi:1979ka,Kurchan1992,AronBiroliCugliandolo2010,Marguet_2021},
from the closed-time-path formalism \cite{WangHeinz2002}, and from the
Schwinger--Keldysh BRST superalgebra discussed in Subsec.~\ref{subsec:24},
and yield higher-order fluctuation--dissipation relations with the standard
theorem as the $n=2$ case. For $n=2$ they have also been solved: the most
general supersymmetric, causal, zero-ghost-number two-point superspace
function is known in closed form
\cite[Eq.~(89)]{Kurchan1992,10.1093/oso/9780198834625.001.0001}. What follows
extends that solution to arbitrary $n$. The invariants $T_a$ of
Eq.~\eqref{eq:Ta-invariant} generate the entire solution space, so that the
most general admissible $n$-point function is parametrized by
bidegree-$(k,k)$ forms on the space of relative times.

Consider $n$ superspace coordinates
\begin{equation}
Z_i=(t_i,\bm x_i,\theta_i,\bar\theta_i),
\qquad i=1,\ldots,n.
\end{equation}
We use the diagonal supersymmetry generators
\begin{equation}
Q=\sum_{i=1}^{n}\partial_{\theta_i},
\qquad
\widebar Q
=
\sum_{i=1}^{n}
\left(
\partial_{\bar\theta_i}
+
\theta_i\partial_{t_i}
\right),
\end{equation}
which satisfy
\begin{equation}
\{Q,\widebar Q\}
=
\sum_{i=1}^{n}\partial_{t_i}.
\end{equation}
Consequently, a function annihilated by both \(Q\) and \(\widebar Q\) is also invariant under a common time translation.
We choose \(Z_n\) as a reference point and introduce the relative coordinates
\begin{equation}
t_{an}=t_a-t_n,
\qquad
\theta_{an}=\theta_a-\theta_n,
\qquad
\bar\theta_{an}=\bar\theta_a-\bar\theta_n,
\end{equation}
with $a=1,\ldots,n-1$.
The following \(n-1\) time variables 
\begin{equation}
T_a
=
t_{an}
+
\theta_{an}\bar\theta_n
\label{eq:Ta-invariant}
\end{equation}
are invariant under both supersymmetry generators. Indeed,
\begin{equation}
Q T_a=0,
\end{equation}
while, using left Grassmann derivatives,
\begin{equation}
\widebar Q t_{an}=\theta_{an},
\qquad
\widebar Q\left(\theta_{an}\bar\theta_n\right)
=
-\theta_{an},
\end{equation}
and therefore
$
\widebar Q T_a=0.
$ To connect this construction with the symmetric superspace interval used in
Subsec.~\ref{subsec:24}, we define
\begin{equation}
    \mathsf {t}_{ab}
    \equiv
    t_{ab}
    +\frac{1}{2}\theta_{ab}\bar\theta_{ab}^{+},
    \qquad
    \bar\theta_{ab}^{+}
    \equiv
    \bar\theta_a+\bar\theta_b.
\end{equation}
This interval is invariant under both diagonal supersymmetry generators.
For \(b=n\), it is related to the asymmetric invariant \(T_a\) by
\begin{equation}
    \mathsf{t}_{an}
    =
    T_a+\frac{1}{2}\theta_{an}\bar\theta_{an}.
\end{equation}
Assuming spatial translation invariance, the most general solution of the Ward identities can therefore be written as
\begin{equation}
F(Z_1,\ldots,Z_n)
=
F\left(
T_a,\bm x_{an};
\theta_{an},\bar\theta_{an}
\right),
\qquad
a=1,\ldots,n-1.
\label{eq:npoint-invariants}
\end{equation}
In particular, the dependence on the reference coordinate \(\bar\theta_n\) is not independent: it enters only through the invariant times \(T_a\).

To make this dependence explicit, we define the restriction
\begin{equation}
\widehat F
\left(
t_{an},\bm x_{an};
\theta_{an},\bar\theta_{an}
\right)
=
\left.
F\left(
T_a,\bm x_{an};
\theta_{an},\bar\theta_{an}
\right)
\right|_{\bar\theta_n=0}.
\end{equation}
Since $
T_a=t_{an}+\theta_{an}\bar\theta_n,
$
the full function is reconstructed as
\begin{equation}
F
=
\exp\left(
\sum_{a=1}^{n-1}
\theta_{an}\bar\theta_n\partial_a
\right)
\widehat F,
\qquad
\partial_a
\equiv
\frac{\partial}{\partial t_{an}}.
\label{eq:F-from-Fhat}
\end{equation}
Because every term in the exponent contains the same Grassmann variable
\(\bar\theta_n\), the exponential truncates after the first order:
\begin{equation}
F
=
\widehat F
+
\sum_{a=1}^{n-1}
\theta_{an}\bar\theta_n\partial_a\widehat F.
\label{eq:F-expansion-base}
\end{equation}
For a Grassmann-even function with vanishing ghost number, the most general
expansion of \(\widehat F\) is
\begin{equation}
\widehat F
=
\sum_{k=0}^{n-1}
\frac{1}{(k!)^2}
\theta_{a_1n}\cdots\theta_{a_kn}
\bar\theta_{b_1n}\cdots\bar\theta_{b_kn}
\omega^{(k,k)}_{
a_1\cdots a_k\,;\,
b_1\cdots b_k
}.
\label{eq:Fhat-form-expansion}
\end{equation}
The coefficients
\(\omega^{(k,k)}\) are antisymmetric separately in the \(a\)-indices and
in the \(b\)-indices. They may therefore be regarded as a differential form of
bidegree \((k,k)\) on the space of relative times.

Acting on the term of bidegree \((k,k)\), the correction in
Eq.~\eqref{eq:F-expansion-base} is
\begin{equation}
\begin{aligned}
F_k-\widehat F_k
=
\frac{1}{(k!)^2}
\theta_{a_0n}
\theta_{a_1n}\cdots\theta_{a_kn}
\bar\theta_{b_1n}\cdots\bar\theta_{b_kn}
\partial_{a_0}
\omega^{(k,k)}_{
a_1\cdots a_k\,;\,
b_1\cdots b_k
}
\bar\theta_n.
\end{aligned}
\label{eq:Fk-correction}
\end{equation}
Moving \(\bar\theta_n\) to the right produces no sign, since it crosses
\(2k\) relative Grassmann variables. Moreover, the product
\(\theta_{a_0n}\cdots\theta_{a_kn}\) projects the coefficient onto its
completely antisymmetric part in the indices
\(a_0,\ldots,a_k\). This introduces the exterior derivative acting on the
first index block:
\begin{equation}
\begin{aligned}
\left(
d\omega^{(k,k)}
\right)_{
a_0\cdots a_k\,;\,
b_1\cdots b_k
}
&=
\sum_{i=0}^{k}
(-1)^i
\partial_{a_i}
\omega^{(k,k)}_{
a_0\cdots\widehat{a_i}\cdots a_k\,;\,
b_1\cdots b_k
}.
\end{aligned}
\label{eq:exterior-derivative}
\end{equation}
Thus,
\begin{equation}
d:
\Omega^{(k,k)}
\longrightarrow
\Omega^{(k+1,k)}.
\end{equation}
Using
\begin{equation}
\theta_{a_0n}\cdots\theta_{a_kn}
\partial_{a_0}
\omega^{(k,k)}_{
a_1\cdots a_k\,;\,
b_1\cdots b_k
}
=
\frac{1}{k+1}
\theta_{a_0n}\cdots\theta_{a_kn}
\left(
d\omega^{(k,k)}
\right)_{
a_0\cdots a_k\,;\,
b_1\cdots b_k
},
\end{equation}
we obtain the complete supersymmetric \(n\)-point function:
\begin{equation}
\begin{aligned}
F
& =
\sum_{k=0}^{n-1}
\frac{1}{(k!)^2}
\theta_{a_1n}\cdots\theta_{a_kn}
\bar\theta_{b_1n}\cdots\bar\theta_{b_kn}
\omega^{(k,k)}_{
a_1\cdots a_k\,;\,
b_1\cdots b_k
}
\\
&+
\sum_{k=0}^{n-2}
\frac{1}{k!(k+1)!}
\theta_{a_0n}\cdots\theta_{a_kn}
\bar\theta_{b_1n}\cdots\bar\theta_{b_kn}
\left(
d\omega^{(k,k)}
\right)_{
a_0\cdots a_k\,;\,
b_1\cdots b_k
}
\bar\theta_n.
\label{eq:general-npoint-kms}
\end{aligned}
\end{equation}

Equation~\eqref{eq:general-npoint-kms} shows that the components containing
the base coordinate \(\bar\theta_n\) are not independent. They are fixed by
the exterior derivative of the components of \(\widehat F\). In component
language, these relations encode the Ward identities associated with the
supersymmetry generated by \(Q\) and \(\widebar Q\).

\subsection{Reduction to the two-point function}
As a check, consider \(n=2\). The general expression becomes
\begin{equation}
F
=
\omega^{(0,0)}
+\theta_{12}\bar\theta_{12}\,\omega^{(1,1)}
+\theta_{12}\bar\theta_2\,
\partial_{t_{12}}\omega^{(0,0)}.
\end{equation}
Using
\begin{equation}
\bar\theta_2
=
\frac12\left(\bar\theta_{12}^{+}-\bar\theta_{12}\right),
\qquad
\bar\theta_{12}^{+}=\bar\theta_1+\bar\theta_2,
\end{equation}
this can be rewritten as
\begin{align}
F
={}&
\omega^{(0,0)}
+\frac12\theta_{12}\bar\theta_{12}^{+}
\partial_{t_{12}}\omega^{(0,0)}
+
\theta_{12}\bar\theta_{12}
\left[
\omega^{(1,1)}
-\frac12\partial_{t_{12}}\omega^{(0,0)}
\right].
\end{align}
Identifying \(F_1=\omega^{(0,0)}\), we recover the two-point Ward identity in
Eq.~\eqref{eq:G2-G1-susy-relation}.

\bibliographystyle{JHEP}
\bibliography{ref_lib_verified}

\providecommand{\href}[2]{#2}\begingroup\raggedright\begin{thebibliography}{10}

\bibitem{HohenbergHalperin1977}
P.C.~Hohenberg and B.I.~Halperin, \emph{Theory of dynamic critical phenomena},
  \href{https://doi.org/10.1103/RevModPhys.49.435}{\emph{Rev. Mod. Phys.}
  {\bfseries 49} (1977) 435}.

\bibitem{FolkMoser2006}
R.~Folk and G.~Moser, \emph{Critical dynamics: A field-theoretical approach},
  \href{https://doi.org/10.1088/0305-4470/39/24/R01}{\emph{J. Phys. A}
  {\bfseries 39} (2006) R207}.

\bibitem{Tauber2014}
U.C.~T{\"a}uber, \emph{Critical Dynamics: A Field Theory Approach to
  Equilibrium and Non-Equilibrium Scaling Behavior}, Cambridge University
  Press, Cambridge (2014),
  \href{https://doi.org/10.1017/CBO9781139046213}{10.1017/CBO9781139046213}.

\bibitem{ZengZhong2023}
S.~Zeng and F.~Zhong, \emph{Theory of critical phenomena with long-range
  temporal interaction},
  \href{https://doi.org/10.1088/1402-4896/acdcc0}{\emph{Phys. Scr.} {\bfseries
  98} (2023) 075017} [\href{https://arxiv.org/abs/2212.11076}{{\ttfamily
  2212.11076}}].

\bibitem{SiebererBuchholdDiehl2016}
L.M.~Sieberer, M.~Buchhold and S.~Diehl, \emph{{Keldysh} field theory for
  driven open quantum systems},
  \href{https://doi.org/10.1088/0034-4885/79/9/096001}{\emph{Rep. Prog. Phys.}
  {\bfseries 79} (2016) 096001}
  [\href{https://arxiv.org/abs/1512.00637}{{\ttfamily 1512.00637}}].

\bibitem{SorienteEtAl2021}
M.~Soriente, T.L.~Heugel, K.~Omiya, R.~Chitra and O.~Zilberberg,
  \emph{Distinctive class of dissipation-induced phase transitions and their
  universal characteristics},
  \href{https://doi.org/10.1103/PhysRevResearch.3.023100}{\emph{Phys. Rev.
  Research} {\bfseries 3} (2021) 023100}
  [\href{https://arxiv.org/abs/2101.12227}{{\ttfamily 2101.12227}}].

\bibitem{KhedriHornZilberberg2022}
A.~Khedri, D.~Horn and O.~Zilberberg, \emph{Fate of exceptional points in the
  presence of nonlinearities},
  \href{https://arxiv.org/abs/2208.11205}{{\ttfamily 2208.11205}}.

\bibitem{Rajagopal:1992qz}
K.~Rajagopal and F.~Wilczek, \emph{{Static and dynamic critical phenomena at a
  second order QCD phase transition}},
  \href{https://doi.org/10.1016/0550-3213(93)90502-G}{\emph{Nucl. Phys. B}
  {\bfseries 399} (1993) 395}
  [\href{https://arxiv.org/abs/hep-ph/9210253}{{\ttfamily hep-ph/9210253}}].

\bibitem{FlorioGrossiSolovievTeaney2022}
A.~Florio, E.~Grossi, A.~Soloviev and D.~Teaney, \emph{Dynamics of the {$O(4)$}
  critical point in {QCD}},
  \href{https://doi.org/10.1103/PhysRevD.105.054512}{\emph{Phys. Rev. D}
  {\bfseries 105} (2022) 054512}
  [\href{https://arxiv.org/abs/2111.03640}{{\ttfamily 2111.03640}}].

\bibitem{Berges2015}
J.~Berges, \emph{Nonequilibrium quantum fields: From cold atoms to cosmology},
  in \emph{Strongly Interacting Quantum Systems out of Equilibrium},
  T.~Giamarchi, A.J.~Millis, O.~Parcollet, H.~Saleur and L.F.~Cugliandolo,
  eds., vol.~99 of \emph{Lecture Notes of the Les Houches Summer School},
  (Oxford), pp.~69--206, Oxford University Press (2016)
  [\href{https://arxiv.org/abs/1503.02907}{{\ttfamily 1503.02907}}].

\bibitem{Moreau:2019jpn}
G.~Moreau and J.~Serreau, \emph{Unequal-time correlators of stochastic scalar
  fields in de sitter space},
  \href{https://doi.org/10.1103/PhysRevD.101.045015}{\emph{Phys. Rev. D}
  {\bfseries 101} (2020) 045015}
  [\href{https://arxiv.org/abs/1912.05358}{{\ttfamily 1912.05358}}].

\bibitem{BauschJanssenWagner1976}
R.~Bausch, H.-K.~Janssen and H.~Wagner, \emph{Renormalized field theory of
  critical dynamics}, \href{https://doi.org/10.1007/BF01312880}{\emph{Z. Phys.
  B} {\bfseries 24} (1976) 113}.

\bibitem{OhnishiKunihiro2006}
K.~Ohnishi and T.~Kunihiro, \emph{Overdamping phenomena near the critical point
  in {$O(N)$} model},
  \href{https://doi.org/10.1016/j.physletb.2005.10.049}{\emph{Physics Letters
  B} {\bfseries 632} (2006) 252}
  [\href{https://arxiv.org/abs/nucl-th/0503017}{{\ttfamily nucl-th/0503017}}].

\bibitem{BoyanovskyDeVega2002}
D.~Boyanovsky and H.J.~de~Vega, \emph{Dynamics near the critical point: The hot
  renormalization group in quantum field theory},
  \href{https://doi.org/10.1103/PhysRevD.65.085038}{\emph{Phys. Rev. D}
  {\bfseries 65} (2002) 085038}
  [\href{https://arxiv.org/abs/hep-ph/0110012}{{\ttfamily hep-ph/0110012}}].

\bibitem{SchweitzerSchlichtingVonSmekal2020}
D.~Schweitzer, S.~Schlichting and L.~von Smekal, \emph{Spectral functions and
  dynamic critical behavior of relativistic {$Z_2$} theories},
  \href{https://doi.org/10.1016/j.nuclphysb.2020.115165}{\emph{Nuclear Physics
  B} {\bfseries 960} (2020) 115165}
  [\href{https://arxiv.org/abs/2007.03374}{{\ttfamily 2007.03374}}].

\bibitem{Kibble:1976sj}
T.W.B.~Kibble, \emph{{Topology of Cosmic Domains and Strings}},
  \href{https://doi.org/10.1088/0305-4470/9/8/029}{\emph{J. Phys. A} {\bfseries
  9} (1976) 1387}.

\bibitem{Kibble:1980mv}
T.W.B.~Kibble, \emph{{Some Implications of a Cosmological Phase Transition}},
  \href{https://doi.org/10.1016/0370-1573(80)90091-5}{\emph{Phys. Rept.}
  {\bfseries 67} (1980) 183}.

\bibitem{Zurek:1985qw}
W.H.~Zurek, \emph{{Cosmological Experiments in Superfluid Helium?}},
  \href{https://doi.org/10.1038/317505a0}{\emph{Nature} {\bfseries 317} (1985)
  505}.

\bibitem{Son:2004iv}
D.T.~Son and M.A.~Stephanov, \emph{{Dynamic universality class of the QCD
  critical point}},
  \href{https://doi.org/10.1103/PhysRevD.70.056001}{\emph{Phys. Rev. D}
  {\bfseries 70} (2004) 056001}
  [\href{https://arxiv.org/abs/hep-ph/0401052}{{\ttfamily hep-ph/0401052}}].

\bibitem{Chattopadhyay:2023jfm}
C.~Chattopadhyay, J.~Ott, T.~Schaefer and V.~Skokov, \emph{{Dynamic scaling of
  order parameter fluctuations in model B}},
  \href{https://doi.org/10.1103/PhysRevD.108.074004}{\emph{Phys. Rev. D}
  {\bfseries 108} (2023) 074004}
  [\href{https://arxiv.org/abs/2304.07279}{{\ttfamily 2304.07279}}].

\bibitem{Chattopadhyay:2024jlh}
C.~Chattopadhyay, J.~Ott, T.~Schaefer and V.V.~Skokov, \emph{{Simulations of
  Stochastic Fluid Dynamics near a Critical Point in the Phase Diagram}},
  \href{https://doi.org/10.1103/PhysRevLett.133.032301}{\emph{Phys. Rev. Lett.}
  {\bfseries 133} (2024) 032301}
  [\href{https://arxiv.org/abs/2403.10608}{{\ttfamily 2403.10608}}].

\bibitem{Chattopadhyay:2025uqo}
C.~Chattopadhyay, J.~Ott, T.~Schaefer and V.V.~Skokov, \emph{{Transport
  properties of stochastic fluids}},
  \href{https://doi.org/10.1103/kltq-qb4t}{\emph{Phys. Rev. D} {\bfseries 112}
  (2025) 114026} [\href{https://arxiv.org/abs/2510.12557}{{\ttfamily
  2510.12557}}].

\bibitem{Chattopadhyay:2025zac}
C.~Chattopadhyay, J.~Ott, T.~Schaefer and V.~Skokov, \emph{{Simulating
  stochastic fluid dynamics}},
  \href{https://doi.org/10.1051/epjconf/202636415002}{\emph{EPJ Web Conf.}
  {\bfseries 364} (2026) 15002}
  [\href{https://arxiv.org/abs/2509.00545}{{\ttfamily 2509.00545}}].

\bibitem{Chattopadhyay:2026dyd}
C.~Chattopadhyay, R.~Maguire, J.~Ott, T.~Schaefer and V.V.~Skokov,
  \emph{{Critical dynamics of the superfluid phase transition in model F}},
  \href{https://doi.org/10.1103/q49b-yygq}{\emph{Phys. Rev. A} {\bfseries 114}
  (2026) 013312} [\href{https://arxiv.org/abs/2603.21479}{{\ttfamily
  2603.21479}}].

\bibitem{Sieke:2026ozy}
L.J.~Sieke, J.~Fuchs and L.~von Smekal, \emph{{Non-equilibrium scaling across
  first-order transitions with self-interacting scalar fields}},
  \href{https://doi.org/10.1016/j.nuclphysb.2026.117585}{\emph{Nucl. Phys. B}
  {\bfseries 1029} (2026) 117585}
  [\href{https://arxiv.org/abs/2605.10346}{{\ttfamily 2605.10346}}].

\bibitem{FlorioGrossiTeaney2024}
A.~Florio, E.~Grossi and D.~Teaney, \emph{Dynamics of the {$O(4)$} critical
  point in {QCD}: Critical pions and diffusion in {Model G}},
  \href{https://doi.org/10.1103/PhysRevD.109.054037}{\emph{Phys. Rev. D}
  {\bfseries 109} (2024) 054037}
  [\href{https://arxiv.org/abs/2306.06887}{{\ttfamily 2306.06887}}].

\bibitem{FlorioEtAl2025Quenching}
A.~Florio, E.~Grossi, A.~Mazeliauskas, A.~Soloviev and D.~Teaney,
  \emph{Quenching through the {QCD} chiral phase transition},
  \href{https://doi.org/10.1103/plfm-z5xx}{\emph{Phys. Rev. D} {\bfseries 112}
  (2025) 114019} [\href{https://arxiv.org/abs/2504.03514}{{\ttfamily
  2504.03514}}].

\bibitem{FlorioEtAl2025Goldstones}
A.~Florio, E.~Grossi, A.~Mazeliauskas, A.~Soloviev and D.~Teaney,
  \emph{Supercooled goldstone bosons at the {QCD} chiral phase transition},
  \href{https://doi.org/10.1103/wyn4-ncdc}{\emph{Phys. Rev. Lett.} {\bfseries
  135} (2025) 242303} [\href{https://arxiv.org/abs/2504.03516}{{\ttfamily
  2504.03516}}].

\bibitem{Dupuis:2020fhh}
N.~Dupuis, L.~Canet, A.~Eichhorn, W.~Metzner, J.M.~Pawlowski, M.~Tissier
  et~al., \emph{{The nonperturbative functional renormalization group and its
  applications}},
  \href{https://doi.org/10.1016/j.physrep.2021.01.001}{\emph{Phys. Rept.}
  {\bfseries 910} (2021) 1} [\href{https://arxiv.org/abs/2006.04853}{{\ttfamily
  2006.04853}}].

\bibitem{RothVonSmekal2023}
J.V.~Roth and L.~von Smekal, \emph{Critical dynamics in a real-time formulation
  of the functional renormalization group},
  \href{https://doi.org/10.1007/JHEP10(2023)065}{\emph{JHEP} {\bfseries 10}
  (2023) 065} [\href{https://arxiv.org/abs/2303.11817}{{\ttfamily
  2303.11817}}].

\bibitem{Batini:2023nan}
L.~Batini, E.~Grossi and N.~Wink, \emph{Dissipation dynamics of a scalar
  field}, \href{https://doi.org/10.1103/PhysRevD.108.125021}{\emph{Phys. Rev.
  D} {\bfseries 108} (2023) 125021}
  [\href{https://arxiv.org/abs/2309.06586}{{\ttfamily 2309.06586}}].

\bibitem{RothEtAl2025ModelG}
J.V.~Roth, Y.~Ye, S.~Schlichting and L.~von Smekal, \emph{Dynamic critical
  behavior of the chiral phase transition from the real-time functional
  renormalization group},
  \href{https://doi.org/10.1007/JHEP01(2025)118}{\emph{JHEP} {\bfseries 01}
  (2025) 118} [\href{https://arxiv.org/abs/2403.04573}{{\ttfamily
  2403.04573}}].

\bibitem{RothEtAl2025ModelsGH}
J.V.~Roth, Y.~Ye, S.~Schlichting and L.~von Smekal, \emph{Universal critical
  dynamics near the chiral phase transition and the {QCD} critical point},
  \href{https://doi.org/10.1103/PhysRevD.111.L111901}{\emph{Phys. Rev. D}
  {\bfseries 111} (2025) L111901}
  [\href{https://arxiv.org/abs/2409.14470}{{\ttfamily 2409.14470}}].

\bibitem{SaitoFujiiItakuraMorimatsu2015}
Y.~Saito, H.~Fujii, K.~Itakura and O.~Morimatsu, \emph{Microscopic
  identification of dissipative modes in relativistic field theories},
  \href{https://doi.org/10.1093/ptep/ptv065}{\emph{Prog. Theor. Exp. Phys.}
  {\bfseries 2015} (2015) 053A02}
  [\href{https://arxiv.org/abs/1309.4892}{{\ttfamily 1309.4892}}].

\bibitem{BorsanyiEtAl2000}
S.~Bors{\'a}nyi, A.~Patk{\'o}s, J.~Polonyi and Z.~Sz{\'e}p, \emph{Fate of the
  classical false vacuum},
  \href{https://doi.org/10.1103/PhysRevD.62.085013}{\emph{Phys. Rev. D}
  {\bfseries 62} (2000) 085013}
  [\href{https://arxiv.org/abs/hep-th/0004059}{{\ttfamily hep-th/0004059}}].

\bibitem{BatiniChatrchyanBerges2024}
L.~Batini, A.~Chatrchyan and J.~Berges, \emph{Real-time dynamics of false
  vacuum decay}, \href{https://doi.org/10.1103/PhysRevD.109.023502}{\emph{Phys.
  Rev. D} {\bfseries 109} (2024) 023502}
  [\href{https://arxiv.org/abs/2310.04206}{{\ttfamily 2310.04206}}].

\bibitem{HattaKunihiro}
Y.~Hatta and T.~Kunihiro, \emph{Renormalization group method applied to kinetic
  equations: Roles of initial values and time},
  \href{https://doi.org/10.1006/aphy.2002.6234}{\emph{Ann. Phys.} {\bfseries
  298} (2002) 24} [\href{https://arxiv.org/abs/hep-th/0108159}{{\ttfamily
  hep-th/0108159}}].

\bibitem{TsumuraKunihiroOhnishi}
K.~Tsumura, T.~Kunihiro and K.~Ohnishi, \emph{Derivation of covariant
  dissipative fluid dynamics in the renormalization-group method},
  \href{https://doi.org/10.1016/j.physletb.2006.12.074}{\emph{Phys. Lett. B}
  {\bfseries 646} (2007) 134}
  [\href{https://arxiv.org/abs/hep-ph/0609056}{{\ttfamily hep-ph/0609056}}].

\bibitem{kamenev2005manybodytheorynonequilibriumsystems}
A.~Kamenev, \emph{Many-body theory of non-equilibrium systems},
  \href{https://doi.org/10.1016/S0924-8099(05)80045-9}{\emph{Les Houches Summer
  School Proceedings} {\bfseries 81} (2005) 177}
  [\href{https://arxiv.org/abs/cond-mat/0412296}{{\ttfamily
  cond-mat/0412296}}].

\bibitem{Floerchinger_2012}
S.~Floerchinger, \emph{Analytic continuation of functional renormalization
  group equations}, \href{https://doi.org/10.1007/JHEP05(2012)021}{\emph{JHEP}
  {\bfseries 05} (2012) 021} [\href{https://arxiv.org/abs/1112.4374}{{\ttfamily
  1112.4374}}].

\bibitem{Floerchinger_2016}
S.~Floerchinger, \emph{Variational principle for theories with dissipation from
  analytic continuation},
  \href{https://doi.org/10.1007/JHEP09(2016)099}{\emph{JHEP} {\bfseries 09}
  (2016) 099} [\href{https://arxiv.org/abs/1603.07148}{{\ttfamily
  1603.07148}}].

\bibitem{Braun:2022mgx}
J.~Braun, Y.-r.~Chen, W.-j.~Fu, A.~Gei{\ss}el, J.~Horak, C.~Huang et~al.,
  \emph{Renormalised spectral flows},
  \href{https://doi.org/10.21468/SciPostPhysCore.6.3.061}{\emph{SciPost Phys.
  Core} {\bfseries 6} (2023) 061}
  [\href{https://arxiv.org/abs/2206.10232}{{\ttfamily 2206.10232}}].

\bibitem{Frangi:2025xss}
G.~Frangi and S.~Grozdanov, \emph{Wilsonian renormalization group and thermal
  field theory in the schwinger--keldysh closed-time-path formalism},
  \href{https://doi.org/10.1103/PhysRevD.111.085034}{\emph{Phys. Rev. D}
  {\bfseries 111} (2025) 085034}
  [\href{https://arxiv.org/abs/2501.16441}{{\ttfamily 2501.16441}}].

\bibitem{StoetzelFloerchinger2025}
T.~Stoetzel and S.~Floerchinger, \emph{Shear viscosity of a relativistic scalar
  field from functional renormalization},
  \href{https://arxiv.org/abs/2512.18740}{{\ttfamily 2512.18740}}.

\bibitem{Canet:2011wf}
L.~Canet, H.~Chat{\'e} and B.~Delamotte, \emph{General framework of the
  non-perturbative renormalization group for non-equilibrium steady states},
  \href{https://doi.org/10.1088/1751-8113/44/49/495001}{\emph{J. Phys. A}
  {\bfseries 44} (2011) 495001}
  [\href{https://arxiv.org/abs/1106.4129}{{\ttfamily 1106.4129}}].

\bibitem{Floerchinger:2026pwi}
S.~Floerchinger, \emph{{Response theory for quantum fields in isolation}},
  \href{https://arxiv.org/abs/2604.13637}{{\ttfamily 2604.13637}}.

\bibitem{AronBiroliCugliandolo2010}
C.~Aron, G.~Biroli and L.F.~Cugliandolo, \emph{Symmetries of generating
  functionals of {Langevin} processes with colored multiplicative noise},
  \href{https://doi.org/10.1088/1742-5468/2010/11/P11018}{\emph{J. Stat. Mech.}
  {\bfseries 2010} (2010) P11018}
  [\href{https://arxiv.org/abs/1007.5059}{{\ttfamily 1007.5059}}].

\bibitem{Marguet_2021}
B.~Marguet, E.~Agoritsas, L.~Canet and V.~Lecomte, \emph{Supersymmetries in
  nonequilibrium langevin dynamics},
  \href{https://doi.org/10.1103/PhysRevE.104.044120}{\emph{Phys. Rev. E}
  {\bfseries 104} (2021) 044120}
  [\href{https://arxiv.org/abs/2101.08766}{{\ttfamily 2101.08766}}].

\bibitem{Gao:2017bqf}
P.~Gao and H.~Liu, \emph{Emergent supersymmetry in local equilibrium systems},
  \href{https://doi.org/10.1007/JHEP01(2018)040}{\emph{JHEP} {\bfseries 01}
  (2018) 040} [\href{https://arxiv.org/abs/1701.07445}{{\ttfamily
  1701.07445}}].

\bibitem{MartinSiggiaRose1973}
P.C.~Martin, E.D.~Siggia and H.A.~Rose, \emph{Statistical dynamics of classical
  systems}, \href{https://doi.org/10.1103/PhysRevA.8.423}{\emph{Phys. Rev. A}
  {\bfseries 8} (1973) 423}.

\bibitem{Janssen1976}
H.-K.~Janssen, \emph{On a lagrangean for classical field dynamics and
  renormalization group calculations of dynamical critical properties},
  \href{https://doi.org/10.1007/BF01316547}{\emph{Z. Phys. B} {\bfseries 23}
  (1976) 377}.

\bibitem{DeDominicis1976}
C.~De~Dominicis, \emph{Techniques de renormalisation de la th{\'e}orie des
  champs et dynamique des ph{\'e}nom{\`e}nes critiques},
  \href{https://doi.org/10.1051/jphyscol:1976138}{\emph{J. Phys. Colloques}
  {\bfseries 37} (1976) C1}.

\bibitem{Parisi:1979ka}
G.~Parisi and N.~Sourlas, \emph{{Random Magnetic Fields, Supersymmetry and
  Negative Dimensions}},
  \href{https://doi.org/10.1103/PhysRevLett.43.744}{\emph{Phys. Rev. Lett.}
  {\bfseries 43} (1979) 744}.

\bibitem{Kurchan1992}
J.~Kurchan, \emph{Supersymmetry in spin glass dynamics},
  \href{https://doi.org/10.1051/jp1:1992214}{\emph{J. Phys. I France}
  {\bfseries 2} (1992) 1333}.

\bibitem{Tyutin:1975qk}
I.V.~Tyutin, \emph{{Gauge Invariance in Field Theory and Statistical Physics in
  Operator Formalism}},  \href{https://arxiv.org/abs/0812.0580}{{\ttfamily
  0812.0580}}.

\bibitem{DeDominicis:1976}
C.~De~Dominicis, \emph{{Techniques de renormalisation de la theorie des champs
  et dynamique des phenomenes critiques}},
  \href{https://doi.org/10.1051/jphyscol:1976138}{\emph{J. Phys. Colloques}
  {\bfseries 37} (1976) C1}.

\bibitem{HaehlLoganayagamRangamani2017}
F.M.~Haehl, R.~Loganayagam and M.~Rangamani, \emph{Schwinger-{K}eldysh
  formalism. {P}art {I}: {BRST} symmetries and superspace},
  \href{https://doi.org/10.1007/JHEP06(2017)069}{\emph{JHEP} {\bfseries 06}
  (2017) 069} [\href{https://arxiv.org/abs/1610.01940}{{\ttfamily
  1610.01940}}].

\bibitem{Kubo:1957mj}
R.~Kubo, \emph{{Statistical-Mechanical Theory of Irreversible Processes. I.
  General Theory and Simple Applications to Magnetic and Conduction Problems}},
  \href{https://doi.org/10.1143/JPSJ.12.570}{\emph{J. Phys. Soc. Jap.}
  {\bfseries 12} (1957) 570}.

\bibitem{Martin:1959jp}
P.C.~Martin and J.~Schwinger, \emph{{Theory of Many-Particle Systems. I}},
  \href{https://doi.org/10.1103/PhysRev.115.1342}{\emph{Phys. Rev.} {\bfseries
  115} (1959) 1342}.

\bibitem{10.1093/oso/9780198834625.001.0001}
J.~Zinn-Justin, \emph{Quantum Field Theory and Critical Phenomena}, vol.~171 of
  \emph{International Series of Monographs on Physics}, Oxford University
  Press, Oxford, 5~ed. (2021),
  \href{https://doi.org/10.1093/oso/9780198834625.001.0001}{10.1093/oso/9780198834625.001.0001}.

\bibitem{GloriosoCrossleyLiu2017}
P.~Glorioso, M.~Crossley and H.~Liu, \emph{Effective field theory of
  dissipative fluids ({II}): classical limit, dynamical {KMS} symmetry and
  entropy current}, \href{https://doi.org/10.1007/JHEP09(2017)096}{\emph{JHEP}
  {\bfseries 09} (2017) 096}
  [\href{https://arxiv.org/abs/1701.07817}{{\ttfamily 1701.07817}}].

\bibitem{Gao:2018bxz}
P.~Gao, P.~Glorioso and H.~Liu, \emph{{Ghostbusters: Unitarity and Causality of
  Non-equilibrium Effective Field Theories}},
  \href{https://doi.org/10.1007/JHEP03(2020)040}{\emph{JHEP} {\bfseries 03}
  (2020) 040} [\href{https://arxiv.org/abs/1803.10778}{{\ttfamily
  1803.10778}}].

\bibitem{Intriligator1997}
K.~Intriligator and N.~Seiberg, \emph{Lectures on supersymmetric gauge theories
  and electric-magnetic duality},  in \emph{Low-Dimensional Applications of
  Quantum Field Theory}, L.~Baulieu, V.~Kazakov, M.~Picco and P.~Windey, eds.,
  (Boston, MA), pp.~161--199, Springer US (1997),
  \href{https://doi.org/10.1007/978-1-4899-1919-9_8}{DOI}.

\bibitem{Crossley:2015evo}
M.~Crossley, P.~Glorioso and H.~Liu, \emph{{Effective field theory of
  dissipative fluids}},
  \href{https://doi.org/10.1007/JHEP09(2017)095}{\emph{JHEP} {\bfseries 09}
  (2017) 095} [\href{https://arxiv.org/abs/1511.03646}{{\ttfamily
  1511.03646}}].

\bibitem{Hertz_2016}
J.A.~Hertz, Y.~Roudi and P.~Sollich, \emph{Path integral methods for the
  dynamics of stochastic and disordered systems},
  \href{https://doi.org/10.1088/1751-8121/50/3/033001}{\emph{Journal of Physics
  A: Mathematical and Theoretical} {\bfseries 50} (2016) 033001}.

\bibitem{Rychkov:2023rgq}
S.~Rychkov, \emph{{Four Lectures on the Random Field Ising Model,
  Parisi-Sourlas Supersymmetry, and Dimensional Reduction}} (3, 2023),
  \href{https://doi.org/10.1007/978-3-031-42000-9}{10.1007/978-3-031-42000-9},
  [\href{https://arxiv.org/abs/2303.09654}{{\ttfamily 2303.09654}}].

\bibitem{HaehlLoganayagamRangamani2017b}
F.M.~Haehl, R.~Loganayagam and M.~Rangamani, \emph{Schwinger-{K}eldysh
  formalism. {P}art {II}: thermal equivariant cohomology},
  \href{https://doi.org/10.1007/JHEP06(2017)070}{\emph{JHEP} {\bfseries 06}
  (2017) 070} [\href{https://arxiv.org/abs/1610.01941}{{\ttfamily
  1610.01941}}].

\bibitem{JensenPinzaniFokeevaYarom2018}
K.~Jensen, N.~Pinzani-Fokeeva and A.~Yarom, \emph{Dissipative hydrodynamics in
  superspace}, \href{https://doi.org/10.1007/JHEP09(2018)127}{\emph{JHEP}
  {\bfseries 09} (2018) 127}
  [\href{https://arxiv.org/abs/1701.07436}{{\ttfamily 1701.07436}}].

\bibitem{JensenMarjiehPinzaniFokeevaYarom2018}
K.~Jensen, R.~Marjieh, N.~Pinzani-Fokeeva and A.~Yarom, \emph{A panoply of
  {S}chwinger-{K}eldysh transport},
  \href{https://doi.org/10.21468/SciPostPhys.5.5.053}{\emph{SciPost Phys.}
  {\bfseries 5} (2018) 053} [\href{https://arxiv.org/abs/1804.04654}{{\ttfamily
  1804.04654}}].

\bibitem{Floerchinger:2021uyo}
S.~Floerchinger and E.~Grossi, \emph{{Conserved and nonconserved Noether
  currents from the quantum effective action}},
  \href{https://doi.org/10.1103/PhysRevD.105.085015}{\emph{Phys. Rev. D}
  {\bfseries 105} (2022) 085015}
  [\href{https://arxiv.org/abs/2102.11098}{{\ttfamily 2102.11098}}].

\bibitem{DeDominicisPeliti1978}
C.~De~Dominicis and L.~Peliti, \emph{Field-theory renormalization and critical
  dynamics above {$T_c$}: Helium, antiferromagnets, and liquid-gas systems},
  \href{https://doi.org/10.1103/PhysRevB.18.353}{\emph{Physical Review B}
  {\bfseries 18} (1978) 353}.

\bibitem{Berges2004}
J.~Berges, \emph{Introduction to nonequilibrium quantum field theory},
  \href{https://doi.org/10.1063/1.1843591}{\emph{AIP Conf. Proc.} {\bfseries
  739} (2004) 3} [\href{https://arxiv.org/abs/hep-ph/0409233}{{\ttfamily
  hep-ph/0409233}}].

\bibitem{WilsonFisher1972}
K.G.~Wilson and M.E.~Fisher, \emph{Critical exponents in 3.99 dimensions},
  \href{https://doi.org/10.1103/PhysRevLett.28.240}{\emph{Phys. Rev. Lett.}
  {\bfseries 28} (1972) 240}.

\bibitem{Wilson:1973jj}
K.G.~Wilson and J.B.~Kogut, \emph{The renormalization group and the epsilon
  expansion}, \href{https://doi.org/10.1016/0370-1573(74)90023-4}{\emph{Phys.
  Rep.} {\bfseries 12} (1974) 75}.

\bibitem{HalperinHohenbergMa1972}
B.I.~Halperin, P.C.~Hohenberg and S.-k.~Ma, \emph{Calculation of dynamic
  critical properties using {W}ilson's expansion methods},
  \href{https://doi.org/10.1103/PhysRevLett.29.1548}{\emph{Phys. Rev. Lett.}
  {\bfseries 29} (1972) 1548}.

\bibitem{HalperinHohenbergMa1974}
B.I.~Halperin, P.C.~Hohenberg and S.-k.~Ma, \emph{Renormalization-group methods
  for critical dynamics: {I}. recursion relations and effects of energy
  conservation}, \href{https://doi.org/10.1103/PhysRevB.10.139}{\emph{Phys.
  Rev. B} {\bfseries 10} (1974) 139}.

\bibitem{AdzhemyanEtAl2022}
L.T.~Adzhemyan, D.A.~Evdokimov, M.~Hnati{\v{c}}, E.V.~Ivanova, M.V.~Kompaniets,
  A.~Kudlis et~al., \emph{Model a of critical dynamics: Five-loop
  {$\varepsilon$}-expansion study},
  \href{https://doi.org/10.1016/j.physa.2022.127530}{\emph{Physica A:
  Statistical Mechanics and its Applications} {\bfseries 600} (2022) 127530}
  [\href{https://arxiv.org/abs/2201.12640}{{\ttfamily 2201.12640}}].

\bibitem{SchweitzerSchlichtingVonSmekal2022}
D.~Schweitzer, S.~Schlichting and L.~von Smekal, \emph{Critical dynamics of
  relativistic diffusion},
  \href{https://doi.org/10.1016/j.nuclphysb.2022.115944}{\emph{Nuclear Physics
  B} {\bfseries 984} (2022) 115944}
  [\href{https://arxiv.org/abs/2110.01696}{{\ttfamily 2110.01696}}].

\bibitem{Watson1944}
G.N.~Watson, \emph{A Treatise on the Theory of Bessel Functions}, Cambridge
  University Press, Cambridge, 2~ed. (1944).

\bibitem{AbramowitzStegun1964}
M.~Abramowitz and I.A.~Stegun, eds., \emph{Handbook of Mathematical Functions
  with Formulas, Graphs, and Mathematical Tables}, no.~55 in National Bureau of
  Standards Applied Mathematics Series, U.S. Government Printing Office,
  Washington, DC (1964).

\bibitem{FolkMoser2004}
R.~Folk and G.~Moser, \emph{Critical dynamics of stochastic models with energy
  conservation (model c)},
  \href{https://doi.org/10.1103/PhysRevE.69.036101}{\emph{Physical Review E}
  {\bfseries 69} (2004) 036101}.

\bibitem{MesterhazyEtAl2013}
D.~Mesterh{\'a}zy, J.H.~Stockemer, L.F.~Palhares and J.~Berges, \emph{Dynamic
  universality class of model c from the functional renormalization group},
  \href{https://doi.org/10.1103/PhysRevB.88.174301}{\emph{Physical Review B}
  {\bfseries 88} (2013) 174301}
  [\href{https://arxiv.org/abs/1307.1700}{{\ttfamily 1307.1700}}].

\bibitem{Israel:1979wp}
W.~Israel and J.M.~Stewart, \emph{{Transient relativistic thermodynamics and
  kinetic theory}},
  \href{https://doi.org/10.1016/0003-4916(79)90130-1}{\emph{Annals Phys.}
  {\bfseries 118} (1979) 341}.

\bibitem{Kovtun:2012rj}
P.~Kovtun, \emph{{Lectures on hydrodynamic fluctuations in relativistic
  theories}}, \href{https://doi.org/10.1088/1751-8113/45/47/473001}{\emph{J.
  Phys. A} {\bfseries 45} (2012) 473001}
  [\href{https://arxiv.org/abs/1205.5040}{{\ttfamily 1205.5040}}].

\bibitem{Bonart:2012}
J.~Bonart, L.F.~Cugliandolo and A.~Gambassi, \emph{{Critical Langevin dynamics
  of the $O(N)$ Ginzburg--Landau model with correlated noise}},
  \href{https://doi.org/10.1088/1742-5468/2012/01/P01014}{\emph{J. Stat. Mech.}
  (2012) P01014} [\href{https://arxiv.org/abs/1109.4107}{{\ttfamily
  1109.4107}}].

\bibitem{WangHeinz2002}
E.~Wang and U.~Heinz, \emph{Generalized fluctuation-dissipation theorem for
  nonlinear response functions},
  \href{https://doi.org/10.1103/PhysRevD.66.025008}{\emph{Phys. Rev. D}
  {\bfseries 66} (2002) 025008}
  [\href{https://arxiv.org/abs/hep-th/9809016}{{\ttfamily hep-th/9809016}}].

\end{thebibliography}\endgroup

\end{document}